\documentclass[preprint, aps,amsmath,amssymb,longbibliography,prfluids]{revtex4-2} 
\usepackage{graphicx}
\usepackage{float}
\usepackage{hyperref}
\hypersetup{colorlinks=true, citecolor=blue, urlcolor=blue, linkcolor=blue}

\begin{document}

\title{Interface instability of two-phase flow in a three-dimensional porous medium}

\author{Joachim Falck Brodin$^1$}
\email[]{j.f.brodin@fys.uio.no}
\author{Kevin Pierce$^{1}$} 
\author{Paula Reis$^1$}
\author{Per Arne Rikvold$^{1,3}$}
\author{Marcel Moura$^1$} 
\author{Mihailo Jankov$^{1}$}
\author{Knut J\o{}rgen M\aa{}l\o{}y$^{1,2}$}
\affiliation{$^1$PoreLab, The NJORD Centre, Department of Physics, University of Oslo, 0371 Oslo, Norway.\\
$^2$PoreLab, Department of Geoscience and Petroleum, Norwegian University of Science and Technology, 7034 Trondheim, Norway.\\
$^3$Department of Physics, Florida State University, Tallahassee, FL 32306-4350, USA}

\date{\today}
\begin{abstract}

\noindent We present an experimental study of immiscible, two-phase fluid flow through a three-dimensional porous medium consisting of randomly-packed, monodisperse glass spheres.
Our experiments combine refractive-index matching and laser-induced fluorescence imaging to resolve the morphology and stability of the moving interface resulting from the injection of one fluid into another.
The imposed injection rate sets a balance between gravitational and viscous forces, producing interface morphologies which range from unstable collections of tangled fingers at low rates to stable sheets at high rates.
The image data are complemented by time-resolved pressure measurements.
We develop a stability criterion for the fluid interface based on the analysis of the 3D images and the pressure data.
This criterion involves the Darcy permeability in each of the two phases and the time derivative of the pressure drop across the medium.
We observe that the relative permeability encountered by the invading fluid is modified by the imposed flow rate in our experiment, which impacts the two-phase flow dynamics.
We show that, in addition to the balance between the relevant forces driving the dynamics, local regions of crystalline order in the beadpack (crystallites) affect the stability of the invading front.
This work provides insights into how disorder on multiple length scales in porous media can interact with viscous, capillary, and gravitational forces to determine the stability and dynamics of immiscible fluid interfaces.
\end{abstract}

\maketitle
\section{Introduction}
The dynamics of multi-phase flow through porous materials is relevant for numerous topics in energy and environmental science, including carbon dioxide sequestration \cite{mcgrail2006,benson2005}, emerging fuel cell technologies \cite{anderson2010, weber2014}, and soil remediation efforts \cite{soga2004, seol2003}.
The invasion of one fluid (e.g., water, air, oil) into a porous material initially saturated with another immiscible fluid is of particular interest for its relevance to soil drainage \cite{toussaint2005, assouline2021} and hydrocarbon migration in the subsurface \cite{birdsell2015,molofsky2021}, among other topics.
While fluid invasion has long been studied in two-dimensional (2D) porous-media experiments, only recently have detailed three-dimensional (3D) imaging experiments of multiphase porous media flows become possible \cite{Harshani2017,Moroni2007,stohr2003,roth2015,Holzner2011,kang2010,Ovdat2006,sharma2011,datta2014,nascimento2019,Dalbe-Morphodynamics}.
Accordingly, while successful descriptions of fluid interface dynamics and stability have been developed from and successfully tested on the 2D experiments \cite{wilkinson1983invasion, glass1989wetting, flekkoy2002flow,birovljev1991gravity,meheust2002, birovljev1991gravity,frette1992a,Lovoll2005,toussaint2012}, these descriptions remain to be tested on 3D invasion data \cite{maaloy2021burst}.
In this study, we utilize the 3D optical scanner developed in \cite{brodin2019new,brodin2020visualization,brodin2022} to undertake a new study of fluid invasion in 3D porous media, with intent to (1) visualize the mechanisms by which interfaces destabilize in 3D and (2) evaluate to what extent existing theories, developed using data from 2D experiments \cite{maaloy2021burst}, can predict interface stability in our 3D experiments.

Fluid invasion describes the process by which a pressure gradient impels one fluid to displace another within a porous medium.
The stability and evolution of the resulting fluid interface depends on the balance of gravitational, viscous, and capillary forces, as well as wetting properties and any changes in the solid structure of the porous medium \cite{saffman1958penetration, glass1989wetting,  birovljev1991gravity,frette1992a,flekkoy2002flow,meheust2002,Lovoll2005,cinar2009experimental,toussaint2012}.
Most of the early experiments on fluid invasion imaged fluid displacement patterns in quasi-2D systems, such as etched glass networks \cite{lenormand1983, lenormand1985} and beads sandwiched between glass plates \cite{maloy1985,birovljev1991gravity} in a porous analogue of the Hele-Shaw cell \cite{Hele-Shaw1898}.
Other early studies imaged 2D slices through 3D displacement patterns \cite{van1957use,stokes1986,frette1990,frette1992a,frette1994}, at a time when computational power and imaging techniques did not allow detailed resolution of the full 3D fluid-invasion bodies.
These experiments produced fluid interfaces whose geometries varied with the applied pressure gradient from compact and non-fractal to ramified and fractal \cite{birovljev1991gravity, lenormand1988,maloy1985,meheust2002,holtzman2015}.
Depending on the balance of the controlling forces, the immiscible-fluid interfaces or ``fronts" can be either stable, with a relatively compact invasion structure and a width that stabilizes at a constant value, or unstable, with a width that grows continually and a fractal morphology, appearing as a collection of overlaid fingers \cite{auradou1999competition,maaloy2021burst,birovljev1991gravity,meheust2002, ayaz2020gravitational}.
The dimensionless fluctuation number $F$ was introduced in \cite{auradou1999competition} within a modified percolation theory of front geometry, and it weighs the relative importance of viscous and gravitational forces against the capillary pressure-threshold fluctuations to characterize front width and stability \cite{maaloy2021burst}.
Assuming that the width of an invasion front scales with the typical size of defending-fluid clusters, one predicts a relation between the front width and the fluctuation number, with an unstable interface for $F<0$ and a stable one for $F>0$ \cite{birovljev1991gravity,meheust2002, ayaz2020gravitational,maaloy2021burst, vincent-dospital2022}.
Although this fluctuation-number theory nicely describes invasion-front stability and width in 2D experiments, verifying whether the theory is easily adapted to describe invasion fronts from 3D experiments remains an important task \cite{maaloy2021burst}.

The combination of refractive-index matching and laser-induced fluorescence (RIM-LIF) imaging \cite{roth2021methods} has emerged in recent decades as a powerful method to visualize fluid flows at the pore scale in 3D systems \cite{Harshani2017,Moroni2007,stohr2003,roth2015,Holzner2011,kang2010,Ovdat2006,sharma2011,datta2014,nascimento2019,Dalbe-Morphodynamics}.
The RIM-LIF method provides lower cost and simpler implementation than alternative 3D imaging techniques, such as X-ray \cite{berg2013_complete,tekseth2024} and NMR tomography \cite{allen1997morphology,yan2012experimental}.
The method improves upon earlier RIM-based experiments that imaged plane projections of 3D fluid flows \cite{frette1990}.
Several works have applied RIM-LIF imaging to characterize two-phase flows through random glass beadpacks \cite{stohr2003, sharma2011, brodin2019new, brodin2020visualization, brodin2022}.
The fluid phases are dyed with different fluorescent compounds and index-matched to the solid.
Index matching minimizes light distortion by reflection and refraction, allowing a laser sheet to induce fluorescence in a thin slice.
Combining a sequence of these slices produces a 3D image from which the fluid and granular bodies can be segmented \cite{brodin2019new,brodin2020visualization}.

Using this method, Ovdat and Berkowitz studied drainage through beadpacks and analyzed the differences between experiments in 2D and 3D geometries \cite{Ovdat2006}.
They noted that, although the density and number of fingers protruding from invasion fronts scaled similarly with flow rate in both 2D and 3D, the 3D experiments showed considerable variation from one experiment to the next, even under otherwise identical conditions.
More recent studies have used the RIM-LIF method to study the formation and mobilization of trapped fluid clusters with confocal microscopy \cite{datta2014,nascimento2019}, and several works have evaluated single-phase flow properties, including local flow velocities \cite{Harshani2017, souzy2020} and the characteristics of scalar mixing \cite{heyman2020, souzy2020, heyman2021}.
Our earlier work details the 3D imaging and laser-scanning methodology as applied to two-phase flows \cite{brodin2019new, brodin2020visualization,brodin2022}.

In the present work, we apply RIM-LIF imaging to study the geometry and stability of an initially planar front progressing through a beadpack.
Experiments in which the front width reaches a quasi-constant value in time are considered \textit{stable}, while those in which the width grows continually over the resolved length- and time-scales are considered \textit{unstable}.
While our earlier work evaluates a point-source injection and thereby samples a wide range of front velocities as the front expands and slows down \cite{brodin2019new, brodin2020visualization, brodin2022}, this work evaluates a planar injection, selected to investigate the front width evolution as a function of global flow rate and to describe an expected transition from stable to unstable invasion.
We evaluate the stability of the immiscible fluid interfaces as they move during the invasion using time-resolved three-dimensional images of the dynamics.
Specifically, we consider a fluid with higher density and viscosity invading one with lower density and viscosity from the top. 
Depending on the flow rate, viscosity either stabilizes the front or gravity destabilizes it, analogous to the Rayleigh-Taylor instability \cite{taylor1950}.
 We regard the results in light of numerous studies of the interplay between capillary, viscous, and gravitational forces \cite{saffman1958penetration,wilkinson1984percolation,frette1990,birovljev1991gravity,frette1992a,frette1994, auradou1999competition,meheust2002,Lovoll2005, Ovdat2006,ayaz2020gravitational, breen2022}.
The image data are complemented by time series of the fluid pressures at the inlet and outlet of the experimental cell, giving a comprehensive picture of interface stability during 3D invasion flows.

The paper is organized as follows.
In Sec.~\ref{sec:met}, we describe the experimental setup, including methods to identify the porous medium and resolve both fluid phases through space and time.
In Sec.~\ref{sec:theory}, we present the theoretical context of the study, including the derivation of front stability in terms of competition between viscosity and gravity and perturbations to the moving interface from disorder in the porous medium.
In Sec.~\ref{sec:res}, we present the results, including measurements of front morphology and stability for different flow rates, front velocity characteristics, fluid-pressure differences, and porous-geometry characteristics.
We develop theoretical relations between the front stability and the flow rate and pressure time series, and we compare these relations with the experimental data.
In Sec.~\ref{sec:disc}, we discuss the observed front stability characteristics and mechanisms of front instability in context of earlier work.
Finally, in Sec.~\ref{sec:conc}, we summarize the key results and present our conclusions.

\section{Experimental methods}\label{sec:met}
\subsection{Three-dimensional imaging system}
\begin{figure*}
	\centering
	\includegraphics[width=\linewidth]{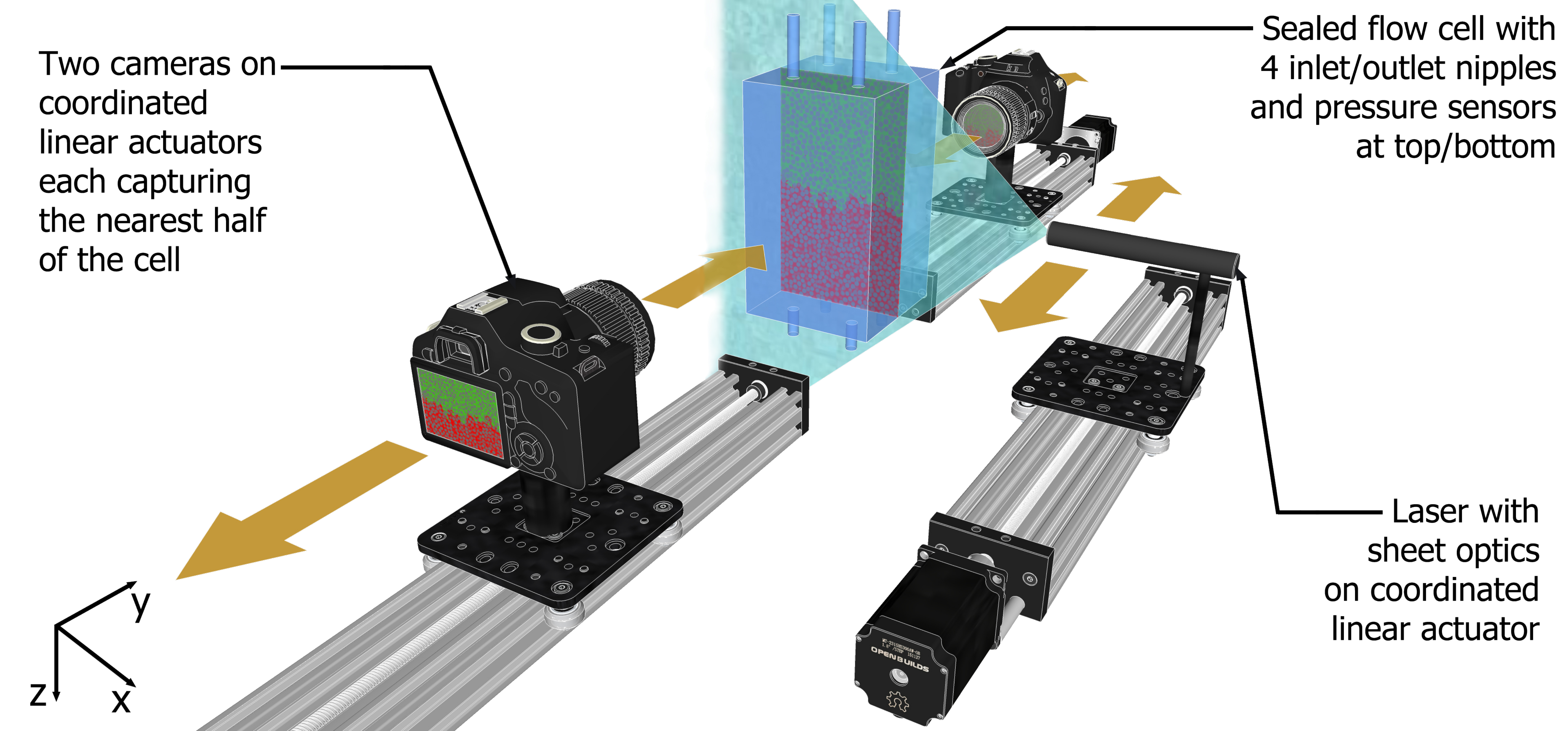}
	\caption{Illustration of the 3D scanner. The central flow cell contains the index-matched fluids and solid. The cell appears transparent in normal light, but when illuminated with a laser sheet, the dyed fluids fluoresce (red/green) and distinguish the two fluids and the solid. Linear actuators move the laser at a velocity $v$ and the cameras at velocity $v n$, where $n$ is the matched refractive index. This velocity difference compensates for the changed optical path length. During scanning, a sequence of 2D images is accumulated into a 3D image. To reduce optical distortion from imperfect refractive-index matching, the cell is imaged from both sides, with each camera imaging its closer half.}
	\label{fig:diagram_set-up}
\end{figure*}
The experiments were imaged in 3D using an optical scanner based on two  cameras and a co-moving laser (Fig.~\ref{fig:diagram_set-up}), see also the Supplemental Video \cite{supMat}.
The laser (40 mW, Z-Laser) produces a 2D vertical sheet of wavelength 532 nm which is scanned through the sample by a linear actuator.
The cameras (8-bit RGB, 2.3 MP, Daheng) on separate actuators follow the movement and map the cell by imaging at a rate of 49 frames per second, plane by plane.
Because the laser light attenuates through the cell due to absorption and imperfect refractive index matching, only one camera is active at a given time during a scan.
The active camera switches half-way through a scan, with each camera imaging only its nearest half of the cell.
The laser light is screened from the resulting images by notch filters on both cameras ($\lambda = (533 \pm 17)$ nm, Thorlabs) to isolate the fluorescence signal.
The timing of the actuators and cameras is organized so that captured voxels have equal resolution in all three spatial dimensions, with $(71\pm2)$ $\mu$m on each side.
The resulting 3D images measure $1200\times 1200\times 1920$ voxels, which we crop to a $1024\times 1024\times 1800$ region of analysis, corresponding to 73 mm  by 73 mm by 129 mm, spaced about one bead diameter from the cell sidewalls.
Each scan has an integration time of approximately 24 seconds, which allows displacement during scanning by typically 1.7 bead diameters (comparable to a single pore) at the highest flow rates we consider. 

\subsection{Porous-medium preparation and fluid characteristics}

The porous medium was prepared as a random packing of borosilicate glass beads of $a = (3.0\pm0.3)$ mm diameter, with refractive index $n=1.47$ (Sigma-Aldrich).
To approximately match the refractive index of these beads and obtain transparency for the imaging, rapeseed oil and glycerol were selected as the two immiscible experimental fluids, with the rapeseed oil (lower viscosity and density) serving as the defending fluid, and glycerol (higher viscosity and density) serving as the invading fluid.
Relevant properties of the fluids are summarized in Table \ref{table:fluidMeas}.
The oil was dyed with 1 mg/l of pyrromethene, while the glycerol was dyed with 20 mg/l of fluorescein (both from Luxottica-Exciton).
Because these dyes have emission peaks at different respective wavelengths 650 nm and 548 nm, the two dyed fluids can be distinguished by color.

\begin{table}[H]
\caption{Reference values and experimental data for refractive index $n$, fluorescence wavelength $\lambda_f$, density $\rho$, viscosity $\mu$, and interfacial tension $\gamma$ of the two dyed fluids. $T$ is the temperature measured in $^\circ$C, as both fluids have temperature-dependent refractive index and viscosity. The viscosity's dependence on temperature was experimentally determined with a temperature-controlled rheometer with one-degree intervals, from 17$^\circ$C to 25$^\circ$C, summarized by fitting the tabulated linear relationships over the data points.}
\centering
\begin{tabular}{|c|c|c|}
\hline
Reference data         &Glycerol (G)      &Rapeseed oil (RO)      \\
\hline
$n$ &    1.46-1.48   &    1.47    \\
$\lambda_f$ & 548 nm & 650 nm \\
\hline
Experimental data&&\\
\hline
$\rho$       &(1.26 $\pm$ 0.02) g/cm$^3$                   &(0.91 $\pm$ 0.02) g/cm$^3$ \\
$\mu$ &   $(3648-113 T /^\circ C \pm$ 200) mPa$\cdot$s                    & $(133-3  T /^\circ C \pm$ 7) mPa$\cdot$s \\
$\gamma$ (G vs RO)     & (16.4 $\pm$ 1.0) mN/m               & (16.4 $\pm$ 1.0)  mN/m\\
\hline
\end{tabular}
\label{table:fluidMeas}
\end{table}

Experiments were prepared by first filling the flow cell with the defending oil, then establishing a random beadpack by pouring glass spheres into the cell while lightly stirring to remove any air bubbles.
The cell was then sealed and the tubing was purged of air in preparation for invading glycerol injection and 3D imaging of the resulting invasion dynamics.

In our two-phase flow experiments, the interfacial tension between the two fluids and the wetting characteristics of the two fluids on the solid material controls the capillary pressures required to invade the individual pores, so these factors could affect the invasion front morphology.
With a tensiometer we measured the fluids' interfacial tension to be $\gamma=(16.4\pm 0.4)$ mN/m, and found neutral fluid-solid wetting angles for both experimental fluids.
In trial experiments, close inspection of the liquid–liquid interface showed that both glycerol-displacing-oil and oil-displacing-glycerol events produced meniscus profiles characteristic of drainage, suggesting that the displacing fluid behaved as the non-wetting phase in both cases.
We therefore conclude that neither fluid exhibited a clearly dominant wetting tendency and that the system exhibits near-neutral wettability, with conditions remaining consistent throughout the experiments.
Given this neutral wettability condition, we chose not to emphasize wetting effects in the analysis that follows, focusing instead on the balance between viscous and gravitational forces at the invasion front.
However, for a comprehensive understanding of invasion morphologies, wettability remains an important factor \cite{stokes1986,zhao2016}.

\begin{figure}[H]
	\centering
	\includegraphics[width=\linewidth]{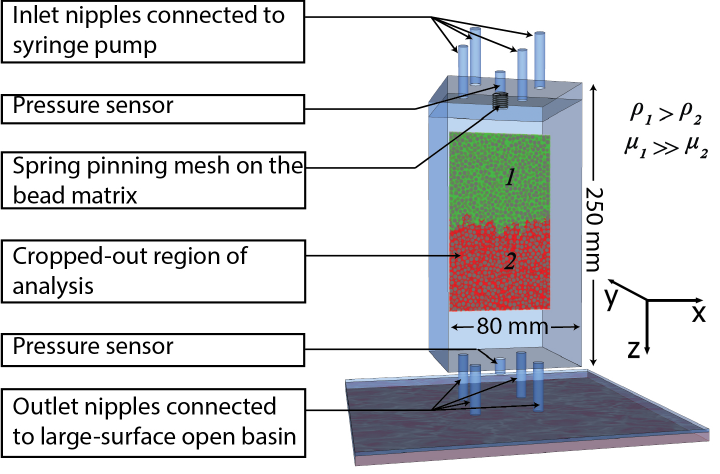}
	\caption{Flow geometry and boundary conditions. The cell is fully sealed, except for the inlet and outlet. Syringe pumps maintain a fixed flow rate at the inlet, while the large outlet basin maintains a quasi-fixed pressure at the outlet. The cell volume from the base to the mesh screen is filled with glass beads. For the image analysis, we cropped out a box with base 73$\times$73 mm$^2$ and height 129 mm, with the top 60 mm below the cell lid.}
	\label{fig:cell_set-up}
\end{figure}

\subsection{Description of the flow cell}

The flow cell was constructed of 6 mm thick PMMA sheets, solvent-welded into a transparent box with a removable lid.
The cell is illustrated in Fig.~\ref{fig:cell_set-up}.
The beadpack extends nearly 10 mm from the top of the cell, where the beads contact a fine steel mesh.
The mesh is loaded with a coil spring that pins against the cell lid.
This mechanism presses the beadpack to ensure that the solid phase does not move during the experiment.
The lid seals with a rubber gasket and bolts to the base on threaded rods.
Pressure sensors (Honeywell - 6PCAFG6G) record the absolute pressures at the top and bottom of the cell.
The lid and outlet each have four flow ports and an additional port for a pressure sensor (see Fig.~\ref{fig:cell_set-up}).
At the lid, the flow ports are connected to two syringe pumps (Harvard PhD Ultra) driving four 140 ml syringes, together providing a total flow rate $Q$.
To maintain a nearly constant outlet pressure, the outlet ports are submerged in rapeseed oil in a relatively wide 30 cm by 40 cm basin.
The basin level changed by at most 3 mm during the experiments, giving outlet pressure variations of up to 30 Pa, depending on the amount of fluid injected throughout the experiment.
Assuming that these pressures equilibrate slowly across the cell, this provides maximum errors in the measured pressure gradients near 10\%.
Room temperatures were measured during the experiments with a table-top thermometer, placed beside the fluid cell.
In several experiments, we used a second probe inside the cell to confirm that the fluid and room temperatures coincided well, always within 0.5$^\circ$ C.
Therefore, we use the room temperature as a proxy for the fluid temperatures inside the cell.

\subsection{Experimental protocol}
\label{sec:protocol}

After preparation of a beadpack immersed in the defending fluid, a preliminary 3D scan was taken from which the solid phase was segmented.
The two-phase flow was then initiated by injecting the invading glycerol fluid at $Q=30$ ml/min, which is the highest flow rate the pumps could sustain.
This initially filled the gap above the mesh screen with glycerol, while maintaining a flat front (see Fig.~\ref{fig:cell_set-up}).
When the gap was fully saturated, glycerol permeated the mesh and invaded downward in a flat front that horizontally spanned the cell.
Once the flat front became fully visible in the imaging region, the flow rate was decreased to the rate chosen for the given experiment, and the imaging sequence was started.
Depending on the chosen flow rate, the scanner was set to one- or two-minute imaging intervals.
Each experiment was stopped when the invasion front first left the imaging region, producing between nine and forty-five 3D images.
The imaging rate was selected to adequately represent front evolution and stability without exceeding the available data storage.

\subsection{Image segmentation of fluid and solid phases}

The solid and fluid phases were segmented by first locating the solid beads and then isolating the invading phase.
The defending phase was segmented as the remainder of the volume.
This process was conducted with the Amira Avizo software, but it corresponds to generic image-analysis steps found in, e.g., openly accessible Python libraries \cite{bradski2000opencv,dey2018hands}.
The segmentation requires special treatment because gradients in light intensity from laser attenuation in the medium and imperfect refractive-index matching modify the intensity across the images, especially near the surfaces of individual beads.

The solid phase was localized from the initial image of the medium fully saturated with the defending fluid, where the beads appeared as dark patches.
A background intensity image was formed by blurring the initial 3D scan over a length scale much larger than the bead diameter.
The initial image was subtracted by this background.
A simple threshold procedure then extracted the solid-phase segmentation map, with some uncertainty caused by imperfect refractive-index matching.
The solid-phase map was refined by isolating bead centroids with Avizo's blob-finding algorithm, then drawing new sphere bodies at these locations.
We have checked that the refined maps contain only a small fraction of spheres ($<1$ \%) which overlap appreciably and represent beadpacks with realistic bulk geometric properties (see Sec.~\ref{sec:result-beadpack}).

The invading fluid was segmented by applying a gradient-map adjusted threshold to the green color channel of the images.
Fluorescence from both the green invading and red defending fluids entered the image green channel, but the invading fluid induced a much stronger response.
The gradient and threshold parameters were manually adjusted in Avizo.
The parameters were then entered into a Python script that creates a 3D thresholding matrix, with linear gradients in the $x$ and $y$ directions.
This matrix filtered the green channel and produced the invading fluid segmentation map.
This map was refined by subtracting its intersection with the solid phase, as green color from the dye tends to bleed into bead margins.
The remaining volume in the cell, with the beads and the invading fluid subtracted, was then defined as the defending fluid.
The entire segmentation process is described in greater detail in \cite{brodin2019new}.

\subsection{Characterizing the fluid invasion front}\label{sec:frontChar}
\begin{figure}[H]
	\centering
	\includegraphics[width=\linewidth]{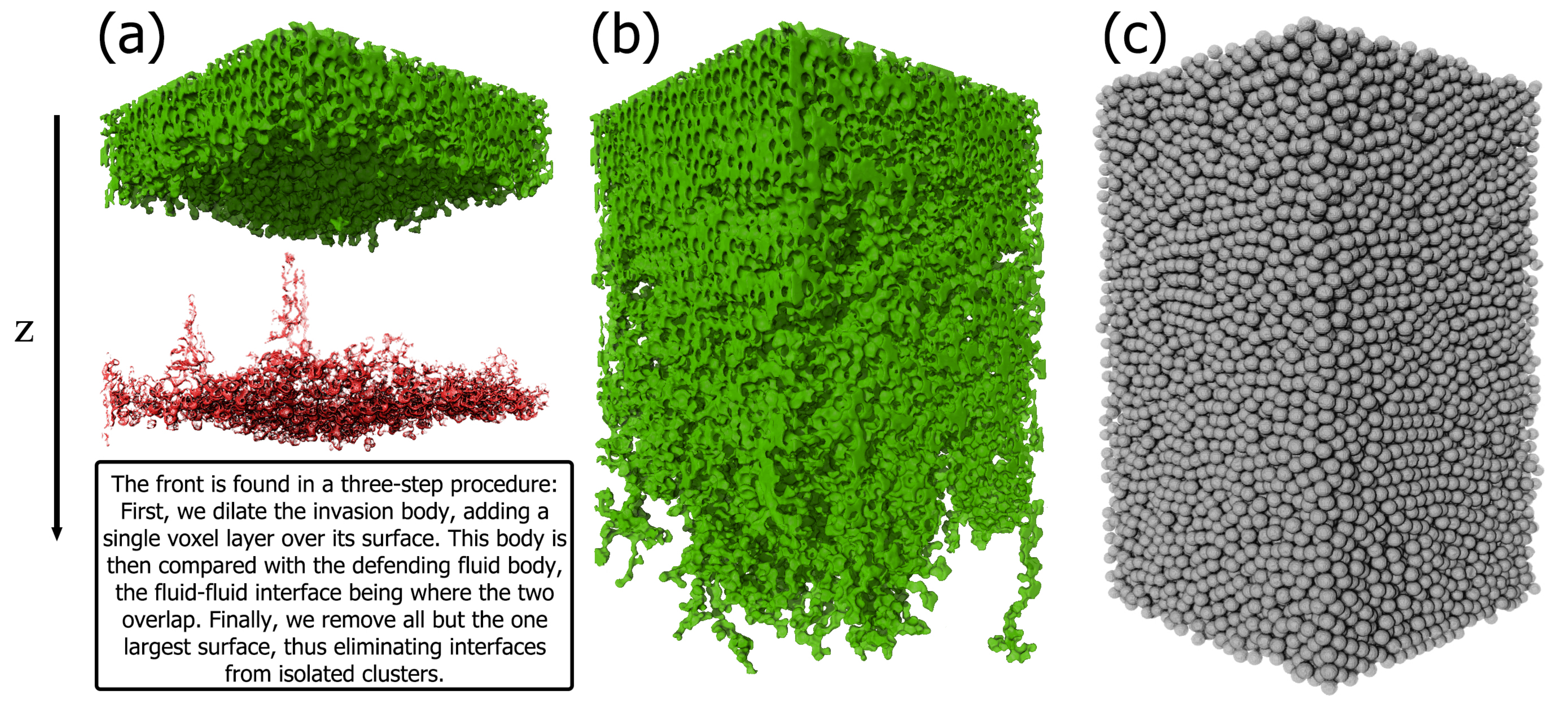}
	\caption{Images from Exp. A ($Q=2$ ml/min) show the segmentation bodies and the evolution from an initially planar front to an unstable, increasingly wide front. In (a), the front (in red) from an early image of the experiment is artificially shifted below the invasion body for visualization. Upward-trailing regions of pinned fluid are visible. In (b), the last image before the front leaves the imaging region is shown, with well-developed fingers extending from the main invasion body. In (c), we show a rendering of the porous medium.}
	\label{fig:evolution}
\end{figure}
Figure \ref{fig:evolution} shows 3D renderings of the segmented invading and solid phases from one experiment (Exp. A, with $Q=2$ ml/min).
The method used to extract the invasion front dividing the fluid phases is described in the text box of Fig.~\ref{fig:evolution}.
In all experiments, the invasion fronts appear as perforated surfaces with holes surrounding bead contacts and extensions that point both upstream and downstream.

We characterize the invasion front and its stability using the ``front distribution", formed as the probability distribution of the vertical coordinates $Z$ of the front voxels [red voxels in Fig.~\ref{fig:evolution}(a)].
In our 3D experiments, front distributions can be highly skewed, as we detail below.
We therefore characterize the front evolution using metrics which include contributions from the extreme tails of the front distributions.
The front width $W$ is a key quantity, defined as the 90\% percentile ($Z_{90}$) minus the 10\% percentile ($Z_{10}$) of front distribution:
\begin{equation}
W=Z_{90}-Z_{10}.
\label{eq:frontWidt}
\end{equation}
This measure resembles the standard deviation of the $z$-coordinates of front voxels, often referred to as the \textit{RMS-roughness} \cite{chakrapani2003scaling}, except the use of quantiles better represents skewed distributions.
We also consider width contributions from ahead and behind the front, represented by the components of $W$ above and below the distribution's median $Z_{50}$:
\begin{align}
    W_+ &= Z_{90}-Z_{50} \label{eq:wp}\\
    W_- &= Z_{50}-Z_{10} \label{eq:wm}.
\end{align}
Due to the equality $W_-+W_+=W$, the proportions $W_+/W$ and $W_-/W$ summarize the respective relative amounts of forward and backward skew in the front distribution.

\section{Theoretical criterion for front instability} \label{sec:theory}
Here we present a first-order prediction, modified from \cite{brodin2022}, for the ``critical flow rate" $Q_c$ corresponding to the transition between stable, viscosity-dominated invasion and unstable, gravity-dominated invasion.
Our prediction involves a two-phase Darcy theory valid only for stable invasion, when the front can be approximated as a sharp interface.

For a single fluid flowing at a rate $Q$, the pressure field can be approximated by Darcy's law in integral form:
\begin{equation}
p(z) = p(0)+\left(\rho g-\frac{\mu Q}{\kappa A}\right)z.
\label{eq:Darcy1}
\end{equation}
Here, $p(z)$ is the pressure at depth $z$, $\rho$ is the fluid density, $g$ is the gravitational acceleration, $\mu$ is the dynamic viscosity, $A$ is the cross-sectional area, and $\kappa$ is the permeability.
In this equation, $z$ increases downward, parallel to $\vec{g}$ (see Fig.~\ref{fig:cell_set-up}).
The local pressure $p(z)$ has both gravitational and viscous contributions, represented in the respectively first and second terms in brackets in Eq.~(\ref{eq:Darcy1}).

Now we consider a two-phase configuration in which one fluid invades a volume initially saturated with another.
We consider an initially flat front at $z=z_0$ and perturb one region of it by a small distance $a$, comparable to the size of an individual pore.
Evaluating the minimal flow rate for which this perturbation grows provides an estimate of $Q_c$.
Using Eq.~(\ref{eq:Darcy1}), the pressure changes in the invading fluid (1)  due to the perturbation can be expressed as
\begin{equation}
\Delta p_1=p_1(z_0+a)-p_1(z_0)=\left(\rho_1 g-\frac{Q \mu_1}{A \kappa_{TP}(Q)}\right)a 
\end{equation}
The corresponding pressure change in the displaced fluid (2) of an unperturbed interface is \begin{equation}
\Delta p_2=p_2(z_0+a)-p_2(z_0)=\left(\rho_2 g-\frac{Q \mu_2}{A \kappa_{SP}}\right)a \; .
\label{eq:pressurechanges}
\end{equation}
Here $\kappa_{SP}$ and $\kappa_{TP}(Q)$ are the respective single- and two-phase permeabilities \cite{brodin2022}.
Here, we allow $\kappa_{TP}(Q)$ to explicitly depend on the flow rate $Q$, reflecting the expectation that clusters of defending fluid trapped behind the main invading front will effectively reduce the permeability for the invading phase, meaning that, in general, $\kappa_{TP}(Q) < \kappa_{SP}$ \cite{feder_flekkoy_hansen_2022, tallakstad2009steady,tallakstad2009sim}.

The increased pressure acting on the fluid-fluid interface induced by the perturbation at a given flow rate $Q$ is $\Delta p^d(Q) = \Delta p_1 - \Delta p_2$, or using Eq.~(\ref{eq:pressurechanges}),
\begin{equation}
    \Delta p^d(Q) = \left[(\rho_1-\rho_2) g-\frac{Q}{A}\left(\frac{\mu_1}{\kappa_{TP}(Q)}-\frac{ \mu_2}{\kappa_{SP}}\right)\right]a \; .
    \label{eq:unstableq}
\end{equation}
If the net pressure $\Delta p^d(Q)$ is positive, any small displacement will be amplified, resulting in instability. In contrast, if $\Delta p^d(Q)$ is negative, the perturbation will be dampened. This holds true regardless of the surface tension value.

This result can be rewritten in terms of an effective capillary number, $\mathrm{Ca}_\text{eff}$ and Bond number, $\mathrm{Bo}$ as
\begin{equation}
    \Delta p^d(Q) = \frac{\gamma}{a} \left[ \mathrm{Bo} - \mathrm{Ca}_\text{eff}(Q) \right],
    \label{eq:Bo_Caeff}
\end{equation}
where
\begin{equation}
    \mathrm{Bo} = \frac{a^2 (\rho_1 - \rho_2) g}{\gamma} \quad\text{ and }\quad    \mathrm{Ca}_\text{eff}(Q) = \frac{a^2}{\gamma} \cdot \frac{Q}{A} \left( \frac{\mu_1}{\kappa_{TP}(Q)} - \frac{\mu_2}{\kappa_{SP}} \right).
\end{equation}
\noindent In this formulation, we expect the transition from stable to unstable to occur when \mbox{$\mathrm{Ca}_\text{eff}=\mathrm{Bo}$}.

In our experiments, because the invading fluid (1) has higher viscosity and density than the defending fluid (2), $\Delta p^d$ increases by the gravitational term and decreases by the viscous term.
The perturbation is unstable and tends to grow when gravity dominates, $\Delta p^d(Q) > 0$, and conversely it is stable and tends to shrink when viscosity dominates, $\Delta p^d(Q) < 0$.
The rate $Q_c$, below which fronts destabilize, therefore follows from $\Delta p^d(Q_c)=0$.
Noting that $\mu_1\gg \mu_2$ (see Table \ref{table:fluidMeas}), we obtain the condition
\begin{equation}
    \frac{Q_c}{\kappa_{TP}(Q_c)} \approx  (\rho_1-\rho_2)\frac{gA}{\mu_1}
    \label{eq:qccondition}
\end{equation}
for the critical flow rate $Q_c$.
Fronts are expected to be unstable for all $Q \lesssim Q_c$, where $Q_c$ is the solution to Eq.~(\ref{eq:qccondition}). 
This result aligns with the findings from the linear stability analysis conducted by Saffman and Taylor
\cite{saffman1958penetration} and Chuoke et al. \cite{chuoke1959instability}. 
In Sec.~\ref{sec:result-permeability}, we study the dependence of $\kappa_{TP}(Q)$ on $Q$ to predict $Q_c$ from Eq.~(\ref{eq:qccondition}), and we compare the prediction with the observed front-stability trends. In Sec. \ref{sec:disc-2dtheory} we discuss the theory further, and expand on how the capillary threshold distribution affects the front width.

\section{Results}\label{sec:res}

We conducted 17 experiments with flow rates ranging from 2 to 30 ml/min.
For each experiment, we segmented the porous solid, tracked the defending and invading fluid bodies through time, and documented the invasion-front dynamics.
We also recorded pressure time series at the top and bottom of the cell.
While we controlled the flow rate of each experiment, the detailed structure of the porous medium differed from one experiment to the next.
The temperature of the experimental fluids also varied between trials, which affected the fluid viscosities (see Table \ref{table:fluidMeas}).
Key parameters and results from the experiments are listed in Table~\ref{table:expSummary}.
Videos of experimental invasion fronts and graphics of the key results can viewed in the Supplemental Material \cite{supMat}.

\begin{table}[H]
\centering
\caption{Summary of experimental results. The \textit{Stability} column indicates whether the front was classified as stable or unstable for each experiment.
Experiments were considered unstable when the best-fit slope of $W/a$ versus the injected volume exceeded $0.005$ ml$^{-1}$, see Sec.~\ref{sec:res-front}.
Our classification is largely insensitive to this threshold, although changes could modify the assessed stability of some experiments.
The ``Crystal fraction" represents the proportion of beads participating in crystallites in each experiment, as described in Sec.~\ref{sec:result-beadpack}.
$\mathrm{d}(p_b-p_t)/\mathrm{d}t$ is the averaged time derivative of the pressure difference across the cell, as described in 
Sec.~\ref{sec:res-pressure}, with errors estimated as the maximum possible rate-of-change of hydrostatic pressure in the outlet basin ($\rho_2 g Q/A_b$, with $A_b$ the basin area).
Pressure measurements were not performed for Exps.~C and G, denoted by hyphens.}

\begin{tabular}{|c|c|c|c|c|}
\hline
\textbf{ID} & \textbf{$Q$ [ml/min]} &  \textbf{Stability} & \textbf{Crystal fraction} & \textbf{$\mathrm{d}(p_b-p_t)/\mathrm{d}t$ [Pa/s]} \\
\hline
A & 2.0  & Unstable & 0.031 & $0.0132 \pm 0.0025$\\
B & 4.0  & Unstable & 0.030 & $0.005 \pm 0.005$\\
C & 5.0  & Stable & 0.056 & -\\
D & 5.0  & Unstable & 0.079 & $ -0.013 \pm 0.006$\\
E & 7.0  & Stable & 0.022 & $0.059 \pm 0.009$\\
F & 7.0  & Unstable & 0.067 & $-0.062 \pm 0.009$\\
G & 7.0  & Stable & 0.017 & -\\
H & 8.0  & Unstable & 0.080 & $0.004 \pm 0.006$\\
I & 8.0  & Unstable & 0.040 & $-0.04 \pm 0.10$\\
J & 10.0  & Stable & 0.014 & $-0.132 \pm 0.012$\\
K & 10.0  & Stable & 0.027 & $-0.193 \pm 0.012$\\
L & 15.0  & Stable & 0.014 & $-0.435 \pm 0.019$\\
M & 15.0  & Stable & 0.048 & $-0.540 \pm 0.019$\\
N & 20.0  & Stable & 0.045 &$ -0.833 \pm 0.025$\\
O & 20.0  & Stable & 0.034 & $-1.003 \pm 0.025$\\
P & 30.0  & Stable & 0.014 & $-1.603 \pm 0.037$\\
Q & 30.0  & Unstable & 0.37 &  $-2.150 \pm 0.037$\\ \hline
\end{tabular}
\label{table:expSummary}
\end{table}

\subsection{Front instability: contributions of fingering and pinning}
\label{sec:res-front}
\begin{figure}[H]
	\centering
	\includegraphics[width=\linewidth]{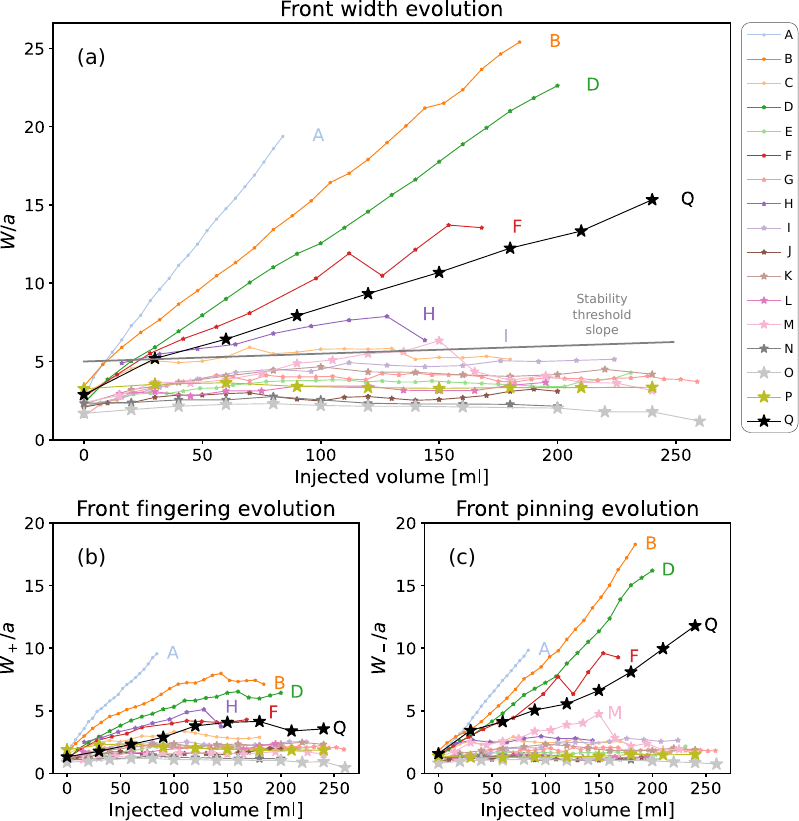}
	\caption{Plots of the front width $W$ (a), forward front width $W_+$ (b), and backward front width $W_-$ (c), all as functions of injected volume (normalized time). The measures are scaled by the bead diameter, $a=3.0$ mm. The plots increase in flow rate with the lettering and symbol size, starting with 2 ml/min for Exp. A, and ending at 30 ml/min for Exp. Q. Continually growing front widths indicate a gravitationally-unstable experiment and correlates with low flow rates. The plotted gray line in (a) indicates the threshold slope for instability, set at $W/a=0.005$ ml$^{-1}$.
 }
	\label{fig:wFingersPinning}
\end{figure}

We classified each experimental invasion front as stable or unstable, using the width time series $W(t)$ shown in Fig.~\ref{fig:wFingersPinning}(a).
Each series $W(t)$ was plotted against the cumulative injected fluid volume and fit with a linear model for the last 85\% of data points [see Figure \ref{fig:wFingersPinning}(a)].
Experiments were considered unstable if the best-fit slope of $W/a$ versus the injected volume exceeded a threshold, chosen at $0.005$ ml$^{-1}$.
This identified experiments A, B, D, F, H, I, and Q as unstable, indicated in the $Stability$ column of Table~\ref{table:expSummary}.
The stability classification was found relatively insensitive to the threshold, although large-enough changes could reclassify several experiments.
As expected, measured front widths $W$ were generally stable for high flow rates, with $W$ taking on values between 2 and 5 bead diameters, while front widths increased across the entire experiment duration for most of the lower flow rate experiments, reaching values in the range of $5$ to $30$ bead diameters by the end of these experiments.
All experiments with flow rates above $Q \approx 8$ ml/min were stable with the exception of Exp.~Q.
Most experiments with flow rates below $Q \lesssim 8$ ml/min were unstable, although Exps.~C, E, and G provide exceptions.
Taken together, these image analysis results suggest a critical flow rate, $Q_c$, near 8 ml/min.
\begin{figure}[H]
	\centering
	\includegraphics[width=1\linewidth]{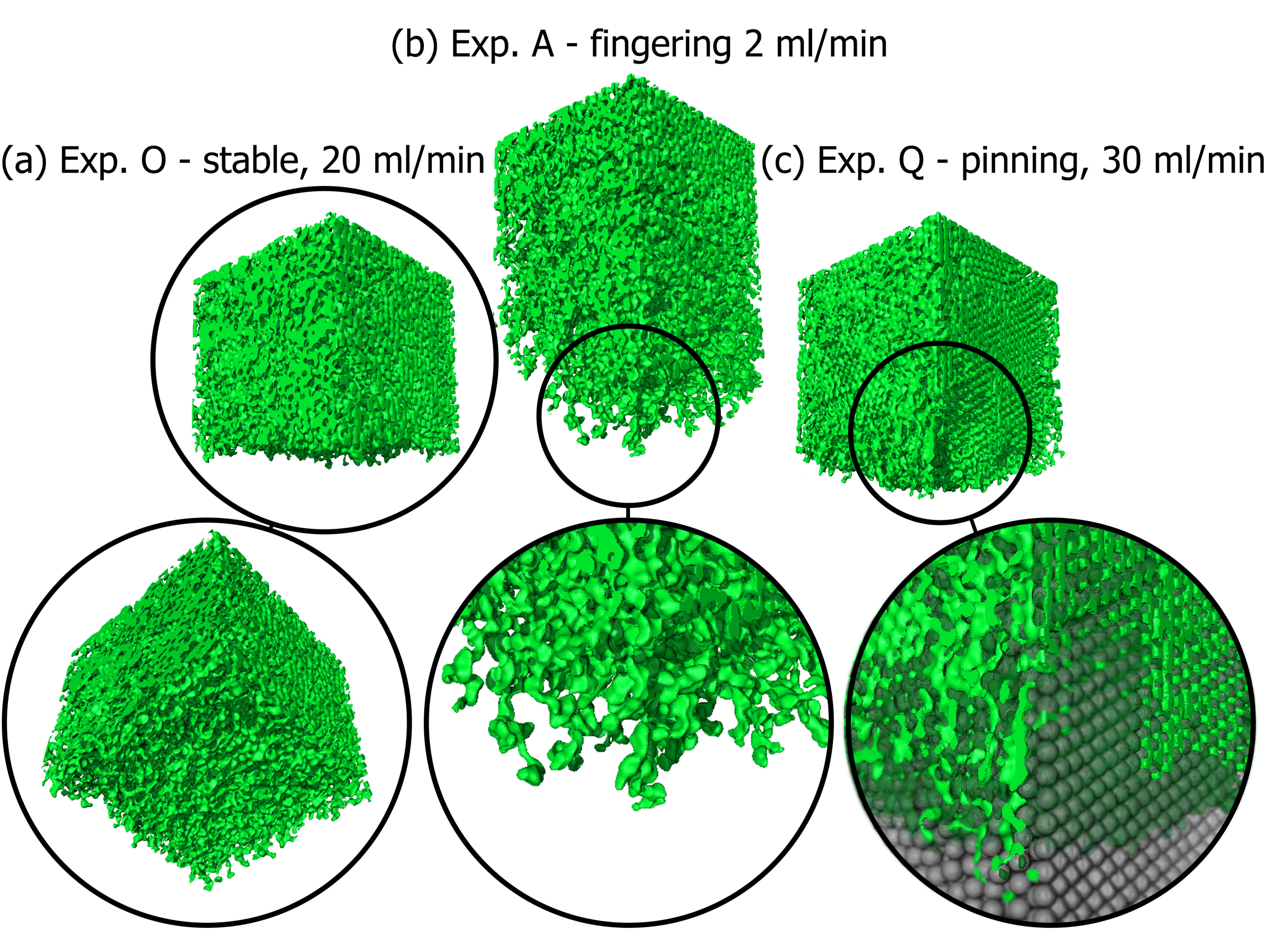}
	\caption{3D renderings from three experiments indicating the two main modes by which fronts destabilize. In (a), at a high flow rate, we have only small perturbations in the front, corresponding to a stable experiment with a relatively small front width (see Fig.~\ref{fig:wFingersPinning}). In (b), at a low flow rate, visible fingers shoot ahead of the bulk of the front, corresponding to an unstable growth (see Fig.~\ref{fig:wFingersPinning}), and forward skew in the front distribution (see Fig.~\ref{fig:distributions}). In (c), also at a high flow rate, we see large sections of ordered beads that the invading flow bypasses. The resulting backward-lagging regions of invading fluid are characteristic of pinning. They produce a high front width (Fig.~\ref{fig:wFingersPinning}) and contribute backward skew to the front distribution 
    (Fig.~\ref{fig:distributions}).}

    \label{fig:flatFingersPinning}
\end{figure}

The experimental images revealed two different mechanisms by which fronts destabilized, which we referred to as fingering and pinning (Sec.~\ref{sec:res-front} and Fig.~\ref{fig:flatFingersPinning}).
Fingering appears as narrow plumes of invading fluid protruding in the direction of the flow.
This instability originates from the interplay between the pressure gradients within the fluid phases due to viscous and gravitational forces, and the capillary-pressure thresholds for interface displacement through individual pores.
If a sufficiently large positive capillary pressure gradient forms along the fluid interface in the flow direction, fingering occurs, and this mechanism has long been understood to control front instability in 2D experiments \cite {maloy1985,maaloy2021burst,meheust2002, toussaint2005, vincent-dospital2022}.
Conversely, a negative capillary pressure gradient in the direction of flow stabilizes the invasion front and suppresses the development of fingering \cite{wilkinson1984percolation,birovljev1991gravity,moura2020,ayaz2020gravitational,maaloy2021burst,breen2022,vincent-dospital2022,reis2023,khobaib2025}.
 
\begin{figure}[H]
	\centering\includegraphics[width=\linewidth]{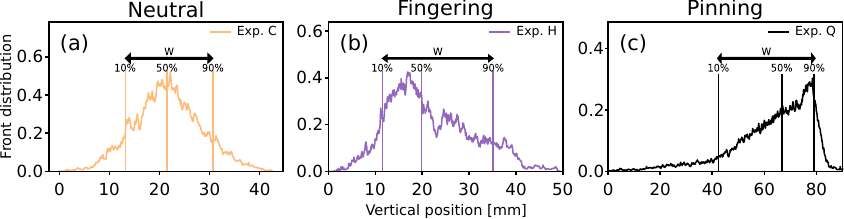}
	\caption{Examples of characteristic front distributions for neutral, fingering, and pinning, from experiments C (a), H (b), and Q (c), respectively. The plots are from the last recorded scan from each experiment.}
	\label{fig:distributions}
\end{figure}

We define pinning as the scenario in which the invading front bypasses a correlated region of the pore space characterized by unusually low permeability [see Fig.~\ref{fig:flatFingersPinning}(c)].
This occurs when the interface encounters an area with significantly smaller pores, where the local capillary pressure is typically insufficient to overcome the higher entry thresholds imposed by the tighter geometry. As a result, the front appears to become pinned in that region, while invasion proceeds preferentially through the surrounding wider pore throats.
Most of our unstable 3D experiments show some evidence of pinning, which we identify through the relatively large contribution of backward-skewed regions in the front distribution to the overall front width (see Fig.~\ref{fig:distributions}).
Previous numerical studies of gravity-driven invasion percolation have similarly demonstrated that introducing large-scale structural correlations in the porous medium can substantially alter front progression and the resulting residual fluid saturation after drainage \cite{ionnidis1996}.
In our case, pinning is associated with regions of correlated crystalline structures much larger than individual pores that impede the progression of the invasion front due to their higher capillary pressure thresholds.
See Sec.~\ref{sec:result-beadpack} for further discussion of the correlated crystalline structures.

When the front distribution is symmetrical, neither fingering nor pinning dominates, and we refer to the front distribution as neutral. For clarity, this does not imply that the front is stable, as the front width might still continue growing indefinitely despite having a symmetrical distribution.
Snapshots of representative neutral, fingered, and pinned fronts are displayed in Figs.~\ref{fig:flatFingersPinning}(a-c).

To quantify the relative influence of fingering and pinning on the observed front dynamics, we evaluate contributions to front-width growth by the positive and negative front widths, $W_+$ and $W_-$, defined in Eqs.~(\ref{eq:wp}) and (\ref{eq:wm}) and plotted in Fig.~\ref{fig:wFingersPinning}(b) and (c).
The different contributions of fingering and pinning to skew in the front width distributions are displayed in Fig.~\ref{fig:distributions}, where a stable front (Exp. C) is compared to fronts predominantly destabilized by fingering (Exp. H) and pinning (Exp. Q), respectively.
In the figure, the front distribution is scaled to show the proportion of front voxels at a given vertical position.

Fingering has long been associated in 2D experiments with unstable displacement processes \cite{maloy1985,chen1985pore,lenormand1990liquids,toussaint2005,meheust2002,vincent-dospital2022}.
To evaluate finger growth in our 3D experiments, we examine the forward front width $W_+$ of Eq.~(\ref{eq:wp}), which represents the amount of forward skew contributed by the growing fingers.
$W_+$ is plotted against the cumulative injected volume for all experiments in Fig.~\ref{fig:wFingersPinning}(b).
All unstable experiments show a rapid growth of fingers for about the first half of each experiment, and the growth of fingers is relatively faster for lower flow rates, consistent with their production by gravity and dissipation by viscosity.
However, the finger growth slows as time progresses, with finger growth stalling toward the end of most unstable experiments (Exps. B, D, and F in particular), even as their overall front widths $W$ continue to increase, Fig.~\ref{fig:wFingersPinning}(a, c).

The observed slowing of finger growth indicates that fingering has a relatively limited impact on interface instability at the late stages of our 3D experiments.
Figure~\ref{fig:wFingersPinning}(c) shows the backward width $W_-$ of Eq.~(\ref{eq:wm}) for all experiments, summarizing the contributions of pinning to the front distributions.
The growth of $W_-$ typically mirrors that of $W$ for all unstable experiments, indicating that growing regions of slow-moving interface pinned behind the main front are the leading source of front growth in unstable 3D experiments. 

\subsection{Velocity characteristics of invasion fronts}
\label{sec:velSec}

\begin{figure}[H]
	\centering
	\includegraphics[width=1\linewidth]{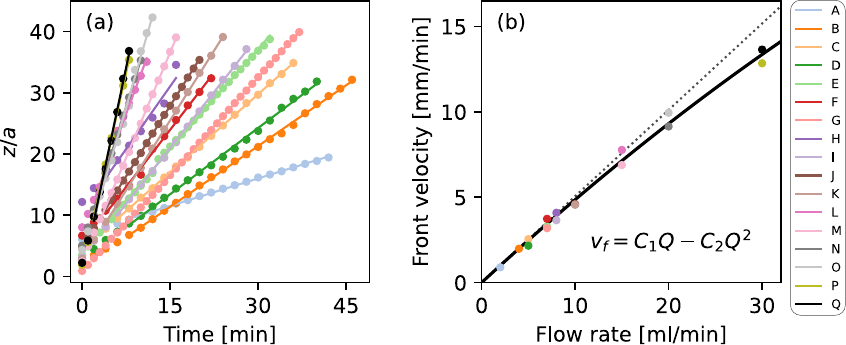}
	\caption{ Panel (a) displays median front positions versus time for all experiments, with a linear fit on the final 85\% of each time series for front velocity determination. Panel (b) displays the flow dependence of the front velocity alongside the best-fit trend $v_f = C_1 Q - C_2 Q^2$ derived in Eq.~(\ref{eq:C1C2}), with $C_1 = (0.506\pm 0.020)$  mm/ml and $C_2 = (0.0020 \pm 0.0008)$ min$\cdot$mm/ml$^2$.}\label{fig:frontVelocity}
\end{figure}

As we externally imposed the injection rate $Q$ of the invading fluid, fronts are expected to move downward with an average velocity determined by the flow rate and available pore volume.
However, due to the trapping of the defending fluid behind the front, the available pore volume for the invading phase is generally less than the total pore volume.
The velocity of fronts during stable invasion is then
\begin{equation}
v_f=\frac{Q}{A \phi\left[1-S(Q)\right]} \;,
\label{eq:sat}
\end{equation}
where $S(Q)$ is the residual saturation of the displaced fluid left behind the front.
Expanding the residual saturation $S$ around the critical flow rate $Q_c$ gives 
\begin{equation}
v_f \approx C_1 Q - C_2 Q^2\; , 
\label{eq:C1C2}
\end{equation}
where $C_1>0$ and $C_2>0$.
These coefficients relate to the flow-rate dependence of the residual saturation as detailed in the Appendix \ref{sec:appen}.

We determined the experimental front velocities by fitting a linear relation to the median front positions versus the time for each experiment, shown in Fig.~\ref{fig:frontVelocity}(a).
The final 85\% of each time series were used to avoid any artificial velocity changes as the front first enters the observation volume.
Errors in median front positions were estimated from the front width and number of front voxels, providing errors $< 1$ mm, which we incorporated into the fitting.

Front velocities for all experiments are plotted against flow rates in Fig.~\ref{fig:frontVelocity}(b), and the constants $C_1$ and $C_2$ found by fitting the measured front velocities to 
Eq.~(\ref{eq:C1C2}), providing $C_1 = (0.506\pm 0.020)$ mm/ml and $C_2 = (0.0020 \pm 0.0008)$ min$\cdot$mm/ml$^2$ with uncertainties calculated as the standard error of the fit parameters.
Despite large uncertainty in the $C_2$ coefficient, the quadratic model shows better agreement with the experimental data in Fig.~\ref{fig:frontVelocity} than a linear model.
This agreement suggests that the residual saturation introduces a weak non-linearity to the dependence of the front velocity on flow rate, in support of Eq.~(\ref{eq:C1C2}).
The weak dependence we observe is consistent with earlier observations that the residual saturation behind a 2D fluid displacement front depends only weakly on the applied pressure gradient \cite{maaloy2021burst}.

\subsection{Analysis of pressure readings}\label{sec:res-pressure}

The pressure readings from the top and bottom of the cell provide an alternative assessment of front stability.
\begin{figure}[H]
    \centering
    \includegraphics[width=0.5\linewidth]{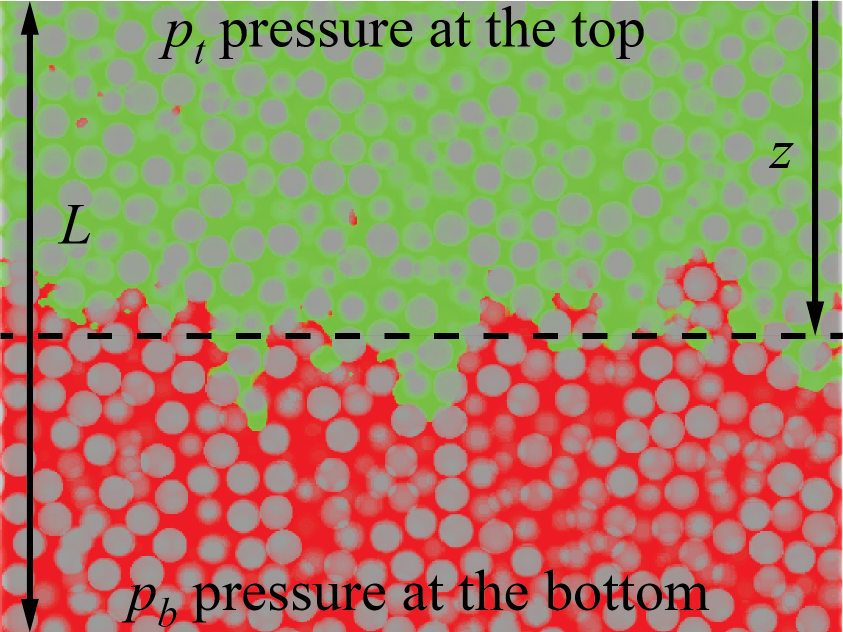}
    \caption{Schematic of a two-phase flow experiment, where the green fluid is displacing the red one from above at a fixed flow rate $Q$. $L$ is the cell height. $z$ the distance from the top of the cell to the front. $p_t$ and $p_b$ are the pressures at the top and bottom of the cell, respectively.}
    \label{fig:theoryIll}
\end{figure}
\noindent Figure \ref{fig:theoryIll} illustrates a stable front during an experiment.
When a front is stable, the difference between the pressure at the bottom ($p_b$) and the top ($p_t$) of the cell can be formulated by using the same approach as in Sec.~\ref{sec:theory}.
For a flat interface at depth $z$, the pressure difference $p_b-p_t$ can be written as
\begin{equation}
    p_b-p_t= \left(\rho_1 g - \frac{Q\mu_1}{A\kappa_{TP}(Q)}\right)z -p_c+\left(\rho_2 g - \frac{Q\mu_2}{A\kappa_{SP}}\right)(L-z) ,
\label{eq:delP}
\end{equation}
where $p_c$ is the capillary pressure at the interface.
The front position advances as $z \propto v_f(Q) t$, with an approximately constant front velocity related to the flow rate via Eq.~(\ref{eq:C1C2}).
Using this relation, we can differentiate $p_b-p_t$ with respect to time $t$, obtaining
\begin{equation}
\frac{\mathrm{d} (p_b-p_t)}{\mathrm{d} t}
\approx v_f(Q) \frac{\mathrm{d} (p_b-p_t)}{\mathrm{d} z}=v_f(Q)\left[(\rho_1-\rho_2)g-\frac{Q}{A}\left(\frac{\mu_1}{\kappa_{TP}(Q)}-\frac{\mu_2}{\kappa_{SP}}\right)\right].
\label{eq:delPdz}
\end{equation}
This result shows the same balance of viscous and gravitational pressures encoded in Eq.~(\ref{eq:unstableq}).
We therefore expect a stable front when $ \mathrm{d}(p_b-p_t)/\mathrm{d}t<0$ and an unstable front otherwise.
When the front is unstable, Eqs.~(\ref{eq:delP}) and (\ref{eq:delPdz}) are no longer valid, since the front can no longer be represented as a sharp interface.

\begin{figure}[H]
	\centering
	\includegraphics[width=1\linewidth]{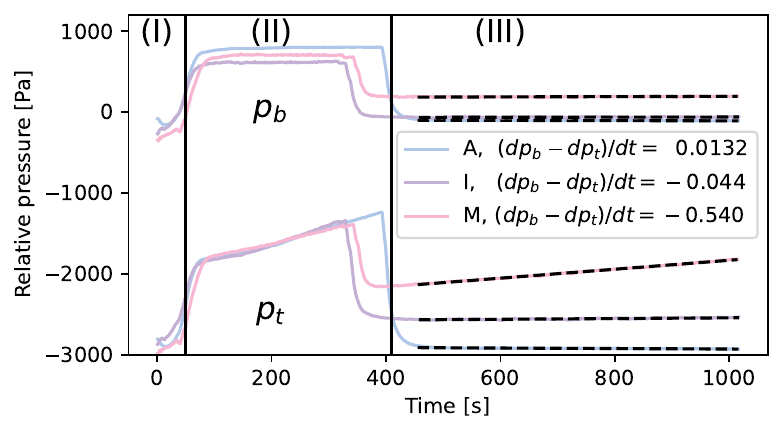}
	\caption{Representative pressure signals from three experiments (A, I, M) at the bottom ($p_b$) and top ($p_t$) of the experimental cell. Stages I and II are prior to altering the flow rate, from the initiation of the experiments. Stable experiments show decreasing $p_b-p_t$ in stage III, when the front progresses through the imaging region at the fixed flow rate $Q$. Values reported in the inset are in Pa/s. All available values are reported together with error estimates in Table~\ref{table:expSummary}.}
	\label{fig:pressure}
\end{figure}

We measured pressure signals in all experiments except Exps.~C and G.
Figure~\ref{fig:pressure} shows $p_t$ and $p_b$ from three representative experiments.
In each experiment, the pressures follow a three-stage response, corresponding to (I) experiment initialization, in which the area above the mesh screen fills with invading fluid; (II) initial invasion, in which the fluid permeates the cell at the maximum injection rate into the imaging region; and (III) interface dynamics, when the pressures evolve under the imposed flow rate, $Q$.
If the interface is stable, the final stage of pressure evolution (III) is approximated by Eqs.~(\ref{eq:delP}) and (\ref{eq:delPdz}). 
According to Eq.~(\ref{eq:delPdz}), the inequality $\mathrm{d}(p_b-p_t)/\mathrm{d}t< 0$ provides a pressure-based criterion for front stability.
By interpreting the time derivative of the pressure difference in Table~\ref{table:expSummary}, we could use this inequality to qualitatively estimate the critical flow rate $Q_c$.
Instead, we evaluate $Q_c$ with a quantitative method in the following section.

\subsection{Flow-dependent two-phase permeability and the critical flow rate}\label{sec:result-permeability}

In Sec.~\ref{sec:theory} we left the possibility that the two-phase permeability $\kappa_{TP}$ encountered by the invading phase was a function of flow rate.
There is extensive literature on two-phase permeability \cite{tallakstad2009steady,tallakstad2009sim,sinha2012effective,grova2011two,anastasiou2024steady,zhang2021quantification}, but the bulk of the work has been done in steady-state flows, where both phases are transported in parallel, which is not the case in the experiments described here.
We expect invasion fronts to leave trapped clusters of the defending fluid behind, which will effectively reduce the porosity and therefore decrease the two-phase permeability.
It stands to reason that a lower flow-rate, which leads to a wider front, would produce larger trapped clusters of defending fluid behind the front.
This has been confirmed in earlier 2D experiments and simulations \cite{birovljev1991gravity, maaloy2021burst,wilkinson1983invasion}, although a theory explaining the relation between two-phase permeability and flow rate remains to be developed.
\begin{figure}[H]
	\centering
	\includegraphics[width=\linewidth]{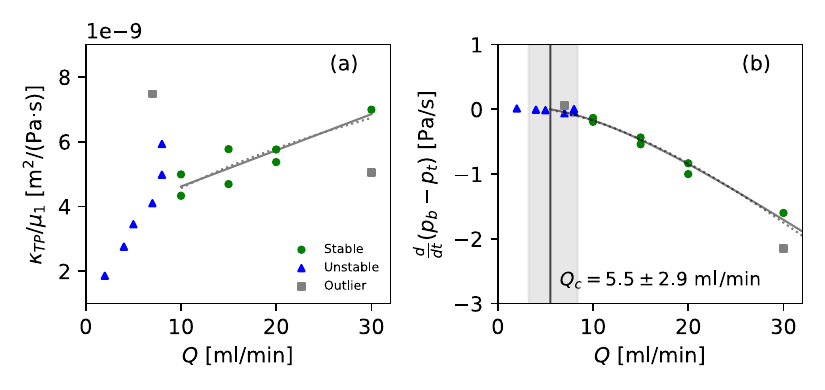}
	\caption{Panel (a) shows the mobility $\kappa_{TP}(Q)/\mu_1$ for different flow rates $Q$. These data are calculated from the experimental data and Eq.~(\ref{eq:delPdz}), as described in Sec.~\ref{sec:result-permeability}. Each experiment provides one data point, and the points have colors representing whether the experiment was classified as stable (green), unstable (blue), or as an outlier as far as the fitting is concerned (gray), as described in the text. A linear fit $\kappa_{TP}/\mu_1 = B_1 + B_2 Q$ is shown as a solid black line, while the power-law fit $\kappa_{TP}/\mu_1 = B_1' + B_2' Q^{0.5}$ is shown for comparison as a dotted gray line.
    The best-fit parameters are provided in the text.
    Panel (b) shows the averaged time derivative of $p_b-p_t$ versus the flow rate, with Eq.~(\ref{eq:delPdz}) evaluated for both linear and power-law fits of $\kappa_{TP}$, providing nearly identical results. The solid black line indicates the solution $Q_c = (5.5 \pm 2.9)$ ml/min of Eq.~(\ref{eq:unstableq}) using the best-fit parameters with the linear model, while the shaded region is the 95\% confidence interval. This solution is equivalent to $\mathrm{d}(p_b-p_t)/\mathrm{d}t=0$. Two experiments were excluded from the fitting (Exps.~G and Q) based on irregularities in their porous geometries, discussed in Sec.~\ref{sec:result-beadpack}.}
	\label{fig:perm-press}
\end{figure}

We can investigate the flow dependence of the two-phase permeability by applying Eq.~(\ref{eq:delPdz}) to the stable experiments.
The experimental data provide all components of this equation except the single-phase permeability $\kappa_{SP}$ and the two-phase permeability $\kappa_{TP}$.
We measured the single-phase permeability for identical beadpacks in \cite{brodin2022}, finding $\kappa_{SP}=(1.7\pm 0.3)\times 10^{-8}$ m$^2$.
We also measured the two-phase permeability, but due to experimental limitations, we only measured it for a high flow rate of 30~ml/min.
The front velocity $v_f(Q)$ is given in Eq.~(\ref{eq:C1C2}) with fit parameters from Sec.~\ref{sec:velSec}, while the fluid densities and viscosities (with their temperature dependences) are provided in Table~\ref{table:fluidMeas}.

Using Eq.~(\ref{eq:delPdz}) with the measured single-phase permeability and the measured front velocity relation in Eq.~(\ref{eq:C1C2}), we have evaluated the mobility $\kappa_{TP}/\mu_1$ as a function of $Q$ and plotted it in Fig.~\ref{fig:perm-press}(a).
We neglected the term involving $\mu_2$, since $\mu_2\ll  \mu_1$.
We plotted the mobility $\kappa_{TP}(Q)/\mu_1$ rather than the two-phase permeability alone because the viscosity depends on temperature, which differs between experiments.
Although Fig.~\ref{fig:perm-press}(a) shows the results for all experiments, the analysis is strictly valid only for the stable experiments.
In the same plot, we have fitted $\kappa_{TP}(Q)/\mu_1$ to the function
\begin{equation}
\kappa_{TP}/\mu_1=B_1+B_2 Q^\alpha \; ,\label{eq:powFit}
\end{equation}
with $B_1$ and $B_2$ as fitting parameters, where $\alpha$ is fixed either to $1$ or $0.5$.
We excluded Exps. G and Q from the fitting [plotted as gray squares in Fig.~\ref{fig:perm-press}(a)] because they have outlier values of $\kappa_{TP}$ accompanied by anomalous porous geometries, as discussed in Sec.~\ref{sec:result-beadpack}.
For the linear model ($\alpha=1$), we find coefficients $B_1 = (3.5 \pm 0.5)\times 10^{-9}$ m$^2$/(Pa$\cdot$s) and $B_2 = (1.1 \pm 0.3)\times 10^{-10}$ m$^2$$\cdot$min /(Pa$\cdot$s$\cdot$ml).
For the power-law model ($\alpha=0.5$) we find $B_1'=(1.6 \pm 1.0)\times 10^{-9}$  m$^2$/(Pa$\cdot$s) and $B_2'=(9.4 \pm 2.3)\times 10^{-10}$ m$^2$$\cdot$min$^{1/2}$ /(Pa$\cdot$s$\cdot$ml$^{1/2}$).
The listed uncertainties are calculated as the standard error of the fit parameters.
The power-law model provides almost identical goodness of fit as the linear model, shown as a dotted line in Fig.~\ref{fig:perm-press}.

A power-law  dependence of $\kappa_{TP}$ on $Q$ with an exponent $\alpha=0.5$ has been measured in steady-state experiments, where both fluids are injected at the same time \cite{tallakstad2009steady,tallakstad2009sim}, although this scenario differs from the experiments described here.
Due to the scatter in the experimental data, we are not able to decide the functional dependence of $\kappa_{TP}/\mu_1$ on Q.
The scatter reflects variations in the porosity and arrangement of grains in the porous structure, giving rise to variations in the permeability and pressure measurements from sample to sample.
This produces relative uncertainties in $\mathrm{d}(p_b-p_t)/\mathrm{d}t$ and $\kappa/\mu_1$ of the order of 20\%. 
Nevertheless, both cases $\alpha=1$ and $\alpha=0.5$ fit the experimental data better than choosing a constant value for $\kappa_{TP}(Q)/\mu_1$, as seen in Fig.~\ref{fig:perm-press}(a).

Using the best fit model to $\kappa_{TP}(Q)$ we plot Eq.~(\ref{eq:delPdz})  in Fig.~\ref{fig:perm-press}(b), showing consistency with the experimental data.
We also use this relation to identify the critical flow rate from the pressure data, which complements our earlier estimate from the image data in Sec.~\ref{sec:res-front}.
Using Eq.~(\ref{eq:delPdz}) to evaluate $\mathrm{d}(p_b-p_t)/\mathrm{d}t=0$, which is equivalent to solving Eq.~(\ref{eq:qccondition}), provides $Q_c = (5.5 \pm 2.9)$ ml/min, with listed uncertainties from 95\% confidence bounds.
This pressure-derived value of $Q_c$ agrees with our image-derived estimate within the experimental uncertainty.

\subsection{Beadpack characterization}\label{sec:result-beadpack}

Several of our experiments show anomalous front stability, including Exps.~C, E, and G, which are stable experiments at low flow rates, and Exp.~$Q$, which was unstable even though it was the highest flow rate studied and the most negative $\mathrm{d}(p_b-p_t)/\mathrm{d}t$ (Table \ref{table:expSummary}).
Here we investigate whether these anomalies originate from differences in these experiments' porous structures.
We consider three main metrics: (1) recognition of small crystalline regions (crystallites), (2) local porosity distributions, and (3) radial correlation functions.

\begin{figure}[H]
	\centering
	\includegraphics[width=\linewidth]{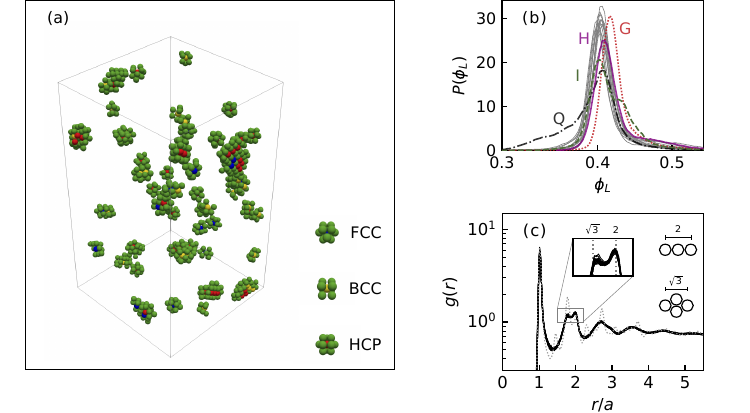}
	\caption{Panel (a) shows crystallites of three types identified by common-neighbor analysis in a typical experiment, Exp.~K. Crystallite centers are colored by type. Panel (b) displays the local porosity distributions calculated in a sphere of diameter  $L=4 a$ from each bead. Outlier experiments with particularly right- (Exps.~G, H, I) and left-skewed (Exp.~Q) porosity distributions are highlighted, reflecting  experiments with particularly loose and tight beadpacks, respectively. Panel (c) shows radial correlation functions $g(r)$ for all beadpacks. Most experiments overlap tightly and exhibit the split second peak characteristic of monodisperse sphere packs \cite{bennett1972}. Exp.~Q deviates from all others (dotted curve), with oscillations characteristic of crystalline order.}
	\label{fig:Geom}
\end{figure}

Crystallites (see \cite{torquato2000}) were identified using adaptive Common-Neighbor Analysis (CNA) as implemented in the Ovito software \cite{ovito}.
CNA examines the local topology around each sphere in the pack for similarity to template crystal structures \cite{honeycutt1987,stukowski2012}.
We used CNA to identify regions with face-centered cubic (FCC), body-centered cubic (BCC), and hexagonal close-packed (HCP) order.
Crystallites were present in all experimental beadpacks, with typically 2\% to 8\% of beads participating.
Figure~\ref{fig:Geom}(a) displays the crystallites identified in a typical beadpack, while Table~\ref{table:expSummary} provides the fraction of beads participating in crystallites for all experiments.

Identified crystallites were primarily of FCC and HCP types, the lowest-porosity arrangements possible in monodisperse beadpacks.
For this reason, we expect that crystallites could affect the dynamics of fluid invasion in the experiments through the coupling of pore size to the capillary and viscous pressures.
In particular, among the ``anomalous" experiments Exp.~Q, 37\% of beads were in crystalline arrangements, representing an abnormally large crystalline region in the pack [see Fig.~\ref{fig:flatFingersPinning}(c)], with an order of magnitude more beads participating in crystallites than any other experiment.

Local porosity distributions were calculated by accumulating the void fraction within spherical domains of diameter $L$ surrounding each detected bead \cite{biswal1998,hilfer1992}.
This produced around 30,000 local porosity observations $\phi_L$ per experiment, and these observations were binned into the histograms $P(\phi_L)$ shown in Fig.~\ref{fig:Geom}(b).
Local porosity distributions depend on the bin size $L$. We selected $L=4a$ to resolve porosity fluctuations at the scale of typical stable front widths.
The obtained local-porosity distributions are largely consistent between experiments, and mean porosity values ranged between $0.390 \leq \varphi \leq 0.420$, with mean $0.406$ and standard error $0.001$.
However, some experiments show particularly left (Exp.~Q) and right-skewed (Exps.~G, H, and I) local-porosity distributions, indicating particularly low- and high-porosity packs, respectively.
In particular, Exp.~G had a flow rate $Q=7.0$ ml/min, lower than several unstable experiments, yet its front remained stable as assessed by the image analysis.
Possibly, the anomalous stability in Exp.~G can be explained by its uniquely high porosity, which differs from all other experiments, as shown in Fig.~\ref{fig:Geom}(b).

Radial correlation functions were calculated to assess the homogeneity of the experimental beadpacks \cite{torquato2002}, again using an implementation from the Ovito software \cite{ovito}.
Figure \ref{fig:Geom}(c) shows the radial correlation functions of the beadpacks of all 17 experiments.
All experiments show a sequence of local peaks separated approximately by the bead diameter ($a$).
These correlation functions tightly overlap for all trials, except for Exp.~Q.
All radial correlation functions show a split second peak at values $\sqrt{3}a$ and $2a$, which originates from the different geometric configurations possible for second-nearest neighbors in a hard-sphere pack [see Fig.~\ref{fig:Geom}(c) inset] \cite{bennett1972}.
Exp.~Q deviates from all others, showing oscillations characteristic of crystalline order \cite{yurchenko2015}, another indication of its irregularity.

\section{Discussion}
\label{sec:disc}

In Section \ref{sec:res}, we evaluated the stability of an initially planar invasion front between two immiscible fluids in a porous medium, using 3D imaging and pressure measurements.
Our study included 17 experiments at 9 different flow rates, allowing us to resolve how the balance of viscous, capillary, and gravitational forces determines front stability as a function of flow rate.

Two independent methods were adopted for front-stability classification.
Time-resolved front widths were classified as stable or unstable with a slope-over-threshold method (Sec.~\ref{sec:res-front}), indicating that fronts usually remain stable for flow rates larger than $Q_c \approx 8$ ml/min. 
Further analysis of the data using the pressure measurements and experimentally-determined front velocities, see Sec. \ref{sec:res-pressure}, predicted a critical flow rate $Q_c = (5.5 \pm 2.9)$ ml/min.
Both of these independent $Q_c$ estimates produce equivalent values within the experimental uncertainty.
Partly, this uncertainty originates from the slow variation of $\mathrm{d}(p_b-p_t)/\mathrm{d}t$ near $Q_c$, see Fig.~\ref{fig:perm-press}.

\subsection{Pinning and fingering mechanisms of instability}
\label{sec:disc-finpin}

The experimental images revealed two different mechanisms by which fronts destabilized, which we referred to as fingering and pinning (Sec.~\ref{sec:res-front} and Fig.~\ref{fig:flatFingersPinning}). 
Fingering appears as narrow plumes of invading fluid protruding in the direction of the flow. This instability originates from the interplay between the pressure gradients within the fluid phases due to viscous and gravitational forces, and the capillary-pressure thresholds for interface displacement through individual pores. If a positive gradient of capillary pressures along the interface between the fluids in the flow direction develops, fingering occurs, and this mechanism has long been understood to control front instability in 2D experiments \cite{maloy1985,auradou1999competition,maaloy2021burst,birovljev1991gravity,meheust2002, ayaz2020gravitational}.

Pinning appears as regions of defending fluid, much larger than a single pore, trailing behind the main front [see Fig.~\ref{fig:flatFingersPinning}(c)].
Most of our unstable 3D experiments exhibit pinning, as identified from relatively large contribution of backward-skewed regions of the front distribution to the front width (See Fig.~\ref{fig:distributions}).
Pinning likely originates from relatively long-ranged correlations in the porous geometry. Earlier numerical simulations of gravity-invasion percolation have noted that introducing large-scale correlations in porous-medium characteristics can fundamentally modify the front progression and residual fluid saturation after drainage \cite{ionnidis1996}. 
Indeed, our detailed analysis of the 3D porous structures revealed crystalline regions (primarily HCP and FCC) of typical extent $4-8a$, dispersed throughout the otherwise randomly-arranged beadpacks (see Fig.~\ref{fig:Geom}).
Considering that our cell was near $24a$ wide (see Fig.~\ref{fig:diagram_set-up}), crystallites represent significant heterogeneity compared with the system size.
Such relatively long-ranged correlations have not been present in most earlier 2D experiments, which are typically much larger porous Hele-Shaw cells filled with randomly-distributed elements (see \cite{vincent-dospital2022}).
For example, \cite{meheust2002} has system size $350a$, compared to ours with size $\sim 24a$. Although most of our beadpacks had less than $4$\% of beads participating in these crystallites (see Table~\ref{table:expSummary}), some had more, such as Exps.~F and D, with 7-8\%, and Exp.~Q, which had 37\%.
Because crystallites have relatively low local porosity, they present locally-higher capillary pressure thresholds for invasion and lower permeabilities for flow.
The elevated capillary pressure thresholds at the crystallites will stall the forward progression of fluid through the crystallites, while the relatively higher permeability of nearby areas will redirect flow around crystallites, providing a likely explanation for the development of backward-trailing regions of defending fluid that we termed pinning.

\subsection{System-size effects on fluid invasion}
\label{sec:disc-syssiz}
Although the majority of our experiments obey the stability criterion for $Q>Q_c$, certain experiments deviate.
Near the transition point $Q_c$, experiments can appear either stable or unstable, independently of the flow rate (Exps.~C to I), and one experiment (Exp.~Q) was unstable despite having the highest experimental flow rate.

We believe these deviations originate from the limited system size, in addition to the long-range spatial correlations in the beadpacks discussed in Sec.~\ref{sec:disc-finpin} and the sensitivity to the specific threshold value chosen for stability classification, see Sec.~\ref{sec:res-front}.
Fronts assessed as unstable may have attained stable widths if allowed to propagate over a longer distance than we could observe in our experimental cell.
For example, viewing Fig.~\ref{fig:wFingersPinning}(a), our slope criterion for stability judges Exps.~F, H, and Q as unstable, whereas these fronts might have stabilized given more time to evolve.

The importance of system size has been emphasized in studies which represent immiscible fluid displacement in porous media as an invasion-percolation process \cite{wilkinson1983invasion,wilkinson1984percolation}.
As pointed out by \citet{wilkinson1984percolation}, the saturation of the invading fluid at breakthrough -- the endpoint of our experiments -- exhibits finite-size scaling, affecting the observed front width.
Consequently, the $Q_c$ values we identified from the pressure and image data were likely affected by the necessarily-limited size of our experimental cell.

\subsection{Residual saturation after drainage in 3D beadpacks}
\label{sec:disc-ressat}

Following the passage of a stable drainage front in a porous medium, the defending phase can be found either in trapped clusters or in wetting films \cite{bryant2003wetting,bryant2004bulk,hoogland2016drainage,moura2019connectivity}.
Numerical and experimental studies have verified that under a stabilizing pressure gradient $\nabla P$ with a spatially-homogeneous distribution of pore sizes, the maximum length of trapped clusters $l_{max}$ exhibits a scaling function $l_{max} \propto |\nabla P|^{-\nu/(1+\nu)}$, where $\nu$ is the critical exponent for the correlation length in 2D or 3D percolation  \cite{wilkinson1984percolation, gouyet1988fractal, birovljev1991gravity, meheust2002, clement1985, hulin1988}. 

While this scaling applies to both 2D and 3D porous media, the structure of the trapped clusters differs between these cases.
Primarily, these differences originate from the higher connectivity between pores in 3D media when compared to 2D.
The low connectivity between pores in 2D porous materials leads to the disconnection of the defending phase at relatively low invading-phase saturations, producing trapped clusters of defending fluid that span across multiple pores and throats \cite{meheust2002,maaloy2021burst}.
In contrast, the high connectivity between pores in 3D porous media ensures that the defending phase cannot be trapped until a significant portion of the porous medium has been occupied by the invading phase, producing lower residual saturations and bodies of trapped fluid having smaller volumes, which may reside in films coating beads or as pendular bridges spanning across the contact points of neighboring grains \cite{nascimento2019}.

For this reason, we expect relatively low residual saturations in our 3D experiments when compared with 2D.
Such low $S(Q)$ values could not be precisely measured with our 3D visualization apparatus, as our spatial resolution was limited by the voxel size (about $a/42$, or $71$ $\mu$m) which may be comparable to the size of trapped defending fluid bodies.
Nevertheless, secondary effects of the residual saturation varying with the imposed flow rate were detected in our data, both in the form of the non-linear function $v_f(Q)$ [Eq.~(\ref{eq:C1C2})] and in $\kappa_{TP}(Q)/\mu_1$ [Eq.~(\ref{eq:powFit})].
A constant $S(Q)$ in our experiments would lead to a linear relationship between $v_f$ and $Q$ and a constant value of $\kappa_{TP}$, inconsistent with our observations.
Instead, we observe a weak but statistically-significant nonlinear dependence of $v_f$ on $Q$ and a value of $\kappa_{TP}$ which grows with increasing $Q$, supporting the hypothesis of a flow-rate-dependent residual saturation.

\subsection{Comparing with existing theory}
\label{sec:disc-2dtheory}

In previous studies of drainage in quasi-2D spatially homogeneous porous media, a dimensionless number $F$ -- termed the fluctuation number -- has been proposed to compare the effective pressure drop developed in the fluids at the pore scale to the width of the fluctuations in the capillary-pressure thresholds to invade pore throats \cite{birovljev1991gravity,auradou1999competition, meheust2002, maaloy2021burst, vincent-dospital2022}.
Positive $F$ values indicate that the invasion front during drainage should reach a stable width, while negative values are associated with unstable fronts, in which the growth of fingers is unbounded.
Besides serving as a front-stability criterion, the fluctuation number can be used to predict the front width, with $W \propto F^{-\nu/(1+\nu)}$, where $\nu$ is the critical exponent for the correlation length in percolation, corresponding to $4/3$ in 2D. In the absence of spatial gradients in capillary pressure thresholds, the $F$ number is effectively proportional to the difference between the Bond and Capillary numbers used in Eq.~(\ref{eq:Bo_Caeff}) as a criterion for front stability \cite{meheust2002}.

While this theory has been verified in numerous studies of displacement flows of immiscible fluids in 2D porous media, \cite{auradou1999competition, meheust2002, birovljev1991gravity, maaloy2021burst, vincent-dospital2022}, its extension to 3D systems is limited.
In 3D systems, negative $F$ values would still be linked to unstable drainage fronts, as they indicate a positive capillary-pressure gradient along the front in the direction of the flow.
Under this condition, the invasion of pores at the tip of the front is favored, and fingers can grow indefinitely.
Conversely, when $F$ is positive, the invasion of pores at the trailing region of the front becomes relatively easier, limiting the width over which the fluid interface can be stretched and producing a stable value over time.

Under stabilizing pressure gradients, $F>0$, the front width $W$ in 3D immiscible-displacement flows should also decrease monotonically with the fluctuation number, but the scaling $W \propto F^{-\nu/(1+\nu)}$ may no longer be valid.
As first pointed out by \citet{wilkinson1984percolation} and later investigated numerically and experimentally \cite{chaouche1994invasion, breen2022}, higher stabilizing pressure gradients lead to narrower saturation-transition zones, in which the defending phase goes from completely saturating the porous medium -- ahead of the front -- to being trapped in clusters at the residual saturation -- behind the front.
However, due to the higher phase connectivity in 3D systems, this saturation-transition region differs substantially from the 2D case \cite{wilkinson1984percolation}.
In 3D, we expect the front to extend over a broad range of saturations, from the tip with a low invading-phase saturation, to a center region where both phases can percolate and very little trapping occurs, to the tail where the invading-phase saturation is high and the defending fluid becomes trapped in isolated clusters.

Based on the analogy of immiscible displacement flows in porous media to percolation on a gradient of occupation probabilities, \cite{gouyet1988fractal,chaouche1994invasion, auradou1999competition, meheust2002}, one could conjecture the expansion of the front width scaling with the fluctuation number in 2D, to a $W_t \propto F^{-\nu/(1+\nu)}$ in 3D, where $W_t$ represents what is defined by \citet{gouyet1988fractal} as the ``front tail width", and $\nu=0.88$.
Still, we have not tested this hypothesis with our experimental data for several reasons.
First, delimiting the ``front tail" is not straightforward as it presents no defining characteristics in our 3D images.
Second, our experimental $F$ values span only approximately a decade, which may be insufficient to substantiate the theoretical prediction.
Finally, in many of our experiments, the volume fraction occupied by crystallites may be too large to ensure spatial homogeneity in the capillary pressure thresholds for throat invasion, fundamental for the scaling to be valid.
For these reasons, the verification of a theoretical scaling relationship between the front width and the fluctuation number based on 3D experiments remains an open research problem.

\section{Conclusion}
\label{sec:conc}
We conducted 3D imaging experiments to assess the stability of fluid invasion fronts in a porous medium for different imposed flow rates.
Above a critical flow rate $Q_c$, fronts stabilized at a well-defined width, whereas below this rate they destabilized.
Two distinct mechanisms of front growth were visible, which we termed ``pinning" and ``fingering".
While fingering describes the primary instability mechanism in earlier 2D experiments, pinning describes a backward-trailing front growth that may predominate in unstable 3D invasion flows, considering the higher connectivity of 3D porous media.
We estimated $Q_c$ using two independent methods, one based on the 3D images, and another based on pressure measurements.
Most of our experiments align with the $Q_c$ predictions, while the experiments which deviate may have been affected by the limited system size and the presence of crystalline regions in the beadpacks.
Our analyses reveal several interesting consequences of trapped defending fluid behind stable invasion fronts which to our knowledge have not been emphasized in earlier studies.
First, the front velocity relates non-linearly to the flow rate, and second, the relative permeability for the invading phase is not a constant, but rather has a weak dependence on the flow rate.
Incorporating these dependences was found necessary to describe the observed front-velocity and pressure-measurement data within a Darcy-based framework.
While our results provide useful insights into the mechanisms of front instability in 3D fluid-invasion processes, future studies will be required to further investigate invasion-front morphology and to explore the application of theories developed for 2D to the description of 3D flows.

\section*{Acknowledgments}

We would like to thank Eirik Grude Flekk{\o}y, Renaud Toussaint, Gloria Buend{\'i}a, Salvatore Torquato, and Paul Meakin for useful discussions.
We gratefully acknowledge support from the University of Oslo and from the Research Council of Norway, through projects 262644 (PoreLab), 325819 (M4), and 324555 (FlowConn).

\appendix
\section*{Appendix: Relation of front velocity to residual saturation}
\label{sec:appen}
The front velocity is given by Eq.~(\ref{eq:sat}) ,
\begin{equation}
v_f(Q)=\frac{Q}{A\phi(1-S(Q))} \; ,
\end{equation}
where $A$  is the cross-section area  of the model, $\phi$ the porosity, and $S(Q)$ the residual saturation of the displaced fluid behind the front.
By expanding the residual saturation around the critical flow rate $Q_c$ to first order in $Q-Q_c$, we get
\begin{multline}
v_f(Q)=\frac{Q}{A\phi(1-S(Q))}
=\frac{Q}{A\phi(1-S(Q_c)-S'(Q_c)(Q-Q_c))}= \\
\frac{Q}{A\phi(1-S(Q_c))(1-\frac{S'(Q_c)}{1-S(Q_c)}(Q-Q_c))}=C_1Q-C_2Q^2 \; ,
\end{multline}
where
\begin{equation}
C_1= \frac{1}{A\phi(1-S(Q_c)}(1-\frac{S'(Q_c)}{1-S(Q_c)} Q_c) \; ,
\end{equation}
and 
\begin{equation}
C_2= - \frac{S'(Q_c)}{A\phi(1-S(Q_c))^2} \; .
\end{equation}
Because increasing the flow rate reduces trapping \cite{maaloy2021burst}, $S'(Q)$ must be negative, implying that $C_2>0$.


\begin{thebibliography}{95}%
\makeatletter
\providecommand \@ifxundefined [1]{%
 \@ifx{#1\undefined}
}%
\providecommand \@ifnum [1]{%
 \ifnum #1\expandafter \@firstoftwo
 \else \expandafter \@secondoftwo
 \fi
}%
\providecommand \@ifx [1]{%
 \ifx #1\expandafter \@firstoftwo
 \else \expandafter \@secondoftwo
 \fi
}%
\providecommand \natexlab [1]{#1}%
\providecommand \enquote  [1]{``#1''}%
\providecommand \bibnamefont  [1]{#1}%
\providecommand \bibfnamefont [1]{#1}%
\providecommand \citenamefont [1]{#1}%
\providecommand \href@noop [0]{\@secondoftwo}%
\providecommand \href [0]{\begingroup \@sanitize@url \@href}%
\providecommand \@href[1]{\@@startlink{#1}\@@href}%
\providecommand \@@href[1]{\endgroup#1\@@endlink}%
\providecommand \@sanitize@url [0]{\catcode `\\12\catcode `\$12\catcode
  `\&12\catcode `\#12\catcode `\^12\catcode `\_12\catcode `\%12\relax}%
\providecommand \@@startlink[1]{}%
\providecommand \@@endlink[0]{}%
\providecommand \url  [0]{\begingroup\@sanitize@url \@url }%
\providecommand \@url [1]{\endgroup\@href {#1}{\urlprefix }}%
\providecommand \urlprefix  [0]{URL }%
\providecommand \Eprint [0]{\href }%
\providecommand \doibase [0]{https://doi.org/}%
\providecommand \selectlanguage [0]{\@gobble}%
\providecommand \bibinfo  [0]{\@secondoftwo}%
\providecommand \bibfield  [0]{\@secondoftwo}%
\providecommand \translation [1]{[#1]}%
\providecommand \BibitemOpen [0]{}%
\providecommand \bibitemStop [0]{}%
\providecommand \bibitemNoStop [0]{.\EOS\space}%
\providecommand \EOS [0]{\spacefactor3000\relax}%
\providecommand \BibitemShut  [1]{\csname bibitem#1\endcsname}%
\let\auto@bib@innerbib\@empty
\bibitem [{\citenamefont {McGrail}\ \emph {et~al.}(2006)\citenamefont
  {McGrail}, \citenamefont {Schaef}, \citenamefont {Ho}, \citenamefont {Chien},
  \citenamefont {Dooley},\ and\ \citenamefont {Davidson}}]{mcgrail2006}%
  \BibitemOpen
  \bibfield  {author} {\bibinfo {author} {\bibfnamefont {B.~P.}\ \bibnamefont
  {McGrail}}, \bibinfo {author} {\bibfnamefont {H.~T.}\ \bibnamefont {Schaef}},
  \bibinfo {author} {\bibfnamefont {A.~M.}\ \bibnamefont {Ho}}, \bibinfo
  {author} {\bibfnamefont {Y.-J.}\ \bibnamefont {Chien}}, \bibinfo {author}
  {\bibfnamefont {J.~J.}\ \bibnamefont {Dooley}},\ and\ \bibinfo {author}
  {\bibfnamefont {C.~L.}\ \bibnamefont {Davidson}},\ }\bibfield  {title}
  {\bibinfo {title} {Potential for carbon dioxide sequestration in flood
  basalts},\ }\href {https://doi.org/10.1029/2005JB004169} {\bibfield
  {journal} {\bibinfo  {journal} {Journal of Geophysical Research: Solid
  Earth}\ }\textbf {\bibinfo {volume} {111}} (\bibinfo {year}
  {2006})}\BibitemShut {NoStop}%
\bibitem [{\citenamefont {Benson}\ \emph {et~al.}(2005)\citenamefont {Benson},
  \citenamefont {Cook}, \citenamefont {Anderson}, \citenamefont {Bachu},
  \citenamefont {Nimir}, \citenamefont {Basu}, \citenamefont {Bradshaw},
  \citenamefont {Deguchi}, \citenamefont {Gale}, \citenamefont {von Goerne},
  \citenamefont {et.~al. Heidug}, \citenamefont {Holloway}, \citenamefont
  {Kamal}, \citenamefont {Keith}, \citenamefont {Loyd}, \citenamefont {Rocha},
  \citenamefont {Senior}, \citenamefont {Thomson}, \citenamefont {Torp},
  \citenamefont {Wildenborg}, \citenamefont {Wilson}, \citenamefont {Zarlenga},
  \citenamefont {Zhou}, \citenamefont {Celia}, \citenamefont {Gunter},
  \citenamefont {King}, \citenamefont {Lindeberg}, \citenamefont {Lombardi},
  \citenamefont {Oldenburg}, \citenamefont {Pruess}, \citenamefont {Rigg},
  \citenamefont {Stevend}, \citenamefont {Wilson},\ and\ \citenamefont
  {Whittaker}}]{benson2005}%
  \BibitemOpen
  \bibfield  {author} {\bibinfo {author} {\bibfnamefont {S.}~\bibnamefont
  {Benson}}, \bibinfo {author} {\bibfnamefont {P.}~\bibnamefont {Cook}},
  \bibinfo {author} {\bibfnamefont {J.}~\bibnamefont {Anderson}}, \bibinfo
  {author} {\bibfnamefont {S.}~\bibnamefont {Bachu}}, \bibinfo {author}
  {\bibfnamefont {H.}~\bibnamefont {Nimir}}, \bibinfo {author} {\bibfnamefont
  {B.}~\bibnamefont {Basu}}, \bibinfo {author} {\bibfnamefont {J.}~\bibnamefont
  {Bradshaw}}, \bibinfo {author} {\bibfnamefont {G.}~\bibnamefont {Deguchi}},
  \bibinfo {author} {\bibfnamefont {J.}~\bibnamefont {Gale}}, \bibinfo {author}
  {\bibfnamefont {G.}~\bibnamefont {von Goerne}}, \bibinfo {author}
  {\bibfnamefont {W.}~\bibnamefont {et.~al. Heidug}}, \bibinfo {author}
  {\bibfnamefont {S.}~\bibnamefont {Holloway}}, \bibinfo {author}
  {\bibfnamefont {R.}~\bibnamefont {Kamal}}, \bibinfo {author} {\bibfnamefont
  {D.}~\bibnamefont {Keith}}, \bibinfo {author} {\bibfnamefont
  {P.}~\bibnamefont {Loyd}}, \bibinfo {author} {\bibfnamefont {P.}~\bibnamefont
  {Rocha}}, \bibinfo {author} {\bibfnamefont {B.}~\bibnamefont {Senior}},
  \bibinfo {author} {\bibfnamefont {J.}~\bibnamefont {Thomson}}, \bibinfo
  {author} {\bibfnamefont {T.}~\bibnamefont {Torp}}, \bibinfo {author}
  {\bibfnamefont {T.}~\bibnamefont {Wildenborg}}, \bibinfo {author}
  {\bibfnamefont {M.}~\bibnamefont {Wilson}}, \bibinfo {author} {\bibfnamefont
  {F.}~\bibnamefont {Zarlenga}}, \bibinfo {author} {\bibfnamefont
  {D.}~\bibnamefont {Zhou}}, \bibinfo {author} {\bibfnamefont {M.}~\bibnamefont
  {Celia}}, \bibinfo {author} {\bibfnamefont {B.}~\bibnamefont {Gunter}},
  \bibinfo {author} {\bibfnamefont {J.~E.}\ \bibnamefont {King}}, \bibinfo
  {author} {\bibfnamefont {E.}~\bibnamefont {Lindeberg}}, \bibinfo {author}
  {\bibfnamefont {S.}~\bibnamefont {Lombardi}}, \bibinfo {author}
  {\bibfnamefont {C.}~\bibnamefont {Oldenburg}}, \bibinfo {author}
  {\bibfnamefont {K.}~\bibnamefont {Pruess}}, \bibinfo {author} {\bibfnamefont
  {A.}~\bibnamefont {Rigg}}, \bibinfo {author} {\bibfnamefont {S.}~\bibnamefont
  {Stevend}}, \bibinfo {author} {\bibfnamefont {E.~S.}\ \bibnamefont
  {Wilson}},\ and\ \bibinfo {author} {\bibfnamefont {S.}~\bibnamefont
  {Whittaker}},\ }\bibfield  {title} {\bibinfo {title} {Underground geological
  storage},\ }in\ \href
  {https://www.ipcc.ch/site/assets/uploads/2018/03/srccs_chapter5-1.pdf} {\emph
  {\bibinfo {booktitle} {IPCC Special Report on Carbon Dioxide Capture and
  Storage}}},\ \bibinfo {editor} {edited by\ \bibinfo {editor} {\bibfnamefont
  {B.}~\bibnamefont {Metz}}, \bibinfo {editor} {\bibfnamefont {O.}~\bibnamefont
  {Davidson}}, \bibinfo {editor} {\bibfnamefont {H.}~\bibnamefont {Coninck}},
  \bibinfo {editor} {\bibfnamefont {M.}~\bibnamefont {Loos}},\ and\ \bibinfo
  {editor} {\bibfnamefont {L.}~\bibnamefont {Meyer}}}\ (\bibinfo  {publisher}
  {Cambridge University Press},\ \bibinfo {year} {2005})\ \bibinfo {edition}
  {1st}\ ed.,\ pp.\ \bibinfo {pages} {195--276}\BibitemShut {NoStop}%
\bibitem [{\citenamefont {Anderson}\ \emph {et~al.}(2010)\citenamefont
  {Anderson}, \citenamefont {Zhang}, \citenamefont {Ding}, \citenamefont
  {Blanco}, \citenamefont {Bi},\ and\ \citenamefont
  {Wilkinson}}]{anderson2010}%
  \BibitemOpen
  \bibfield  {author} {\bibinfo {author} {\bibfnamefont {R.}~\bibnamefont
  {Anderson}}, \bibinfo {author} {\bibfnamefont {L.}~\bibnamefont {Zhang}},
  \bibinfo {author} {\bibfnamefont {Y.}~\bibnamefont {Ding}}, \bibinfo {author}
  {\bibfnamefont {M.}~\bibnamefont {Blanco}}, \bibinfo {author} {\bibfnamefont
  {X.}~\bibnamefont {Bi}},\ and\ \bibinfo {author} {\bibfnamefont {D.~P.}\
  \bibnamefont {Wilkinson}},\ }\bibfield  {title} {\bibinfo {title} {A critical
  review of two-phase flow in gas flow channels of proton exchange membrane
  fuel cells},\ }\href {https://doi.org/10.1016/j.jpowsour.2009.12.123}
  {\bibfield  {journal} {\bibinfo  {journal} {Journal of Power Sources}\
  }\textbf {\bibinfo {volume} {195}},\ \bibinfo {pages} {4531} (\bibinfo {year}
  {2010})}\BibitemShut {NoStop}%
\bibitem [{\citenamefont {Weber}\ \emph {et~al.}(2014)\citenamefont {Weber},
  \citenamefont {Borup}, \citenamefont {Darling}, \citenamefont {Das},
  \citenamefont {Dursch}, \citenamefont {Gu}, \citenamefont {Harvey},
  \citenamefont {Kusoglu}, \citenamefont {Litster}, \citenamefont {Mench},
  \citenamefont {Mukundan}, \citenamefont {Owejan}, \citenamefont {Pharoah},
  \citenamefont {Secanell},\ and\ \citenamefont {Zenyuk}}]{weber2014}%
  \BibitemOpen
  \bibfield  {author} {\bibinfo {author} {\bibfnamefont {A.~Z.}\ \bibnamefont
  {Weber}}, \bibinfo {author} {\bibfnamefont {R.~L.}\ \bibnamefont {Borup}},
  \bibinfo {author} {\bibfnamefont {R.~M.}\ \bibnamefont {Darling}}, \bibinfo
  {author} {\bibfnamefont {P.~K.}\ \bibnamefont {Das}}, \bibinfo {author}
  {\bibfnamefont {T.~J.}\ \bibnamefont {Dursch}}, \bibinfo {author}
  {\bibfnamefont {W.}~\bibnamefont {Gu}}, \bibinfo {author} {\bibfnamefont
  {D.}~\bibnamefont {Harvey}}, \bibinfo {author} {\bibfnamefont
  {A.}~\bibnamefont {Kusoglu}}, \bibinfo {author} {\bibfnamefont
  {S.}~\bibnamefont {Litster}}, \bibinfo {author} {\bibfnamefont {M.~M.}\
  \bibnamefont {Mench}}, \bibinfo {author} {\bibfnamefont {R.}~\bibnamefont
  {Mukundan}}, \bibinfo {author} {\bibfnamefont {J.~P.}\ \bibnamefont
  {Owejan}}, \bibinfo {author} {\bibfnamefont {J.~G.}\ \bibnamefont {Pharoah}},
  \bibinfo {author} {\bibfnamefont {M.}~\bibnamefont {Secanell}},\ and\
  \bibinfo {author} {\bibfnamefont {I.~V.}\ \bibnamefont {Zenyuk}},\ }\bibfield
   {title} {\bibinfo {title} {A critical review of modeling transport phenomena
  in polymer-electrolyte fuel cells},\ }\href
  {https://doi.org/10.1149/2.0751412jes} {\bibfield  {journal} {\bibinfo
  {journal} {Journal of the Electrochemical Society}\ }\textbf {\bibinfo
  {volume} {161}},\ \bibinfo {pages} {F1254} (\bibinfo {year}
  {2014})}\BibitemShut {NoStop}%
\bibitem [{\citenamefont {Soga}\ \emph {et~al.}(2004)\citenamefont {Soga},
  \citenamefont {Page},\ and\ \citenamefont {Illangasekare}}]{soga2004}%
  \BibitemOpen
  \bibfield  {author} {\bibinfo {author} {\bibfnamefont {K.}~\bibnamefont
  {Soga}}, \bibinfo {author} {\bibfnamefont {J.}~\bibnamefont {Page}},\ and\
  \bibinfo {author} {\bibfnamefont {T.}~\bibnamefont {Illangasekare}},\
  }\bibfield  {title} {\bibinfo {title} {A review of {NAPL} source zone
  remediation efficiency and the mass flux approach},\ }\href
  {https://doi.org/10.1016/j.jhazmat.2004.02.034} {\bibfield  {journal}
  {\bibinfo  {journal} {Journal of Hazardous Materials}\ }\textbf {\bibinfo
  {volume} {110}},\ \bibinfo {pages} {13} (\bibinfo {year} {2004})}\BibitemShut
  {NoStop}%
\bibitem [{\citenamefont {Seol}\ \emph {et~al.}(2003)\citenamefont {Seol},
  \citenamefont {Zhang},\ and\ \citenamefont {Schwartz}}]{seol2003}%
  \BibitemOpen
  \bibfield  {author} {\bibinfo {author} {\bibfnamefont {Y.}~\bibnamefont
  {Seol}}, \bibinfo {author} {\bibfnamefont {H.}~\bibnamefont {Zhang}},\ and\
  \bibinfo {author} {\bibfnamefont {F.~W.}\ \bibnamefont {Schwartz}},\
  }\bibfield  {title} {\bibinfo {title} {{A review of in situ chemical
  oxidation and heterogeneity}},\ }\href {https://doi.org/10.2113/9.1.37}
  {\bibfield  {journal} {\bibinfo  {journal} {Environmental \& Engineering
  Geoscience}\ }\textbf {\bibinfo {volume} {9}},\ \bibinfo {pages} {37}
  (\bibinfo {year} {2003})}\BibitemShut {NoStop}%
\bibitem [{\citenamefont {Toussaint}\ \emph {et~al.}(2005)\citenamefont
  {Toussaint}, \citenamefont {L{\o}voll}, \citenamefont {M{\'e}heust},
  \citenamefont {M{\aa}l{\o}y},\ and\ \citenamefont
  {Schmittbuhl}}]{toussaint2005}%
  \BibitemOpen
  \bibfield  {author} {\bibinfo {author} {\bibfnamefont {R.}~\bibnamefont
  {Toussaint}}, \bibinfo {author} {\bibfnamefont {G.}~\bibnamefont
  {L{\o}voll}}, \bibinfo {author} {\bibfnamefont {Y.}~\bibnamefont
  {M{\'e}heust}}, \bibinfo {author} {\bibfnamefont {K.~J.}\ \bibnamefont
  {M{\aa}l{\o}y}},\ and\ \bibinfo {author} {\bibfnamefont {J.}~\bibnamefont
  {Schmittbuhl}},\ }\bibfield  {title} {\bibinfo {title} {Influence of
  pore-scale disorder on viscous fingering during drainage},\ }\href
  {https://doi.org/10.1209/epl/i2005-10136-9} {\bibfield  {journal} {\bibinfo
  {journal} {Europhysics Letters}\ }\textbf {\bibinfo {volume} {71}},\ \bibinfo
  {pages} {583} (\bibinfo {year} {2005})}\BibitemShut {NoStop}%
\bibitem [{\citenamefont {Assouline}(2021)}]{assouline2021}%
  \BibitemOpen
  \bibfield  {author} {\bibinfo {author} {\bibfnamefont {S.}~\bibnamefont
  {Assouline}},\ }\bibfield  {title} {\bibinfo {title} {What can we learn from
  the water retention characteristic of a soil regarding its hydrological and
  agricultural functions? {R}eview and analysis of actual knowledge},\ }\href
  {https://doi.org/10.1029/2021WR031026} {\bibfield  {journal} {\bibinfo
  {journal} {Water Resources Research}\ }\textbf {\bibinfo {volume} {57}}
  (\bibinfo {year} {2021})}\BibitemShut {NoStop}%
\bibitem [{\citenamefont {Birdsell}\ \emph {et~al.}(2015)\citenamefont
  {Birdsell}, \citenamefont {Rajaram}, \citenamefont {Dempsey},\ and\
  \citenamefont {Viswanathan}}]{birdsell2015}%
  \BibitemOpen
  \bibfield  {author} {\bibinfo {author} {\bibfnamefont {D.~T.}\ \bibnamefont
  {Birdsell}}, \bibinfo {author} {\bibfnamefont {H.}~\bibnamefont {Rajaram}},
  \bibinfo {author} {\bibfnamefont {D.}~\bibnamefont {Dempsey}},\ and\ \bibinfo
  {author} {\bibfnamefont {H.~S.}\ \bibnamefont {Viswanathan}},\ }\bibfield
  {title} {\bibinfo {title} {Hydraulic fracturing fluid migration in the
  subsurface: A review and expanded modeling results},\ }\href
  {https://doi.org/10.1002/2015WR017810} {\bibfield  {journal} {\bibinfo
  {journal} {Water Resources Research}\ }\textbf {\bibinfo {volume} {51}},\
  \bibinfo {pages} {7159} (\bibinfo {year} {2015})}\BibitemShut {NoStop}%
\bibitem [{\citenamefont {Molofsky}\ \emph {et~al.}(2021)\citenamefont
  {Molofsky}, \citenamefont {Connor}, \citenamefont {{Van De Ven}},
  \citenamefont {Hemingway}, \citenamefont {Richardson}, \citenamefont
  {Strasert}, \citenamefont {McGuire},\ and\ \citenamefont
  {Paquette}}]{molofsky2021}%
  \BibitemOpen
  \bibfield  {author} {\bibinfo {author} {\bibfnamefont {L.}~\bibnamefont
  {Molofsky}}, \bibinfo {author} {\bibfnamefont {J.~A.}\ \bibnamefont
  {Connor}}, \bibinfo {author} {\bibfnamefont {C.~J.}\ \bibnamefont {{Van De
  Ven}}}, \bibinfo {author} {\bibfnamefont {M.~P.}\ \bibnamefont {Hemingway}},
  \bibinfo {author} {\bibfnamefont {S.~D.}\ \bibnamefont {Richardson}},
  \bibinfo {author} {\bibfnamefont {B.~A.}\ \bibnamefont {Strasert}}, \bibinfo
  {author} {\bibfnamefont {T.~M.}\ \bibnamefont {McGuire}},\ and\ \bibinfo
  {author} {\bibfnamefont {S.~M.}\ \bibnamefont {Paquette}},\ }\bibfield
  {title} {\bibinfo {title} {A review of physical, chemical, and hydrogeologic
  characteristics of stray gas migration: Implications for investigation and
  remediation},\ }\href {https://doi.org/10.1016/j.scitotenv.2021.146234}
  {\bibfield  {journal} {\bibinfo  {journal} {Science of The Total
  Environment}\ }\textbf {\bibinfo {volume} {779}},\ \bibinfo {pages} {146234}
  (\bibinfo {year} {2021})}\BibitemShut {NoStop}%
\bibitem [{\citenamefont {Harshani}\ \emph {et~al.}(2016)\citenamefont
  {Harshani}, \citenamefont {Galindo-Torres}, \citenamefont {Scheuermann},\
  and\ \citenamefont {Muhlhaus}}]{Harshani2017}%
  \BibitemOpen
  \bibfield  {author} {\bibinfo {author} {\bibfnamefont {H.~M.}\ \bibnamefont
  {Harshani}}, \bibinfo {author} {\bibfnamefont {S.~A.}\ \bibnamefont
  {Galindo-Torres}}, \bibinfo {author} {\bibfnamefont {A.}~\bibnamefont
  {Scheuermann}},\ and\ \bibinfo {author} {\bibfnamefont {H.~B.}\ \bibnamefont
  {Muhlhaus}},\ }\bibfield  {title} {\bibinfo {title} {Experimental study of
  porous media flow using hydro-gel beads and {LED} based {PIV}},\ }\href
  {https://doi.org/10.1088/1361-6501/28/1/015902} {\bibfield  {journal}
  {\bibinfo  {journal} {Measurement Science and Technology}\ }\textbf {\bibinfo
  {volume} {28}},\ \bibinfo {pages} {015902} (\bibinfo {year}
  {2016})}\BibitemShut {NoStop}%
\bibitem [{\citenamefont {Moroni}\ \emph {et~al.}(2007)\citenamefont {Moroni},
  \citenamefont {Kleinfelter},\ and\ \citenamefont {Cushman}}]{Moroni2007}%
  \BibitemOpen
  \bibfield  {author} {\bibinfo {author} {\bibfnamefont {M.}~\bibnamefont
  {Moroni}}, \bibinfo {author} {\bibfnamefont {N.}~\bibnamefont
  {Kleinfelter}},\ and\ \bibinfo {author} {\bibfnamefont {J.~H.}\ \bibnamefont
  {Cushman}},\ }\bibfield  {title} {\bibinfo {title} {Analysis of dispersion in
  porous media via matched-index particle tracking velocimetry experiments},\
  }\href {https://doi.org/10.1016/j.advwatres.2006.02.005} {\bibfield
  {journal} {\bibinfo  {journal} {Advances in Water Resources}\ }\textbf
  {\bibinfo {volume} {30}},\ \bibinfo {pages} {1} (\bibinfo {year}
  {2007})}\BibitemShut {NoStop}%
\bibitem [{\citenamefont {St{\"{o}}hr}\ \emph {et~al.}(2003)\citenamefont
  {St{\"{o}}hr}, \citenamefont {Roth},\ and\ \citenamefont
  {J{\"{a}}hne}}]{stohr2003}%
  \BibitemOpen
  \bibfield  {author} {\bibinfo {author} {\bibfnamefont {M.}~\bibnamefont
  {St{\"{o}}hr}}, \bibinfo {author} {\bibfnamefont {K.}~\bibnamefont {Roth}},\
  and\ \bibinfo {author} {\bibfnamefont {B.}~\bibnamefont {J{\"{a}}hne}},\
  }\bibfield  {title} {\bibinfo {title} {Measurement of {3D} pore-scale flow in
  index-matched porous media},\ }\href
  {https://doi.org/10.1007/s00348-003-0641-x} {\bibfield  {journal} {\bibinfo
  {journal} {Experiments in Fluids}\ }\textbf {\bibinfo {volume} {35}},\
  \bibinfo {pages} {159} (\bibinfo {year} {2003})}\BibitemShut {NoStop}%
\bibitem [{\citenamefont {Roth}\ \emph {et~al.}(2015)\citenamefont {Roth},
  \citenamefont {Mont-Eton}, \citenamefont {Gilbert}, \citenamefont {Lei},\
  and\ \citenamefont {Mays}}]{roth2015}%
  \BibitemOpen
  \bibfield  {author} {\bibinfo {author} {\bibfnamefont {E.~J.}\ \bibnamefont
  {Roth}}, \bibinfo {author} {\bibfnamefont {M.~E.}\ \bibnamefont {Mont-Eton}},
  \bibinfo {author} {\bibfnamefont {B.}~\bibnamefont {Gilbert}}, \bibinfo
  {author} {\bibfnamefont {T.~C.}\ \bibnamefont {Lei}},\ and\ \bibinfo {author}
  {\bibfnamefont {D.~C.}\ \bibnamefont {Mays}},\ }\bibfield  {title} {\bibinfo
  {title} {Measurement of colloidal phenomena during flow through refractive
  index matched porous media},\ }\href {https://doi.org/10.1063/1.4935576}
  {\bibfield  {journal} {\bibinfo  {journal} {Review of Scientific
  Instruments}\ }\textbf {\bibinfo {volume} {86}},\ \bibinfo {pages} {1}
  (\bibinfo {year} {2015})}\BibitemShut {NoStop}%
\bibitem [{\citenamefont {Kong}\ \emph {et~al.}(2011)\citenamefont {Kong},
  \citenamefont {Holzner}, \citenamefont {Stauffer},\ and\ \citenamefont
  {Kinzelbach}}]{Holzner2011}%
  \BibitemOpen
  \bibfield  {author} {\bibinfo {author} {\bibfnamefont {X.-Z.}\ \bibnamefont
  {Kong}}, \bibinfo {author} {\bibfnamefont {M.}~\bibnamefont {Holzner}},
  \bibinfo {author} {\bibfnamefont {F.}~\bibnamefont {Stauffer}},\ and\
  \bibinfo {author} {\bibfnamefont {W.}~\bibnamefont {Kinzelbach}},\ }\bibfield
   {title} {\bibinfo {title} {Time-resolved {3D} visualization of air injection
  in a liquid-saturated refractive-index-matched porous medium},\ }\href
  {https://doi.org/10.1007/s00348-010-1018-6} {\bibfield  {journal} {\bibinfo
  {journal} {Experiments in Fluids}\ }\textbf {\bibinfo {volume} {50}},\
  \bibinfo {pages} {1659} (\bibinfo {year} {2011})}\BibitemShut {NoStop}%
\bibitem [{\citenamefont {Kang}\ \emph {et~al.}(2010)\citenamefont {Kang},
  \citenamefont {Lee}, \citenamefont {Nam}, \citenamefont {Kim}, \citenamefont
  {Park}, \citenamefont {Lee},\ and\ \citenamefont {Kwang}}]{kang2010}%
  \BibitemOpen
  \bibfield  {author} {\bibinfo {author} {\bibfnamefont {J.~H.}\ \bibnamefont
  {Kang}}, \bibinfo {author} {\bibfnamefont {K.~J.}\ \bibnamefont {Lee}},
  \bibinfo {author} {\bibfnamefont {J.~H.}\ \bibnamefont {Nam}}, \bibinfo
  {author} {\bibfnamefont {C.~J.}\ \bibnamefont {Kim}}, \bibinfo {author}
  {\bibfnamefont {H.~S.}\ \bibnamefont {Park}}, \bibinfo {author}
  {\bibfnamefont {S.}~\bibnamefont {Lee}},\ and\ \bibinfo {author}
  {\bibfnamefont {I.}~\bibnamefont {Kwang}},\ }\bibfield  {title} {\bibinfo
  {title} {Visualization of invasion-percolation drainage process in porous
  media using density-matched immiscible fluids and refractive-index-matched
  solid structures},\ }\href {https://doi.org/10.1016/j.jpowsour.2009.11.087}
  {\bibfield  {journal} {\bibinfo  {journal} {Journal of Power Sources}\
  }\textbf {\bibinfo {volume} {195}},\ \bibinfo {pages} {2608} (\bibinfo {year}
  {2010})}\BibitemShut {NoStop}%
\bibitem [{\citenamefont {Ovdat}\ and\ \citenamefont
  {Berkowitz}(2006)}]{Ovdat2006}%
  \BibitemOpen
  \bibfield  {author} {\bibinfo {author} {\bibfnamefont {H.}~\bibnamefont
  {Ovdat}}\ and\ \bibinfo {author} {\bibfnamefont {B.}~\bibnamefont
  {Berkowitz}},\ }\bibfield  {title} {\bibinfo {title} {Pore-scale study of
  drainage displacement under combined capillary and gravity effects in
  index-matched porous media},\ }\href {https://doi.org/10.1029/2005WR004553}
  {\bibfield  {journal} {\bibinfo  {journal} {Water Resources Research}\
  }\textbf {\bibinfo {volume} {42}} (\bibinfo {year} {2006})}\BibitemShut
  {NoStop}%
\bibitem [{\citenamefont {Sharma}\ \emph {et~al.}(2011)\citenamefont {Sharma},
  \citenamefont {Aswathi}, \citenamefont {Sane}, \citenamefont {Ghosh},\ and\
  \citenamefont {Bhattacharya}}]{sharma2011}%
  \BibitemOpen
  \bibfield  {author} {\bibinfo {author} {\bibfnamefont {P.}~\bibnamefont
  {Sharma}}, \bibinfo {author} {\bibfnamefont {P.}~\bibnamefont {Aswathi}},
  \bibinfo {author} {\bibfnamefont {A.}~\bibnamefont {Sane}}, \bibinfo {author}
  {\bibfnamefont {S.}~\bibnamefont {Ghosh}},\ and\ \bibinfo {author}
  {\bibfnamefont {S.}~\bibnamefont {Bhattacharya}},\ }\bibfield  {title}
  {\bibinfo {title} {{Three-dimensional real-time imaging of bi-phasic flow
  through porous media}},\ }\href {https://doi.org/10.1063/1.3658822}
  {\bibfield  {journal} {\bibinfo  {journal} {Review of Scientific
  Instruments}\ }\textbf {\bibinfo {volume} {82}} (\bibinfo {year}
  {2011})}\BibitemShut {NoStop}%
\bibitem [{\citenamefont {Datta}\ \emph {et~al.}(2014)\citenamefont {Datta},
  \citenamefont {Dupin},\ and\ \citenamefont {Weitz}}]{datta2014}%
  \BibitemOpen
  \bibfield  {author} {\bibinfo {author} {\bibfnamefont {S.~S.}\ \bibnamefont
  {Datta}}, \bibinfo {author} {\bibfnamefont {J.~B.}\ \bibnamefont {Dupin}},\
  and\ \bibinfo {author} {\bibfnamefont {D.~A.}\ \bibnamefont {Weitz}},\
  }\bibfield  {title} {\bibinfo {title} {Fluid breakup during simultaneous
  two-phase flow through a three-dimensional porous medium},\ }\href
  {https://doi.org/10.1063/1.4884955} {\bibfield  {journal} {\bibinfo
  {journal} {Physics of Fluids}\ }\textbf {\bibinfo {volume} {26}} (\bibinfo
  {year} {2014})}\BibitemShut {NoStop}%
\bibitem [{\citenamefont {do~Nascimento}\ \emph {et~al.}(2019)\citenamefont
  {do~Nascimento}, \citenamefont {{Vimieiro Junior}}, \citenamefont
  {Paciornik},\ and\ \citenamefont {Carvalho}}]{nascimento2019}%
  \BibitemOpen
  \bibfield  {author} {\bibinfo {author} {\bibfnamefont {D.~F.}\ \bibnamefont
  {do~Nascimento}}, \bibinfo {author} {\bibfnamefont {J.~R.}\ \bibnamefont
  {{Vimieiro Junior}}}, \bibinfo {author} {\bibfnamefont {S.}~\bibnamefont
  {Paciornik}},\ and\ \bibinfo {author} {\bibfnamefont {M.~S.}\ \bibnamefont
  {Carvalho}},\ }\bibfield  {title} {\bibinfo {title} {Pore scale visualization
  of drainage in {3D} porous media by confocal microscopy},\ }\href
  {https://doi.org/10.1038/s41598-019-48803-z} {\bibfield  {journal} {\bibinfo
  {journal} {Scientific Reports}\ }\textbf {\bibinfo {volume} {9}},\ \bibinfo
  {pages} {1} (\bibinfo {year} {2019})}\BibitemShut {NoStop}%
\bibitem [{\citenamefont {Dalbe}\ and\ \citenamefont
  {Juanes}(2018)}]{Dalbe-Morphodynamics}%
  \BibitemOpen
  \bibfield  {author} {\bibinfo {author} {\bibfnamefont {M.-J.}\ \bibnamefont
  {Dalbe}}\ and\ \bibinfo {author} {\bibfnamefont {R.}~\bibnamefont {Juanes}},\
  }\bibfield  {title} {\bibinfo {title} {Morphodynamics of fluid-fluid
  displacement in three-dimensional deformable granular media},\ }\href
  {https://doi.org/10.1103/PhysRevApplied.9.024028} {\bibfield  {journal}
  {\bibinfo  {journal} {Physical Review Applied}\ }\textbf {\bibinfo {volume}
  {9}},\ \bibinfo {pages} {024028} (\bibinfo {year} {2018})}\BibitemShut
  {NoStop}%
\bibitem [{\citenamefont {Wilkinson}\ and\ \citenamefont
  {Willemsen}(1983)}]{wilkinson1983invasion}%
  \BibitemOpen
  \bibfield  {author} {\bibinfo {author} {\bibfnamefont {D.}~\bibnamefont
  {Wilkinson}}\ and\ \bibinfo {author} {\bibfnamefont {J.~F.}\ \bibnamefont
  {Willemsen}},\ }\bibfield  {title} {\bibinfo {title} {Invasion percolation: A
  new form of percolation theory},\ }\href
  {https://doi.org/10.1088/0305-4470/16/14/028} {\bibfield  {journal} {\bibinfo
   {journal} {Journal of Physics A: Mathematical and General}\ }\textbf
  {\bibinfo {volume} {16}},\ \bibinfo {pages} {3365} (\bibinfo {year}
  {1983})}\BibitemShut {NoStop}%
\bibitem [{\citenamefont {Glass}\ \emph {et~al.}(1989)\citenamefont {Glass},
  \citenamefont {Steenhuis},\ and\ \citenamefont
  {Parlange}}]{glass1989wetting}%
  \BibitemOpen
  \bibfield  {author} {\bibinfo {author} {\bibfnamefont {R.}~\bibnamefont
  {Glass}}, \bibinfo {author} {\bibfnamefont {T.}~\bibnamefont {Steenhuis}},\
  and\ \bibinfo {author} {\bibfnamefont {J.-Y.}\ \bibnamefont {Parlange}},\
  }\bibfield  {title} {\bibinfo {title} {Wetting front instability 2:
  Experimental determination of relationships between system parameters and
  two-dimensional unstable flow field behavior in initially dry porous media},\
  }\href {https://doi.org/10.1029/WR025i006p01195} {\bibfield  {journal}
  {\bibinfo  {journal} {Water Resources Research}\ }\textbf {\bibinfo {volume}
  {25}},\ \bibinfo {pages} {1195} (\bibinfo {year} {1989})}\BibitemShut
  {NoStop}%
\bibitem [{\citenamefont {Flekk{\o}y}\ \emph {et~al.}(2002)\citenamefont
  {Flekk{\o}y}, \citenamefont {Schmittbuhl}, \citenamefont {L{\o}vholt},
  \citenamefont {Oxaal}, \citenamefont {M{\aa}l{\o}y},\ and\ \citenamefont
  {Aagaard}}]{flekkoy2002flow}%
  \BibitemOpen
  \bibfield  {author} {\bibinfo {author} {\bibfnamefont {E.~G.}\ \bibnamefont
  {Flekk{\o}y}}, \bibinfo {author} {\bibfnamefont {J.}~\bibnamefont
  {Schmittbuhl}}, \bibinfo {author} {\bibfnamefont {F.}~\bibnamefont
  {L{\o}vholt}}, \bibinfo {author} {\bibfnamefont {U.}~\bibnamefont {Oxaal}},
  \bibinfo {author} {\bibfnamefont {K.~J.}\ \bibnamefont {M{\aa}l{\o}y}},\ and\
  \bibinfo {author} {\bibfnamefont {P.}~\bibnamefont {Aagaard}},\ }\bibfield
  {title} {\bibinfo {title} {Flow paths in wetting unsaturated flow:
  Experiments and simulations},\ }\href
  {https://doi.org/10.1103/PhysRevE.65.036312} {\bibfield  {journal} {\bibinfo
  {journal} {Physical Review E}\ }\textbf {\bibinfo {volume} {65}},\ \bibinfo
  {pages} {036312} (\bibinfo {year} {2002})}\BibitemShut {NoStop}%
\bibitem [{\citenamefont {Birovljev}\ \emph {et~al.}(1991)\citenamefont
  {Birovljev}, \citenamefont {Furuberg}, \citenamefont {Feder}, \citenamefont
  {J{\o}ssang}, \citenamefont {M{\aa}l{\o}y},\ and\ \citenamefont
  {Aharony}}]{birovljev1991gravity}%
  \BibitemOpen
  \bibfield  {author} {\bibinfo {author} {\bibfnamefont {A.}~\bibnamefont
  {Birovljev}}, \bibinfo {author} {\bibfnamefont {L.}~\bibnamefont {Furuberg}},
  \bibinfo {author} {\bibfnamefont {J.}~\bibnamefont {Feder}}, \bibinfo
  {author} {\bibfnamefont {T.}~\bibnamefont {J{\o}ssang}}, \bibinfo {author}
  {\bibfnamefont {K.}~\bibnamefont {M{\aa}l{\o}y}},\ and\ \bibinfo {author}
  {\bibfnamefont {A.}~\bibnamefont {Aharony}},\ }\bibfield  {title} {\bibinfo
  {title} {Gravity invasion percolation in two dimensions: Experiment and
  simulation},\ }\href {https://doi.org/10.1103/PhysRevLett.67.584} {\bibfield
  {journal} {\bibinfo  {journal} {Physical Review Letters}\ }\textbf {\bibinfo
  {volume} {67}},\ \bibinfo {pages} {584} (\bibinfo {year} {1991})}\BibitemShut
  {NoStop}%
\bibitem [{\citenamefont {M\'eheust}\ \emph {et~al.}(2002)\citenamefont
  {M\'eheust}, \citenamefont {L\o{}voll}, \citenamefont {M\aa{}l\o{}y},\ and\
  \citenamefont {Schmittbuhl}}]{meheust2002}%
  \BibitemOpen
  \bibfield  {author} {\bibinfo {author} {\bibfnamefont {Y.}~\bibnamefont
  {M\'eheust}}, \bibinfo {author} {\bibfnamefont {G.}~\bibnamefont
  {L\o{}voll}}, \bibinfo {author} {\bibfnamefont {K.~J.}\ \bibnamefont
  {M\aa{}l\o{}y}},\ and\ \bibinfo {author} {\bibfnamefont {J.}~\bibnamefont
  {Schmittbuhl}},\ }\bibfield  {title} {\bibinfo {title} {Interface scaling in
  a two-dimensional porous medium under combined viscous, gravity, and
  capillary effects},\ }\href {https://doi.org/10.1103/PhysRevE.66.051603}
  {\bibfield  {journal} {\bibinfo  {journal} {Physical Review E}\ }\textbf
  {\bibinfo {volume} {66}},\ \bibinfo {pages} {051603} (\bibinfo {year}
  {2002})}\BibitemShut {NoStop}%
\bibitem [{\citenamefont {Frette}\ \emph {et~al.}(1992)\citenamefont {Frette},
  \citenamefont {Feder}, \citenamefont {J\o{}ssang},\ and\ \citenamefont
  {Meakin}}]{frette1992a}%
  \BibitemOpen
  \bibfield  {author} {\bibinfo {author} {\bibfnamefont {V.}~\bibnamefont
  {Frette}}, \bibinfo {author} {\bibfnamefont {J.}~\bibnamefont {Feder}},
  \bibinfo {author} {\bibfnamefont {T.}~\bibnamefont {J\o{}ssang}},\ and\
  \bibinfo {author} {\bibfnamefont {P.}~\bibnamefont {Meakin}},\ }\bibfield
  {title} {\bibinfo {title} {Buoyancy-driven fluid migration in porous media},\
  }\href {https://doi.org/10.1103/PhysRevLett.68.3164} {\bibfield  {journal}
  {\bibinfo  {journal} {Physical Review Letters}\ }\textbf {\bibinfo {volume}
  {68}},\ \bibinfo {pages} {3164} (\bibinfo {year} {1992})}\BibitemShut
  {NoStop}%
\bibitem [{\citenamefont {L{\o}voll}\ \emph {et~al.}(2005)\citenamefont
  {L{\o}voll}, \citenamefont {M{\'e}heust}, \citenamefont {M{\aa}l{\o}y},
  \citenamefont {Aker},\ and\ \citenamefont {Schmittbuhl}}]{Lovoll2005}%
  \BibitemOpen
  \bibfield  {author} {\bibinfo {author} {\bibfnamefont {G.}~\bibnamefont
  {L{\o}voll}}, \bibinfo {author} {\bibfnamefont {Y.}~\bibnamefont
  {M{\'e}heust}}, \bibinfo {author} {\bibfnamefont {K.~J.}\ \bibnamefont
  {M{\aa}l{\o}y}}, \bibinfo {author} {\bibfnamefont {E.}~\bibnamefont {Aker}},\
  and\ \bibinfo {author} {\bibfnamefont {J.}~\bibnamefont {Schmittbuhl}},\
  }\bibfield  {title} {\bibinfo {title} {Competition of gravity, capillary and
  viscous forces during drainage in a two-dimensional porous medium, a pore
  scale study},\ }\href {https://doi.org/10.1016/j.energy.2004.03.100}
  {\bibfield  {journal} {\bibinfo  {journal} {Energy}\ }\textbf {\bibinfo
  {volume} {30}},\ \bibinfo {pages} {861} (\bibinfo {year} {2005})}\BibitemShut
  {NoStop}%
\bibitem [{\citenamefont {Toussaint}\ \emph {et~al.}(2012)\citenamefont
  {Toussaint}, \citenamefont {M\aa{}l\o{}y}, \citenamefont {M\'eheust},
  \citenamefont {L\o{}voll}, \citenamefont {Jankov}, \citenamefont
  {Sch{\"a}fer},\ and\ \citenamefont {Schmittbuhl}}]{toussaint2012}%
  \BibitemOpen
  \bibfield  {author} {\bibinfo {author} {\bibfnamefont {R.}~\bibnamefont
  {Toussaint}}, \bibinfo {author} {\bibfnamefont {K.}~\bibnamefont
  {M\aa{}l\o{}y}}, \bibinfo {author} {\bibfnamefont {Y.}~\bibnamefont
  {M\'eheust}}, \bibinfo {author} {\bibfnamefont {G.}~\bibnamefont
  {L\o{}voll}}, \bibinfo {author} {\bibfnamefont {M.}~\bibnamefont {Jankov}},
  \bibinfo {author} {\bibfnamefont {G.}~\bibnamefont {Sch{\"a}fer}},\ and\
  \bibinfo {author} {\bibfnamefont {J.}~\bibnamefont {Schmittbuhl}},\
  }\bibfield  {title} {\bibinfo {title} {Two-phase flow: Structure, upscaling,
  and consequences for macroscopic transport properties},\ }\href
  {https://doi.org/10.2136/vzj2011.0123} {\bibfield  {journal} {\bibinfo
  {journal} {Vadose Zone Journal}\ }\textbf {\bibinfo {volume} {11}} (\bibinfo
  {year} {2012})}\BibitemShut {NoStop}%
\bibitem [{\citenamefont {M{\aa}l{\o}y}\ \emph {et~al.}(2021)\citenamefont
  {M{\aa}l{\o}y}, \citenamefont {Moura}, \citenamefont {Hansen}, \citenamefont
  {Flekk{\o}y},\ and\ \citenamefont {Toussaint}}]{maaloy2021burst}%
  \BibitemOpen
  \bibfield  {author} {\bibinfo {author} {\bibfnamefont {K.~J.}\ \bibnamefont
  {M{\aa}l{\o}y}}, \bibinfo {author} {\bibfnamefont {M.}~\bibnamefont {Moura}},
  \bibinfo {author} {\bibfnamefont {A.}~\bibnamefont {Hansen}}, \bibinfo
  {author} {\bibfnamefont {E.~G.}\ \bibnamefont {Flekk{\o}y}},\ and\ \bibinfo
  {author} {\bibfnamefont {R.}~\bibnamefont {Toussaint}},\ }\bibfield  {title}
  {\bibinfo {title} {Burst dynamics, upscaling and dissipation of slow drainage
  in porous media},\ }\href {https://doi.org/10.3389/fphy.2021.796019}
  {\bibfield  {journal} {\bibinfo  {journal} {Frontiers in Physics}\ }\textbf
  {\bibinfo {volume} {9}},\ \bibinfo {pages} {796019} (\bibinfo {year}
  {2021})}\BibitemShut {NoStop}%
\bibitem [{\citenamefont {Brodin}(2019)}]{brodin2019new}%
  \BibitemOpen
  \bibfield  {author} {\bibinfo {author} {\bibfnamefont {J.~F.}\ \bibnamefont
  {Brodin}},\ }\emph {\bibinfo {title} {A New Vision for {3D} Experiments on
  Flow in Porous Media}},\ \href {https://doi.org/10.31219/osf.io/yfq74}
  {Master's thesis},\ \bibinfo  {school} {University of Oslo} (\bibinfo {year}
  {2019})\BibitemShut {NoStop}%
\bibitem [{\citenamefont {Brodin}\ \emph
  {et~al.}(2022{\natexlab{a}})\citenamefont {Brodin}, \citenamefont {Moura},
  \citenamefont {Toussaint}, \citenamefont {M\aa{}l\o{}y},\ and\ \citenamefont
  {Rikvold}}]{brodin2020visualization}%
  \BibitemOpen
  \bibfield  {author} {\bibinfo {author} {\bibfnamefont {J.~F.}\ \bibnamefont
  {Brodin}}, \bibinfo {author} {\bibfnamefont {M.}~\bibnamefont {Moura}},
  \bibinfo {author} {\bibfnamefont {R.}~\bibnamefont {Toussaint}}, \bibinfo
  {author} {\bibfnamefont {K.~J.}\ \bibnamefont {M\aa{}l\o{}y}},\ and\ \bibinfo
  {author} {\bibfnamefont {P.~A.}\ \bibnamefont {Rikvold}},\ }\bibfield
  {title} {\bibinfo {title} {Visualization by optical fluorescence of two-phase
  flow in a three-dimensional porous medium},\ }\href
  {https://doi.org/10.1088/1742-6596/2241/1/012004} {\bibfield  {journal}
  {\bibinfo  {journal} {Journal of Physics: Conference Series}\ }\textbf
  {\bibinfo {volume} {2241}},\ \bibinfo {pages} {012004} (\bibinfo {year}
  {2022}{\natexlab{a}})}\BibitemShut {NoStop}%
\bibitem [{\citenamefont {Brodin}\ \emph
  {et~al.}(2022{\natexlab{b}})\citenamefont {Brodin}, \citenamefont {Rikvold},
  \citenamefont {Moura}, \citenamefont {Toussaint},\ and\ \citenamefont
  {M{\aa}l{\o}y}}]{brodin2022}%
  \BibitemOpen
  \bibfield  {author} {\bibinfo {author} {\bibfnamefont {J.~F.}\ \bibnamefont
  {Brodin}}, \bibinfo {author} {\bibfnamefont {P.~A.}\ \bibnamefont {Rikvold}},
  \bibinfo {author} {\bibfnamefont {M.}~\bibnamefont {Moura}}, \bibinfo
  {author} {\bibfnamefont {R.}~\bibnamefont {Toussaint}},\ and\ \bibinfo
  {author} {\bibfnamefont {K.~J.}\ \bibnamefont {M{\aa}l{\o}y}},\ }\bibfield
  {title} {\bibinfo {title} {Competing gravitational and viscous effects in
  {3D} two-phase flow investigated with a table-top optical scanner},\ }\href
  {https://doi.org/10.3389/fphy.2022.936915} {\bibfield  {journal} {\bibinfo
  {journal} {Frontiers in Physics}\ }\textbf {\bibinfo {volume} {10}} (\bibinfo
  {year} {2022}{\natexlab{b}})}\BibitemShut {NoStop}%
\bibitem [{\citenamefont {Saffman}\ and\ \citenamefont
  {Taylor}(1958)}]{saffman1958penetration}%
  \BibitemOpen
  \bibfield  {author} {\bibinfo {author} {\bibfnamefont {P.~G.}\ \bibnamefont
  {Saffman}}\ and\ \bibinfo {author} {\bibfnamefont {G.~I.}\ \bibnamefont
  {Taylor}},\ }\bibfield  {title} {\bibinfo {title} {The penetration of a fluid
  into a porous medium or {Hele-Shaw} cell containing a more viscous liquid},\
  }\href {https://doi.org/10.1098/rspa.1958.0085} {\bibfield  {journal}
  {\bibinfo  {journal} {Proceedings of the Royal Society of London. Series A.
  Mathematical and Physical Sciences}\ }\textbf {\bibinfo {volume} {245}},\
  \bibinfo {pages} {312} (\bibinfo {year} {1958})}\BibitemShut {NoStop}%
\bibitem [{\citenamefont {Cinar}\ \emph {et~al.}(2009)\citenamefont {Cinar},
  \citenamefont {Riaz},\ and\ \citenamefont
  {Tchelepi}}]{cinar2009experimental}%
  \BibitemOpen
  \bibfield  {author} {\bibinfo {author} {\bibfnamefont {Y.}~\bibnamefont
  {Cinar}}, \bibinfo {author} {\bibfnamefont {A.}~\bibnamefont {Riaz}},\ and\
  \bibinfo {author} {\bibfnamefont {H.~A.}\ \bibnamefont {Tchelepi}},\
  }\bibfield  {title} {\bibinfo {title} {Experimental study of co$_2$ injection
  into saline formations},\ }\href {https://doi.org/10.2118/110628-PA}
  {\bibfield  {journal} {\bibinfo  {journal} {{SPE} Journal}\ }\textbf
  {\bibinfo {volume} {14}},\ \bibinfo {pages} {588} (\bibinfo {year}
  {2009})}\BibitemShut {NoStop}%
\bibitem [{\citenamefont {Lenormand}\ \emph {et~al.}(1983)\citenamefont
  {Lenormand}, \citenamefont {Zarcone},\ and\ \citenamefont
  {Sarr}}]{lenormand1983}%
  \BibitemOpen
  \bibfield  {author} {\bibinfo {author} {\bibfnamefont {R.}~\bibnamefont
  {Lenormand}}, \bibinfo {author} {\bibfnamefont {C.}~\bibnamefont {Zarcone}},\
  and\ \bibinfo {author} {\bibfnamefont {A.}~\bibnamefont {Sarr}},\ }\bibfield
  {title} {\bibinfo {title} {Mechanisms of the displacement of one fluid by
  another in a network of capillary ducts},\ }\href
  {https://doi.org/10.1017/S0022112083003110} {\bibfield  {journal} {\bibinfo
  {journal} {Journal of Fluid Mechanics}\ }\textbf {\bibinfo {volume} {135}},\
  \bibinfo {pages} {337} (\bibinfo {year} {1983})}\BibitemShut {NoStop}%
\bibitem [{\citenamefont {Lenormand}\ and\ \citenamefont
  {Zarcone}(1985)}]{lenormand1985}%
  \BibitemOpen
  \bibfield  {author} {\bibinfo {author} {\bibfnamefont {R.}~\bibnamefont
  {Lenormand}}\ and\ \bibinfo {author} {\bibfnamefont {C.}~\bibnamefont
  {Zarcone}},\ }\bibfield  {title} {\bibinfo {title} {Invasion percolation in
  an etched network: Measurement of a fractal dimension},\ }\href
  {https://doi.org/10.1103/PhysRevLett.54.2226} {\bibfield  {journal} {\bibinfo
   {journal} {Physical Review Letters}\ }\textbf {\bibinfo {volume} {54}},\
  \bibinfo {pages} {2226} (\bibinfo {year} {1985})}\BibitemShut {NoStop}%
\bibitem [{\citenamefont {M\aa{}l\o{}y}\ \emph {et~al.}(1985)\citenamefont
  {M\aa{}l\o{}y}, \citenamefont {Feder},\ and\ \citenamefont
  {J\o{}ssang}}]{maloy1985}%
  \BibitemOpen
  \bibfield  {author} {\bibinfo {author} {\bibfnamefont {K.~J.}\ \bibnamefont
  {M\aa{}l\o{}y}}, \bibinfo {author} {\bibfnamefont {J.}~\bibnamefont
  {Feder}},\ and\ \bibinfo {author} {\bibfnamefont {T.}~\bibnamefont
  {J\o{}ssang}},\ }\bibfield  {title} {\bibinfo {title} {Viscous fingering
  fractals in porous media},\ }\href
  {https://doi.org/10.1103/PhysRevLett.55.2688} {\bibfield  {journal} {\bibinfo
   {journal} {Physical Review Letters}\ }\textbf {\bibinfo {volume} {55}},\
  \bibinfo {pages} {2688} (\bibinfo {year} {1985})}\BibitemShut {NoStop}%
\bibitem [{\citenamefont {Hele-Shaw}(1898)}]{Hele-Shaw1898}%
  \BibitemOpen
  \bibfield  {author} {\bibinfo {author} {\bibfnamefont {H.~S.}\ \bibnamefont
  {Hele-Shaw}},\ }\bibfield  {title} {\bibinfo {title} {The flow of water},\
  }\href {https://doi.org/10.1038/058034a0} {\bibfield  {journal} {\bibinfo
  {journal} {Nature}\ }\textbf {\bibinfo {volume} {58}},\ \bibinfo {pages} {34}
  (\bibinfo {year} {1898})}\BibitemShut {NoStop}%
\bibitem [{\citenamefont {Van~Meurs}(1957)}]{van1957use}%
  \BibitemOpen
  \bibfield  {author} {\bibinfo {author} {\bibfnamefont {P.}~\bibnamefont
  {Van~Meurs}},\ }\bibfield  {title} {\bibinfo {title} {The use of transparent
  three-dimensional models for studying the mechanism of flow processes in oil
  reservoirs},\ }\href {https://doi.org/10.2118/678-G} {\bibfield  {journal}
  {\bibinfo  {journal} {Transactions of the AIME}\ }\textbf {\bibinfo {volume}
  {210}},\ \bibinfo {pages} {295} (\bibinfo {year} {1957})}\BibitemShut
  {NoStop}%
\bibitem [{\citenamefont {Stokes}\ \emph {et~al.}(1986)\citenamefont {Stokes},
  \citenamefont {Weitz}, \citenamefont {Gollub}, \citenamefont {Dougherty},
  \citenamefont {Robbins}, \citenamefont {Chaikin},\ and\ \citenamefont
  {Lindsay}}]{stokes1986}%
  \BibitemOpen
  \bibfield  {author} {\bibinfo {author} {\bibfnamefont {J.~P.}\ \bibnamefont
  {Stokes}}, \bibinfo {author} {\bibfnamefont {D.~A.}\ \bibnamefont {Weitz}},
  \bibinfo {author} {\bibfnamefont {J.~P.}\ \bibnamefont {Gollub}}, \bibinfo
  {author} {\bibfnamefont {A.}~\bibnamefont {Dougherty}}, \bibinfo {author}
  {\bibfnamefont {M.~O.}\ \bibnamefont {Robbins}}, \bibinfo {author}
  {\bibfnamefont {P.~M.}\ \bibnamefont {Chaikin}},\ and\ \bibinfo {author}
  {\bibfnamefont {H.~M.}\ \bibnamefont {Lindsay}},\ }\bibfield  {title}
  {\bibinfo {title} {Interfacial stability of immiscible displacement in a
  porous medium},\ }\href {https://doi.org/10.1103/PhysRevLett.57.1718}
  {\bibfield  {journal} {\bibinfo  {journal} {Phys. Rev. Lett.}\ }\textbf
  {\bibinfo {volume} {57}},\ \bibinfo {pages} {1718} (\bibinfo {year}
  {1986})}\BibitemShut {NoStop}%
\bibitem [{\citenamefont {Frette}\ \emph {et~al.}(1990)\citenamefont {Frette},
  \citenamefont {M{\aa}l{\o}y}, \citenamefont {Boger}, \citenamefont {Feder},
  \citenamefont {J\o{}ssang},\ and\ \citenamefont {Meakin}}]{frette1990}%
  \BibitemOpen
  \bibfield  {author} {\bibinfo {author} {\bibfnamefont {V.}~\bibnamefont
  {Frette}}, \bibinfo {author} {\bibfnamefont {K.~J.}\ \bibnamefont
  {M{\aa}l{\o}y}}, \bibinfo {author} {\bibfnamefont {F.}~\bibnamefont {Boger}},
  \bibinfo {author} {\bibfnamefont {J.}~\bibnamefont {Feder}}, \bibinfo
  {author} {\bibfnamefont {T.}~\bibnamefont {J\o{}ssang}},\ and\ \bibinfo
  {author} {\bibfnamefont {P.}~\bibnamefont {Meakin}},\ }\bibfield  {title}
  {\bibinfo {title} {Diffusion-limited-aggregation-like displacement structures
  in a three-dimensional porous medium},\ }\href
  {https://doi.org/10.1103/PhysRevA.42.3432} {\bibfield  {journal} {\bibinfo
  {journal} {Physical Review A}\ }\textbf {\bibinfo {volume} {42}},\ \bibinfo
  {pages} {3432} (\bibinfo {year} {1990})}\BibitemShut {NoStop}%
\bibitem [{\citenamefont {Frette}\ \emph {et~al.}(1994)\citenamefont {Frette},
  \citenamefont {Feder}, \citenamefont {J\o{}ssang}, \citenamefont {Meakin},\
  and\ \citenamefont {M\aa{}l\o{}y}}]{frette1994}%
  \BibitemOpen
  \bibfield  {author} {\bibinfo {author} {\bibfnamefont {V.}~\bibnamefont
  {Frette}}, \bibinfo {author} {\bibfnamefont {J.}~\bibnamefont {Feder}},
  \bibinfo {author} {\bibfnamefont {T.}~\bibnamefont {J\o{}ssang}}, \bibinfo
  {author} {\bibfnamefont {P.}~\bibnamefont {Meakin}},\ and\ \bibinfo {author}
  {\bibfnamefont {K.~J.}\ \bibnamefont {M\aa{}l\o{}y}},\ }\bibfield  {title}
  {\bibinfo {title} {Fast, immiscible fluid-fluid displacement in
  three-dimensional porous media at finite viscosity contrast},\ }\href
  {https://doi.org/10.1103/PhysRevE.50.2881} {\bibfield  {journal} {\bibinfo
  {journal} {Phys. Rev. E}\ }\textbf {\bibinfo {volume} {50}},\ \bibinfo
  {pages} {2881} (\bibinfo {year} {1994})}\BibitemShut {NoStop}%
\bibitem [{\citenamefont {Lenormand}\ \emph {et~al.}(1988)\citenamefont
  {Lenormand}, \citenamefont {Touboul},\ and\ \citenamefont
  {Zarcone}}]{lenormand1988}%
  \BibitemOpen
  \bibfield  {author} {\bibinfo {author} {\bibfnamefont {R.}~\bibnamefont
  {Lenormand}}, \bibinfo {author} {\bibfnamefont {E.}~\bibnamefont {Touboul}},\
  and\ \bibinfo {author} {\bibfnamefont {C.}~\bibnamefont {Zarcone}},\
  }\bibfield  {title} {\bibinfo {title} {Numerical models and experiments on
  immiscible displacements in porous media},\ }\href
  {https://doi.org/10.1017/S0022112088000953} {\bibfield  {journal} {\bibinfo
  {journal} {Journal of Fluid Mechanics}\ }\textbf {\bibinfo {volume} {189}},\
  \bibinfo {pages} {165} (\bibinfo {year} {1988})}\BibitemShut {NoStop}%
\bibitem [{\citenamefont {Holtzman}\ and\ \citenamefont
  {Segre}(2015)}]{holtzman2015}%
  \BibitemOpen
  \bibfield  {author} {\bibinfo {author} {\bibfnamefont {R.}~\bibnamefont
  {Holtzman}}\ and\ \bibinfo {author} {\bibfnamefont {E.}~\bibnamefont
  {Segre}},\ }\bibfield  {title} {\bibinfo {title} {Wettability stabilizes
  fluid invasion into porous media via nonlocal, cooperative pore filling},\
  }\href {https://doi.org/10.1103/PhysRevLett.115.164501} {\bibfield  {journal}
  {\bibinfo  {journal} {Physical Review Letters}\ }\textbf {\bibinfo {volume}
  {115}},\ \bibinfo {pages} {164501} (\bibinfo {year} {2015})}\BibitemShut
  {NoStop}%
\bibitem [{\citenamefont {Auradou}\ \emph {et~al.}(1999)\citenamefont
  {Auradou}, \citenamefont {M{\aa}l{\o}y}, \citenamefont {Schmittbuhl},
  \citenamefont {Hansen},\ and\ \citenamefont
  {Bideau}}]{auradou1999competition}%
  \BibitemOpen
  \bibfield  {author} {\bibinfo {author} {\bibfnamefont {H.}~\bibnamefont
  {Auradou}}, \bibinfo {author} {\bibfnamefont {K.~J.}\ \bibnamefont
  {M{\aa}l{\o}y}}, \bibinfo {author} {\bibfnamefont {J.}~\bibnamefont
  {Schmittbuhl}}, \bibinfo {author} {\bibfnamefont {A.}~\bibnamefont
  {Hansen}},\ and\ \bibinfo {author} {\bibfnamefont {D.}~\bibnamefont
  {Bideau}},\ }\bibfield  {title} {\bibinfo {title} {Competition between
  correlated buoyancy and uncorrelated capillary effects during drainage},\
  }\href {https://doi.org/10.1103/PhysRevE.60.7224} {\bibfield  {journal}
  {\bibinfo  {journal} {Physical Review E}\ }\textbf {\bibinfo {volume} {60}},\
  \bibinfo {pages} {7224} (\bibinfo {year} {1999})}\BibitemShut {NoStop}%
\bibitem [{\citenamefont {Ayaz}\ \emph {et~al.}(2020)\citenamefont {Ayaz},
  \citenamefont {Toussaint}, \citenamefont {Sch{\"a}fer},\ and\ \citenamefont
  {M{\aa}l{\o}y}}]{ayaz2020gravitational}%
  \BibitemOpen
  \bibfield  {author} {\bibinfo {author} {\bibfnamefont {M.}~\bibnamefont
  {Ayaz}}, \bibinfo {author} {\bibfnamefont {R.}~\bibnamefont {Toussaint}},
  \bibinfo {author} {\bibfnamefont {G.}~\bibnamefont {Sch{\"a}fer}},\ and\
  \bibinfo {author} {\bibfnamefont {K.~J.}\ \bibnamefont {M{\aa}l{\o}y}},\
  }\bibfield  {title} {\bibinfo {title} {Gravitational and finite-size effects
  on pressure saturation curves during drainage},\ }\href
  {https://doi.org/10.1029/2019WR026279} {\bibfield  {journal} {\bibinfo
  {journal} {Water Resources Research}\ }\textbf {\bibinfo {volume} {56}},\
  \bibinfo {pages} {e2019WR026279} (\bibinfo {year} {2020})}\BibitemShut
  {NoStop}%
\bibitem [{\citenamefont {Vincent-Dospital}\ \emph {et~al.}(2022)\citenamefont
  {Vincent-Dospital}, \citenamefont {Moura}, \citenamefont {Toussaint},\ and\
  \citenamefont {Måløy}}]{vincent-dospital2022}%
  \BibitemOpen
  \bibfield  {author} {\bibinfo {author} {\bibfnamefont {T.}~\bibnamefont
  {Vincent-Dospital}}, \bibinfo {author} {\bibfnamefont {M.}~\bibnamefont
  {Moura}}, \bibinfo {author} {\bibfnamefont {R.}~\bibnamefont {Toussaint}},\
  and\ \bibinfo {author} {\bibfnamefont {K.~J.}\ \bibnamefont {Måløy}},\
  }\bibfield  {title} {\bibinfo {title} {Stable and unstable capillary
  fingering in porous media with a gradient in grain size},\ }\href
  {https://doi.org/10.1038/s42005-022-01072-1} {\bibfield  {journal} {\bibinfo
  {journal} {Communications Physics}\ }\textbf {\bibinfo {volume} {5}},\
  \bibinfo {pages} {306} (\bibinfo {year} {2022})}\BibitemShut {NoStop}%
\bibitem [{\citenamefont {Roth}\ \emph {et~al.}(2021)\citenamefont {Roth},
  \citenamefont {Mays}, \citenamefont {Neupauer}, \citenamefont {Sather},\ and\
  \citenamefont {Crimaldi}}]{roth2021methods}%
  \BibitemOpen
  \bibfield  {author} {\bibinfo {author} {\bibfnamefont {E.~J.}\ \bibnamefont
  {Roth}}, \bibinfo {author} {\bibfnamefont {D.~C.}\ \bibnamefont {Mays}},
  \bibinfo {author} {\bibfnamefont {R.~M.}\ \bibnamefont {Neupauer}}, \bibinfo
  {author} {\bibfnamefont {L.~J.}\ \bibnamefont {Sather}},\ and\ \bibinfo
  {author} {\bibfnamefont {J.~P.}\ \bibnamefont {Crimaldi}},\ }\bibfield
  {title} {\bibinfo {title} {Methods for laser-induced fluorescence imaging of
  solute plumes at the {Darcy} scale in quasi-two-dimensional, refractive
  index-matched porous media},\ }\href
  {https://doi.org/10.1007/s11242-021-01545-x} {\bibfield  {journal} {\bibinfo
  {journal} {Transport in Porous Media}\ }\textbf {\bibinfo {volume} {136}},\
  \bibinfo {pages} {879} (\bibinfo {year} {2021})}\BibitemShut {NoStop}%
\bibitem [{\citenamefont {Berg}\ \emph {et~al.}(2013)\citenamefont {Berg},
  \citenamefont {Ott}, \citenamefont {Klapp}, \citenamefont {Schwing},
  \citenamefont {Neiteler}, \citenamefont {Brussee}, \citenamefont {Makurat},
  \citenamefont {Leu}, \citenamefont {Enzmann}, \citenamefont {Schwarz},
  \citenamefont {Kersten}, \citenamefont {Irvine},\ and\ \citenamefont
  {Stampanoni}}]{berg2013_complete}%
  \BibitemOpen
  \bibfield  {author} {\bibinfo {author} {\bibfnamefont {S.}~\bibnamefont
  {Berg}}, \bibinfo {author} {\bibfnamefont {H.}~\bibnamefont {Ott}}, \bibinfo
  {author} {\bibfnamefont {S.~A.}\ \bibnamefont {Klapp}}, \bibinfo {author}
  {\bibfnamefont {A.}~\bibnamefont {Schwing}}, \bibinfo {author} {\bibfnamefont
  {R.}~\bibnamefont {Neiteler}}, \bibinfo {author} {\bibfnamefont
  {N.}~\bibnamefont {Brussee}}, \bibinfo {author} {\bibfnamefont
  {A.}~\bibnamefont {Makurat}}, \bibinfo {author} {\bibfnamefont
  {L.}~\bibnamefont {Leu}}, \bibinfo {author} {\bibfnamefont {F.}~\bibnamefont
  {Enzmann}}, \bibinfo {author} {\bibfnamefont {J.-O.}\ \bibnamefont
  {Schwarz}}, \bibinfo {author} {\bibfnamefont {M.}~\bibnamefont {Kersten}},
  \bibinfo {author} {\bibfnamefont {S.}~\bibnamefont {Irvine}},\ and\ \bibinfo
  {author} {\bibfnamefont {M.}~\bibnamefont {Stampanoni}},\ }\bibfield  {title}
  {\bibinfo {title} {Real-time {3D} imaging of {Haines} jumps in porous media
  flow},\ }\href {https://doi.org/10.1073/pnas.1221373110} {\bibfield
  {journal} {\bibinfo  {journal} {Proceedings of the National Academy of
  Sciences}\ }\textbf {\bibinfo {volume} {110}},\ \bibinfo {pages} {3755}
  (\bibinfo {year} {2013})}\BibitemShut {NoStop}%
\bibitem [{\citenamefont {Tekseth}\ \emph {et~al.}(2024)\citenamefont
  {Tekseth}, \citenamefont {Mirzaei}, \citenamefont {Lukic}, \citenamefont
  {Chattopadhyay},\ and\ \citenamefont {Breiby}}]{tekseth2024}%
  \BibitemOpen
  \bibfield  {author} {\bibinfo {author} {\bibfnamefont {K.~R.}\ \bibnamefont
  {Tekseth}}, \bibinfo {author} {\bibfnamefont {F.}~\bibnamefont {Mirzaei}},
  \bibinfo {author} {\bibfnamefont {B.}~\bibnamefont {Lukic}}, \bibinfo
  {author} {\bibfnamefont {B.}~\bibnamefont {Chattopadhyay}},\ and\ \bibinfo
  {author} {\bibfnamefont {D.~W.}\ \bibnamefont {Breiby}},\ }\bibfield  {title}
  {\bibinfo {title} {Multiscale drainage dynamics with {Haines} jumps monitored
  by stroboscopic {4D} {X-ray} microscopy},\ }\href
  {https://doi.org/10.1073/pnas.2305890120} {\bibfield  {journal} {\bibinfo
  {journal} {Proceedings of the National Academy of Sciences}\ }\textbf
  {\bibinfo {volume} {121}},\ \bibinfo {pages} {e2305890120} (\bibinfo {year}
  {2024})}\BibitemShut {NoStop}%
\bibitem [{\citenamefont {Allen}\ \emph {et~al.}(1997)\citenamefont {Allen},
  \citenamefont {Stephenson},\ and\ \citenamefont
  {Strange}}]{allen1997morphology}%
  \BibitemOpen
  \bibfield  {author} {\bibinfo {author} {\bibfnamefont {S.}~\bibnamefont
  {Allen}}, \bibinfo {author} {\bibfnamefont {P.}~\bibnamefont {Stephenson}},\
  and\ \bibinfo {author} {\bibfnamefont {J.~H.}\ \bibnamefont {Strange}},\
  }\bibfield  {title} {\bibinfo {title} {Morphology of porous media studied by
  nuclear magnetic resonance},\ }\href {https://doi.org/10.1063/1.473780}
  {\bibfield  {journal} {\bibinfo  {journal} {Journal of Chemical Physics}\
  }\textbf {\bibinfo {volume} {106}},\ \bibinfo {pages} {7802} (\bibinfo {year}
  {1997})}\BibitemShut {NoStop}%
\bibitem [{\citenamefont {Yan}\ \emph {et~al.}(2012)\citenamefont {Yan},
  \citenamefont {Luo}, \citenamefont {Wang}, \citenamefont {Toussaint},
  \citenamefont {Schmittbuhl}, \citenamefont {Vasseur}, \citenamefont {Chen},
  \citenamefont {Yu},\ and\ \citenamefont {Zhang}}]{yan2012experimental}%
  \BibitemOpen
  \bibfield  {author} {\bibinfo {author} {\bibfnamefont {J.}~\bibnamefont
  {Yan}}, \bibinfo {author} {\bibfnamefont {X.}~\bibnamefont {Luo}}, \bibinfo
  {author} {\bibfnamefont {W.}~\bibnamefont {Wang}}, \bibinfo {author}
  {\bibfnamefont {R.}~\bibnamefont {Toussaint}}, \bibinfo {author}
  {\bibfnamefont {J.}~\bibnamefont {Schmittbuhl}}, \bibinfo {author}
  {\bibfnamefont {G.}~\bibnamefont {Vasseur}}, \bibinfo {author} {\bibfnamefont
  {F.}~\bibnamefont {Chen}}, \bibinfo {author} {\bibfnamefont {A.}~\bibnamefont
  {Yu}},\ and\ \bibinfo {author} {\bibfnamefont {L.}~\bibnamefont {Zhang}},\
  }\bibfield  {title} {\bibinfo {title} {An experimental study of secondary oil
  migration in a three-dimensional tilted porous medium},\ }\href
  {https://doi.org/10.1306/09091110140} {\bibfield  {journal} {\bibinfo
  {journal} {AAPG bulletin}\ }\textbf {\bibinfo {volume} {96}},\ \bibinfo
  {pages} {773} (\bibinfo {year} {2012})}\BibitemShut {NoStop}%
\bibitem [{\citenamefont {Souzy}\ \emph {et~al.}(2020)\citenamefont {Souzy},
  \citenamefont {Lhuissier}, \citenamefont {Méheust}, \citenamefont {{Le
  Borgne}},\ and\ \citenamefont {Metzger}}]{souzy2020}%
  \BibitemOpen
  \bibfield  {author} {\bibinfo {author} {\bibfnamefont {M.}~\bibnamefont
  {Souzy}}, \bibinfo {author} {\bibfnamefont {H.}~\bibnamefont {Lhuissier}},
  \bibinfo {author} {\bibfnamefont {Y.}~\bibnamefont {Méheust}}, \bibinfo
  {author} {\bibfnamefont {T.}~\bibnamefont {{Le Borgne}}},\ and\ \bibinfo
  {author} {\bibfnamefont {B.}~\bibnamefont {Metzger}},\ }\bibfield  {title}
  {\bibinfo {title} {Velocity distributions, dispersion and stretching in
  three-dimensional porous media},\ }\href
  {https://doi.org/10.1017/jfm.2020.113} {\bibfield  {journal} {\bibinfo
  {journal} {Journal of Fluid Mechanics}\ }\textbf {\bibinfo {volume} {891}},\
  \bibinfo {pages} {A16} (\bibinfo {year} {2020})}\BibitemShut {NoStop}%
\bibitem [{\citenamefont {Heyman}\ \emph {et~al.}(2020)\citenamefont {Heyman},
  \citenamefont {Lester}, \citenamefont {Turuban}, \citenamefont
  {M{\'{e}}heust},\ and\ \citenamefont {{Le Borgne}}}]{heyman2020}%
  \BibitemOpen
  \bibfield  {author} {\bibinfo {author} {\bibfnamefont {J.}~\bibnamefont
  {Heyman}}, \bibinfo {author} {\bibfnamefont {D.~R.}\ \bibnamefont {Lester}},
  \bibinfo {author} {\bibfnamefont {R.}~\bibnamefont {Turuban}}, \bibinfo
  {author} {\bibfnamefont {Y.}~\bibnamefont {M{\'{e}}heust}},\ and\ \bibinfo
  {author} {\bibfnamefont {T.}~\bibnamefont {{Le Borgne}}},\ }\bibfield
  {title} {\bibinfo {title} {Stretching and folding sustain microscale chemical
  gradients in porous media},\ }\href {https://doi.org/10.1073/pnas.2002858117}
  {\bibfield  {journal} {\bibinfo  {journal} {Proceedings of the National
  Academy of Sciences}\ }\textbf {\bibinfo {volume} {117}},\ \bibinfo {pages}
  {13359} (\bibinfo {year} {2020})}\BibitemShut {NoStop}%
\bibitem [{\citenamefont {Heyman}\ \emph {et~al.}(2021)\citenamefont {Heyman},
  \citenamefont {Lester},\ and\ \citenamefont {Le~Borgne}}]{heyman2021}%
  \BibitemOpen
  \bibfield  {author} {\bibinfo {author} {\bibfnamefont {J.}~\bibnamefont
  {Heyman}}, \bibinfo {author} {\bibfnamefont {D.~R.}\ \bibnamefont {Lester}},\
  and\ \bibinfo {author} {\bibfnamefont {T.}~\bibnamefont {Le~Borgne}},\
  }\bibfield  {title} {\bibinfo {title} {Scalar signatures of chaotic mixing in
  porous media},\ }\href {https://doi.org/10.1103/PhysRevLett.126.034505}
  {\bibfield  {journal} {\bibinfo  {journal} {Physical Review Letters}\
  }\textbf {\bibinfo {volume} {126}},\ \bibinfo {pages} {034505} (\bibinfo
  {year} {2021})}\BibitemShut {NoStop}%
\bibitem [{\citenamefont {Taylor}(1950)}]{taylor1950}%
  \BibitemOpen
  \bibfield  {author} {\bibinfo {author} {\bibfnamefont {G.~I.}\ \bibnamefont
  {Taylor}},\ }\bibfield  {title} {\bibinfo {title} {The instability of liquid
  surfaces when accelerated in a direction perpendicular to their planes},\
  }\href {https://doi.org/10.1098/rspa.1950.0052} {\bibfield  {journal}
  {\bibinfo  {journal} {Proceedings of the Royal Society of London. Series A.
  Mathematical and Physical Sciences}\ }\textbf {\bibinfo {volume} {201}},\
  \bibinfo {pages} {192} (\bibinfo {year} {1950})}\BibitemShut {NoStop}%
\bibitem [{\citenamefont {Wilkinson}(1984)}]{wilkinson1984percolation}%
  \BibitemOpen
  \bibfield  {author} {\bibinfo {author} {\bibfnamefont {D.}~\bibnamefont
  {Wilkinson}},\ }\bibfield  {title} {\bibinfo {title} {Percolation model of
  immiscible displacement in the presence of buoyancy forces},\ }\href
  {https://doi.org/10.1103/PhysRevA.30.520} {\bibfield  {journal} {\bibinfo
  {journal} {Physical Review A}\ }\textbf {\bibinfo {volume} {30}},\ \bibinfo
  {pages} {520} (\bibinfo {year} {1984})}\BibitemShut {NoStop}%
\bibitem [{\citenamefont {Breen}\ \emph {et~al.}(2022)\citenamefont {Breen},
  \citenamefont {Pride}, \citenamefont {Masson},\ and\ \citenamefont
  {Manga}}]{breen2022}%
  \BibitemOpen
  \bibfield  {author} {\bibinfo {author} {\bibfnamefont {S.~J.}\ \bibnamefont
  {Breen}}, \bibinfo {author} {\bibfnamefont {S.~R.}\ \bibnamefont {Pride}},
  \bibinfo {author} {\bibfnamefont {Y.}~\bibnamefont {Masson}},\ and\ \bibinfo
  {author} {\bibfnamefont {M.}~\bibnamefont {Manga}},\ }\bibfield  {title}
  {\bibinfo {title} {Stable drainage in a gravity field},\ }\href
  {https://doi.org/10.1016/j.advwatres.2022.104150} {\bibfield  {journal}
  {\bibinfo  {journal} {Advances in Water Resources}\ }\textbf {\bibinfo
  {volume} {162}},\ \bibinfo {pages} {104150} (\bibinfo {year}
  {2022})}\BibitemShut {NoStop}%
\bibitem [{sup()}]{supMat}%
  \BibitemOpen
  \href@noop {} {}\bibinfo {note} {See Supplemental Material at [URL will be
  inserted by publisher] for experiment videos and brief summary of methods and
  results}\BibitemShut {NoStop}%
\bibitem [{\citenamefont {Zhao}\ \emph {et~al.}(2016)\citenamefont {Zhao},
  \citenamefont {MacMinn},\ and\ \citenamefont {Juanes}}]{zhao2016}%
  \BibitemOpen
  \bibfield  {author} {\bibinfo {author} {\bibfnamefont {B.}~\bibnamefont
  {Zhao}}, \bibinfo {author} {\bibfnamefont {C.~W.}\ \bibnamefont {MacMinn}},\
  and\ \bibinfo {author} {\bibfnamefont {R.}~\bibnamefont {Juanes}},\
  }\bibfield  {title} {\bibinfo {title} {Wettability control on multiphase flow
  in patterned microfluidics},\ }\href
  {https://doi.org/10.1073/pnas.1603387113} {\bibfield  {journal} {\bibinfo
  {journal} {Proceedings of the National Academy of Sciences}\ }\textbf
  {\bibinfo {volume} {113}},\ \bibinfo {pages} {10251} (\bibinfo {year}
  {2016})}\BibitemShut {NoStop}%
\bibitem [{\citenamefont {Bradski}(2000)}]{bradski2000opencv}%
  \BibitemOpen
  \bibfield  {author} {\bibinfo {author} {\bibfnamefont {G.}~\bibnamefont
  {Bradski}},\ }\bibfield  {title} {\bibinfo {title} {The {OpenCV} library},\
  }\href {https://doi.org/10.4236/sgre.2011.23030} {\bibfield  {journal}
  {\bibinfo  {journal} {Dr. Dobb's Journal: Software Tools for the Professional
  Programmer}\ }\textbf {\bibinfo {volume} {25}},\ \bibinfo {pages} {120}
  (\bibinfo {year} {2000})}\BibitemShut {NoStop}%
\bibitem [{\citenamefont {Dey}(2018)}]{dey2018hands}%
  \BibitemOpen
  \bibfield  {author} {\bibinfo {author} {\bibfnamefont {S.}~\bibnamefont
  {Dey}},\ }\href
  {https://www.packtpub.com/en-us/product/hands-on-image-processing-with-python-9781789343731}
  {\emph {\bibinfo {title} {Hands-On Image Processing with Python: Expert
  Techniques for Advanced Image Analysis and Effective Interpretation of Image
  Data}}}\ (\bibinfo  {publisher} {Packt Publishing Ltd},\ \bibinfo {year}
  {2018})\BibitemShut {NoStop}%
\bibitem [{\citenamefont {Chakrapani}\ \emph {et~al.}(2003)\citenamefont
  {Chakrapani}, \citenamefont {Mitchell}, \citenamefont {Van~Winkle},\ and\
  \citenamefont {Rikvold}}]{chakrapani2003scaling}%
  \BibitemOpen
  \bibfield  {author} {\bibinfo {author} {\bibfnamefont {M.}~\bibnamefont
  {Chakrapani}}, \bibinfo {author} {\bibfnamefont {S.}~\bibnamefont
  {Mitchell}}, \bibinfo {author} {\bibfnamefont {D.}~\bibnamefont
  {Van~Winkle}},\ and\ \bibinfo {author} {\bibfnamefont {P.}~\bibnamefont
  {Rikvold}},\ }\bibfield  {title} {\bibinfo {title} {Scaling analysis of
  polyacrylamide gel surfaces synthesized in the presence of surfactants},\
  }\href {https://doi.org/10.1016/S0021-9797(02)00144-3} {\bibfield  {journal}
  {\bibinfo  {journal} {Journal of Colloid and Interface Science}\ }\textbf
  {\bibinfo {volume} {258}},\ \bibinfo {pages} {186} (\bibinfo {year}
  {2003})}\BibitemShut {NoStop}%
\bibitem [{\citenamefont {Feder}\ \emph {et~al.}(2022)\citenamefont {Feder},
  \citenamefont {Flekk{\o}y},\ and\ \citenamefont
  {Hansen}}]{feder_flekkoy_hansen_2022}%
  \BibitemOpen
  \bibfield  {author} {\bibinfo {author} {\bibfnamefont {J.}~\bibnamefont
  {Feder}}, \bibinfo {author} {\bibfnamefont {E.~G.}\ \bibnamefont
  {Flekk{\o}y}},\ and\ \bibinfo {author} {\bibfnamefont {A.}~\bibnamefont
  {Hansen}},\ }\href {https://doi.org/10.1017/9781009100717} {\emph {\bibinfo
  {title} {Physics of Flow in Porous Media}}}\ (\bibinfo  {publisher}
  {Cambridge University Press},\ \bibinfo {year} {2022})\BibitemShut {NoStop}%
\bibitem [{\citenamefont {Tallakstad}\ \emph
  {et~al.}(2009{\natexlab{a}})\citenamefont {Tallakstad}, \citenamefont
  {Knudsen}, \citenamefont {Ramstad}, \citenamefont {L{\o}voll}, \citenamefont
  {M{\aa}l{\o}y}, \citenamefont {Toussaint},\ and\ \citenamefont
  {Flekk{\o}y}}]{tallakstad2009steady}%
  \BibitemOpen
  \bibfield  {author} {\bibinfo {author} {\bibfnamefont {K.~T.}\ \bibnamefont
  {Tallakstad}}, \bibinfo {author} {\bibfnamefont {H.~A.}\ \bibnamefont
  {Knudsen}}, \bibinfo {author} {\bibfnamefont {T.}~\bibnamefont {Ramstad}},
  \bibinfo {author} {\bibfnamefont {G.}~\bibnamefont {L{\o}voll}}, \bibinfo
  {author} {\bibfnamefont {K.~J.}\ \bibnamefont {M{\aa}l{\o}y}}, \bibinfo
  {author} {\bibfnamefont {R.}~\bibnamefont {Toussaint}},\ and\ \bibinfo
  {author} {\bibfnamefont {E.~G.}\ \bibnamefont {Flekk{\o}y}},\ }\bibfield
  {title} {\bibinfo {title} {Steady-state two-phase flow in porous media:
  Statistics and transport properties},\ }\href
  {https://doi.org/10.1103/PhysRevLett.102.074502} {\bibfield  {journal}
  {\bibinfo  {journal} {Physical Review Letters}\ }\textbf {\bibinfo {volume}
  {102}},\ \bibinfo {pages} {074502} (\bibinfo {year}
  {2009}{\natexlab{a}})}\BibitemShut {NoStop}%
\bibitem [{\citenamefont {Tallakstad}\ \emph
  {et~al.}(2009{\natexlab{b}})\citenamefont {Tallakstad}, \citenamefont
  {L{\o}voll}, \citenamefont {Knudsen}, \citenamefont {Ramstad}, \citenamefont
  {Flekk{\o}y},\ and\ \citenamefont {M{\aa}l{\o}y}}]{tallakstad2009sim}%
  \BibitemOpen
  \bibfield  {author} {\bibinfo {author} {\bibfnamefont {K.~T.}\ \bibnamefont
  {Tallakstad}}, \bibinfo {author} {\bibfnamefont {G.}~\bibnamefont
  {L{\o}voll}}, \bibinfo {author} {\bibfnamefont {H.~A.}\ \bibnamefont
  {Knudsen}}, \bibinfo {author} {\bibfnamefont {T.}~\bibnamefont {Ramstad}},
  \bibinfo {author} {\bibfnamefont {E.~G.}\ \bibnamefont {Flekk{\o}y}},\ and\
  \bibinfo {author} {\bibfnamefont {K.~J.}\ \bibnamefont {M{\aa}l{\o}y}},\
  }\bibfield  {title} {\bibinfo {title} {Steady-state, simultaneous two-phase
  flow in porous media: An experimental study},\ }\href
  {https://doi.org/10.1103/PhysRevE.80.036308} {\bibfield  {journal} {\bibinfo
  {journal} {Physical Review E}\ }\textbf {\bibinfo {volume} {80}},\ \bibinfo
  {pages} {036308} (\bibinfo {year} {2009}{\natexlab{b}})}\BibitemShut
  {NoStop}%
\bibitem [{\citenamefont {Chuoke}\ \emph {et~al.}(1959)\citenamefont {Chuoke},
  \citenamefont {Van~Meurs},\ and\ \citenamefont {van~der
  Poel}}]{chuoke1959instability}%
  \BibitemOpen
  \bibfield  {author} {\bibinfo {author} {\bibfnamefont {R.}~\bibnamefont
  {Chuoke}}, \bibinfo {author} {\bibfnamefont {P.}~\bibnamefont {Van~Meurs}},\
  and\ \bibinfo {author} {\bibfnamefont {C.}~\bibnamefont {van~der Poel}},\
  }\bibfield  {title} {\bibinfo {title} {The instability of slow, immiscible,
  viscous liquid-liquid displacements in permeable media},\ }\href
  {https://doi.org/10.2118/1141-G} {\bibfield  {journal} {\bibinfo  {journal}
  {Transactions of the AIME}\ }\textbf {\bibinfo {volume} {216}},\ \bibinfo
  {pages} {188} (\bibinfo {year} {1959})}\BibitemShut {NoStop}%
\bibitem [{\citenamefont {Moura}\ \emph {et~al.}(2020)\citenamefont {Moura},
  \citenamefont {M{\aa}l{\o}y}, \citenamefont {Flekk{\o}y},\ and\ \citenamefont
  {Toussaint}}]{moura2020}%
  \BibitemOpen
  \bibfield  {author} {\bibinfo {author} {\bibfnamefont {M.}~\bibnamefont
  {Moura}}, \bibinfo {author} {\bibfnamefont {K.~J.}\ \bibnamefont
  {M{\aa}l{\o}y}}, \bibinfo {author} {\bibfnamefont {E.~G.}\ \bibnamefont
  {Flekk{\o}y}},\ and\ \bibinfo {author} {\bibfnamefont {R.}~\bibnamefont
  {Toussaint}},\ }\bibfield  {title} {\bibinfo {title} {Intermittent dynamics
  of slow drainage experiments in porous media: Characterization under
  different boundary conditions},\ }\href
  {https://doi.org/10.3389/fphy.2019.00217} {\bibfield  {journal} {\bibinfo
  {journal} {Frontiers in Physics}\ }\textbf {\bibinfo {volume} {7}},\ \bibinfo
  {pages} {217} (\bibinfo {year} {2020})}\BibitemShut {NoStop}%
\bibitem [{\citenamefont {Reis}\ \emph {et~al.}(2023)\citenamefont {Reis},
  \citenamefont {Moura}, \citenamefont {Linga}, \citenamefont {Rikvold},
  \citenamefont {Toussaint}, \citenamefont {Flekkøy},\ and\ \citenamefont
  {Måløy}}]{reis2023}%
  \BibitemOpen
  \bibfield  {author} {\bibinfo {author} {\bibfnamefont {P.}~\bibnamefont
  {Reis}}, \bibinfo {author} {\bibfnamefont {M.}~\bibnamefont {Moura}},
  \bibinfo {author} {\bibfnamefont {G.}~\bibnamefont {Linga}}, \bibinfo
  {author} {\bibfnamefont {P.~A.}\ \bibnamefont {Rikvold}}, \bibinfo {author}
  {\bibfnamefont {R.}~\bibnamefont {Toussaint}}, \bibinfo {author}
  {\bibfnamefont {E.~G.}\ \bibnamefont {Flekkøy}},\ and\ \bibinfo {author}
  {\bibfnamefont {K.~J.}\ \bibnamefont {Måløy}},\ }\bibfield  {title}
  {\bibinfo {title} {A simplified pore-scale model for slow drainage including
  film-flow effects},\ }\href
  {https://doi.org/https://doi.org/10.1016/j.advwatres.2023.104580} {\bibfield
  {journal} {\bibinfo  {journal} {Advances in Water Resources}\ }\textbf
  {\bibinfo {volume} {182}},\ \bibinfo {pages} {104580} (\bibinfo {year}
  {2023})}\BibitemShut {NoStop}%
\bibitem [{\citenamefont {Khobaib}\ \emph {et~al.}(2025)\citenamefont
  {Khobaib}, \citenamefont {Reis}, \citenamefont {Moura}, \citenamefont
  {Toussaint}, \citenamefont {Flekk\o{}y},\ and\ \citenamefont
  {M\aa{}l\o{}y}}]{khobaib2025}%
  \BibitemOpen
  \bibfield  {author} {\bibinfo {author} {\bibfnamefont {K.}~\bibnamefont
  {Khobaib}}, \bibinfo {author} {\bibfnamefont {P.}~\bibnamefont {Reis}},
  \bibinfo {author} {\bibfnamefont {M.}~\bibnamefont {Moura}}, \bibinfo
  {author} {\bibfnamefont {R.}~\bibnamefont {Toussaint}}, \bibinfo {author}
  {\bibfnamefont {E.~G.}\ \bibnamefont {Flekk\o{}y}},\ and\ \bibinfo {author}
  {\bibfnamefont {K.~J.}\ \bibnamefont {M\aa{}l\o{}y}},\ }\bibfield  {title}
  {\bibinfo {title} {Gravity stabilized drainage in porous media with
  controlled disorder},\ }\href
  {https://doi.org/10.1103/PhysRevResearch.7.023040} {\bibfield  {journal}
  {\bibinfo  {journal} {Phys. Rev. Res.}\ }\textbf {\bibinfo {volume} {7}},\
  \bibinfo {pages} {023040} (\bibinfo {year} {2025})}\BibitemShut {NoStop}%
\bibitem [{\citenamefont {Ioannidis}\ \emph {et~al.}(1996)\citenamefont
  {Ioannidis}, \citenamefont {Chatzis},\ and\ \citenamefont
  {Dullien}}]{ionnidis1996}%
  \BibitemOpen
  \bibfield  {author} {\bibinfo {author} {\bibfnamefont {M.~A.}\ \bibnamefont
  {Ioannidis}}, \bibinfo {author} {\bibfnamefont {I.}~\bibnamefont {Chatzis}},\
  and\ \bibinfo {author} {\bibfnamefont {F.~A.~L.}\ \bibnamefont {Dullien}},\
  }\bibfield  {title} {\bibinfo {title} {Macroscopic percolation model of
  immiscible displacement: Effects of buoyancy and spatial structure},\ }\href
  {https://doi.org/10.1029/95WR02216} {\bibfield  {journal} {\bibinfo
  {journal} {Water Resources Research}\ }\textbf {\bibinfo {volume} {32}},\
  \bibinfo {pages} {3297} (\bibinfo {year} {1996})}\BibitemShut {NoStop}%
\bibitem [{\citenamefont {Chen}\ and\ \citenamefont
  {Wilkinson}(1985)}]{chen1985pore}%
  \BibitemOpen
  \bibfield  {author} {\bibinfo {author} {\bibfnamefont {J.-D.}\ \bibnamefont
  {Chen}}\ and\ \bibinfo {author} {\bibfnamefont {D.}~\bibnamefont
  {Wilkinson}},\ }\bibfield  {title} {\bibinfo {title} {Pore-scale viscous
  fingering in porous media},\ }\href
  {https://doi.org/10.1103/PhysRevLett.55.1892} {\bibfield  {journal} {\bibinfo
   {journal} {Physical Review Letters}\ }\textbf {\bibinfo {volume} {55}},\
  \bibinfo {pages} {1892} (\bibinfo {year} {1985})}\BibitemShut {NoStop}%
\bibitem [{\citenamefont {Lenormand}(1990)}]{lenormand1990liquids}%
  \BibitemOpen
  \bibfield  {author} {\bibinfo {author} {\bibfnamefont {R.}~\bibnamefont
  {Lenormand}},\ }\bibfield  {title} {\bibinfo {title} {Liquids in porous
  media},\ }\href {https://doi.org/10.1088/0953-8984/2/S/008} {\bibfield
  {journal} {\bibinfo  {journal} {Journal of Physics: Condensed Matter}\
  }\textbf {\bibinfo {volume} {2}},\ \bibinfo {pages} {SA79} (\bibinfo {year}
  {1990})}\BibitemShut {NoStop}%
\bibitem [{\citenamefont {Sinha}\ and\ \citenamefont
  {Hansen}(2012)}]{sinha2012effective}%
  \BibitemOpen
  \bibfield  {author} {\bibinfo {author} {\bibfnamefont {S.}~\bibnamefont
  {Sinha}}\ and\ \bibinfo {author} {\bibfnamefont {A.}~\bibnamefont {Hansen}},\
  }\bibfield  {title} {\bibinfo {title} {Effective rheology of immiscible
  two-phase flow in porous media},\ }\href
  {https://doi.org/10.1209/0295-5075/99/44004} {\bibfield  {journal} {\bibinfo
  {journal} {Europhysics Letters}\ }\textbf {\bibinfo {volume} {99}},\ \bibinfo
  {pages} {44004} (\bibinfo {year} {2012})}\BibitemShut {NoStop}%
\bibitem [{\citenamefont {Gr{\o}va}\ and\ \citenamefont
  {Hansen}(2011)}]{grova2011two}%
  \BibitemOpen
  \bibfield  {author} {\bibinfo {author} {\bibfnamefont {M.}~\bibnamefont
  {Gr{\o}va}}\ and\ \bibinfo {author} {\bibfnamefont {A.}~\bibnamefont
  {Hansen}},\ }\bibfield  {title} {\bibinfo {title} {Two-phase flow in porous
  media: Power-law scaling of effective permeability},\ }\href
  {https://doi.org/10.1088/1742-6596/319/1/012009} {\bibfield  {journal}
  {\bibinfo  {journal} {Journal of Physics: Conference Series}\ }\textbf
  {\bibinfo {volume} {319}},\ \bibinfo {pages} {012009} (\bibinfo {year}
  {2011})}\BibitemShut {NoStop}%
\bibitem [{\citenamefont {Anastasiou}\ \emph {et~al.}(2024)\citenamefont
  {Anastasiou}, \citenamefont {Zarikos}, \citenamefont {Yiotis}, \citenamefont
  {Talon},\ and\ \citenamefont {Salin}}]{anastasiou2024steady}%
  \BibitemOpen
  \bibfield  {author} {\bibinfo {author} {\bibfnamefont {A.}~\bibnamefont
  {Anastasiou}}, \bibinfo {author} {\bibfnamefont {I.}~\bibnamefont {Zarikos}},
  \bibinfo {author} {\bibfnamefont {A.}~\bibnamefont {Yiotis}}, \bibinfo
  {author} {\bibfnamefont {L.}~\bibnamefont {Talon}},\ and\ \bibinfo {author}
  {\bibfnamefont {D.}~\bibnamefont {Salin}},\ }\bibfield  {title} {\bibinfo
  {title} {Steady-state dynamics of ganglia populations during immiscible
  two-phase flows in porous micromodels: Effects of the capillary number and
  flow ratio on effective rheology and size distributions},\ }\href
  {https://doi.org/10.1007/s11242-023-02041-0} {\bibfield  {journal} {\bibinfo
  {journal} {Transport in Porous Media}\ }\textbf {\bibinfo {volume} {151}},\
  \bibinfo {pages} {469} (\bibinfo {year} {2024})}\BibitemShut {NoStop}%
\bibitem [{\citenamefont {Zhang}\ \emph {et~al.}(2021)\citenamefont {Zhang},
  \citenamefont {Bijeljic}, \citenamefont {Gao}, \citenamefont {Lin},\ and\
  \citenamefont {Blunt}}]{zhang2021quantification}%
  \BibitemOpen
  \bibfield  {author} {\bibinfo {author} {\bibfnamefont {Y.}~\bibnamefont
  {Zhang}}, \bibinfo {author} {\bibfnamefont {B.}~\bibnamefont {Bijeljic}},
  \bibinfo {author} {\bibfnamefont {Y.}~\bibnamefont {Gao}}, \bibinfo {author}
  {\bibfnamefont {Q.}~\bibnamefont {Lin}},\ and\ \bibinfo {author}
  {\bibfnamefont {M.~J.}\ \bibnamefont {Blunt}},\ }\bibfield  {title} {\bibinfo
  {title} {Quantification of nonlinear multiphase flow in porous media},\
  }\href {https://doi.org/10.1029/2020GL090477} {\bibfield  {journal} {\bibinfo
   {journal} {Geophysical Research Letters}\ }\textbf {\bibinfo {volume}
  {48}},\ \bibinfo {pages} {e2020GL090477} (\bibinfo {year}
  {2021})}\BibitemShut {NoStop}%
\bibitem [{\citenamefont {Torquato}\ \emph {et~al.}(2000)\citenamefont
  {Torquato}, \citenamefont {Truskett},\ and\ \citenamefont
  {Debenedetti}}]{torquato2000}%
  \BibitemOpen
  \bibfield  {author} {\bibinfo {author} {\bibfnamefont {S.}~\bibnamefont
  {Torquato}}, \bibinfo {author} {\bibfnamefont {T.~M.}\ \bibnamefont
  {Truskett}},\ and\ \bibinfo {author} {\bibfnamefont {P.~G.}\ \bibnamefont
  {Debenedetti}},\ }\bibfield  {title} {\bibinfo {title} {Is random close
  packing of spheres well defined?},\ }\href
  {https://doi.org/10.1103/physrevlett.84.2064} {\bibfield  {journal} {\bibinfo
   {journal} {Physical Review Letters}\ }\textbf {\bibinfo {volume} {84}},\
  \bibinfo {pages} {2064} (\bibinfo {year} {2000})}\BibitemShut {NoStop}%
\bibitem [{\citenamefont {Stukowski}(2010)}]{ovito}%
  \BibitemOpen
  \bibfield  {author} {\bibinfo {author} {\bibfnamefont {A.}~\bibnamefont
  {Stukowski}},\ }\bibfield  {title} {\bibinfo {title} {Visualization and
  analysis of atomistic simulation data with {OVITO}-- the open visualization
  tool},\ }\href {https://doi.org/10.1088/0965-0393/18/1/015012} {\bibfield
  {journal} {\bibinfo  {journal} {Modeling and Simulation in Natural Sciences
  and Engineering}\ }\textbf {\bibinfo {volume} {18}} (\bibinfo {year}
  {2010})}\BibitemShut {NoStop}%
\bibitem [{\citenamefont {Honeycutt}\ and\ \citenamefont
  {Andersen}(1987)}]{honeycutt1987}%
  \BibitemOpen
  \bibfield  {author} {\bibinfo {author} {\bibfnamefont {J.~D.}\ \bibnamefont
  {Honeycutt}}\ and\ \bibinfo {author} {\bibfnamefont {H.~C.}\ \bibnamefont
  {Andersen}},\ }\bibfield  {title} {\bibinfo {title} {Molecular dynamics study
  of melting and freezing of small {Lennard-Jones} clusters},\ }\href
  {https://doi.org/10.1021/j100303a014} {\bibfield  {journal} {\bibinfo
  {journal} {Journal of Physical Chemistry}\ }\textbf {\bibinfo {volume}
  {91}},\ \bibinfo {pages} {4950} (\bibinfo {year} {1987})}\BibitemShut
  {NoStop}%
\bibitem [{\citenamefont {Stukowski}(2012)}]{stukowski2012}%
  \BibitemOpen
  \bibfield  {author} {\bibinfo {author} {\bibfnamefont {A.}~\bibnamefont
  {Stukowski}},\ }\bibfield  {title} {\bibinfo {title} {Structure
  identification methods for atomistic simulations of crystalline materials},\
  }\href {https://doi.org/10.1088/0965-0393/20/4/045021} {\bibfield  {journal}
  {\bibinfo  {journal} {Modelling and Simulation in Materials Science and
  Engineering}\ }\textbf {\bibinfo {volume} {20}},\ \bibinfo {pages} {045021}
  (\bibinfo {year} {2012})}\BibitemShut {NoStop}%
\bibitem [{\citenamefont {Bennett}(1972)}]{bennett1972}%
  \BibitemOpen
  \bibfield  {author} {\bibinfo {author} {\bibfnamefont {C.~H.}\ \bibnamefont
  {Bennett}},\ }\bibfield  {title} {\bibinfo {title} {{Serially deposited
  amorphous aggregates of hard spheres}},\ }\href
  {https://doi.org/10.1063/1.1661585} {\bibfield  {journal} {\bibinfo
  {journal} {Journal of Applied Physics}\ }\textbf {\bibinfo {volume} {43}},\
  \bibinfo {pages} {2727} (\bibinfo {year} {1972})}\BibitemShut {NoStop}%
\bibitem [{\citenamefont {Biswal}\ \emph {et~al.}(1998)\citenamefont {Biswal},
  \citenamefont {Manwart},\ and\ \citenamefont {Hilfer}}]{biswal1998}%
  \BibitemOpen
  \bibfield  {author} {\bibinfo {author} {\bibfnamefont {B.}~\bibnamefont
  {Biswal}}, \bibinfo {author} {\bibfnamefont {C.}~\bibnamefont {Manwart}},\
  and\ \bibinfo {author} {\bibfnamefont {R.}~\bibnamefont {Hilfer}},\
  }\bibfield  {title} {\bibinfo {title} {Three-dimensional local porosity
  analysis of porous media},\ }\href
  {https://doi.org/10.1016/S0378-4371(98)00111-3} {\bibfield  {journal}
  {\bibinfo  {journal} {Physica A: Statistical Mechanics and its Applications}\
  }\textbf {\bibinfo {volume} {255}},\ \bibinfo {pages} {221} (\bibinfo {year}
  {1998})}\BibitemShut {NoStop}%
\bibitem [{\citenamefont {Hilfer}(1992)}]{hilfer1992}%
  \BibitemOpen
  \bibfield  {author} {\bibinfo {author} {\bibfnamefont {R.}~\bibnamefont
  {Hilfer}},\ }\bibfield  {title} {\bibinfo {title} {Local-porosity theory for
  flow in porous media},\ }\href {https://doi.org/10.1103/PhysRevB.45.7115}
  {\bibfield  {journal} {\bibinfo  {journal} {Physical Review B}\ }\textbf
  {\bibinfo {volume} {45}},\ \bibinfo {pages} {7115} (\bibinfo {year}
  {1992})}\BibitemShut {NoStop}%
\bibitem [{\citenamefont {Torquato}(2002)}]{torquato2002}%
  \BibitemOpen
  \bibfield  {author} {\bibinfo {author} {\bibfnamefont {S.}~\bibnamefont
  {Torquato}},\ }\href {https://doi.org/10.1007/978-1-4757-6355-3} {\emph
  {\bibinfo {title} {Random Heterogeneous Materials}}}\ (\bibinfo  {publisher}
  {Springer},\ \bibinfo {year} {2002})\BibitemShut {NoStop}%
\bibitem [{\citenamefont {Yurchenko}\ \emph {et~al.}(2015)\citenamefont
  {Yurchenko}, \citenamefont {Kryuchkov},\ and\ \citenamefont
  {Ivlev}}]{yurchenko2015}%
  \BibitemOpen
  \bibfield  {author} {\bibinfo {author} {\bibfnamefont {S.~O.}\ \bibnamefont
  {Yurchenko}}, \bibinfo {author} {\bibfnamefont {N.~P.}\ \bibnamefont
  {Kryuchkov}},\ and\ \bibinfo {author} {\bibfnamefont {A.~V.}\ \bibnamefont
  {Ivlev}},\ }\bibfield  {title} {\bibinfo {title} {Pair correlations in
  classical crystals: The shortest-graph method},\ }\href
  {https://doi.org/10.1063/1.4926945} {\bibfield  {journal} {\bibinfo
  {journal} {Journal of Chemical Physics}\ }\textbf {\bibinfo {volume} {143}},\
  \bibinfo {pages} {034506} (\bibinfo {year} {2015})}\BibitemShut {NoStop}%
\bibitem [{\citenamefont {Bryant}\ and\ \citenamefont
  {Johnson}(2003)}]{bryant2003wetting}%
  \BibitemOpen
  \bibfield  {author} {\bibinfo {author} {\bibfnamefont {S.~L.}\ \bibnamefont
  {Bryant}}\ and\ \bibinfo {author} {\bibfnamefont {A.}~\bibnamefont
  {Johnson}},\ }\bibfield  {title} {\bibinfo {title} {Wetting phase
  connectivity and irreducible saturation in simple granular media},\ }\href
  {https://doi.org/10.1016/S0021-9797(03)00371-0} {\bibfield  {journal}
  {\bibinfo  {journal} {Journal of Colloid and Interface Science}\ }\textbf
  {\bibinfo {volume} {263}},\ \bibinfo {pages} {572} (\bibinfo {year}
  {2003})}\BibitemShut {NoStop}%
\bibitem [{\citenamefont {Bryant}\ and\ \citenamefont
  {Johnson}(2004)}]{bryant2004bulk}%
  \BibitemOpen
  \bibfield  {author} {\bibinfo {author} {\bibfnamefont {S.~L.}\ \bibnamefont
  {Bryant}}\ and\ \bibinfo {author} {\bibfnamefont {A.}~\bibnamefont
  {Johnson}},\ }\bibfield  {title} {\bibinfo {title} {Bulk and film
  contributions to fluid/fluid interfacial area in granular media},\ }\href
  {https://doi.org/10.1080/00986440490472742} {\bibfield  {journal} {\bibinfo
  {journal} {Chemical Engineering Communications}\ }\textbf {\bibinfo {volume}
  {191}},\ \bibinfo {pages} {1660} (\bibinfo {year} {2004})}\BibitemShut
  {NoStop}%
\bibitem [{\citenamefont {Hoogland}\ \emph {et~al.}(2016)\citenamefont
  {Hoogland}, \citenamefont {Lehmann}, \citenamefont {Mokso},\ and\
  \citenamefont {Or}}]{hoogland2016drainage}%
  \BibitemOpen
  \bibfield  {author} {\bibinfo {author} {\bibfnamefont {F.}~\bibnamefont
  {Hoogland}}, \bibinfo {author} {\bibfnamefont {P.}~\bibnamefont {Lehmann}},
  \bibinfo {author} {\bibfnamefont {R.}~\bibnamefont {Mokso}},\ and\ \bibinfo
  {author} {\bibfnamefont {D.}~\bibnamefont {Or}},\ }\bibfield  {title}
  {\bibinfo {title} {Drainage mechanisms in porous media: From piston-like
  invasion to formation of corner flow networks},\ }\href
  {https://doi.org/10.1002/2016WR019299} {\bibfield  {journal} {\bibinfo
  {journal} {Water Resources Research}\ }\textbf {\bibinfo {volume} {52}},\
  \bibinfo {pages} {8413} (\bibinfo {year} {2016})}\BibitemShut {NoStop}%
\bibitem [{\citenamefont {Moura}\ \emph {et~al.}(2019)\citenamefont {Moura},
  \citenamefont {Flekk{\o}y}, \citenamefont {M{\aa}l{\o}y}, \citenamefont
  {Sch{\"a}fer},\ and\ \citenamefont {Toussaint}}]{moura2019connectivity}%
  \BibitemOpen
  \bibfield  {author} {\bibinfo {author} {\bibfnamefont {M.}~\bibnamefont
  {Moura}}, \bibinfo {author} {\bibfnamefont {E.~G.}\ \bibnamefont
  {Flekk{\o}y}}, \bibinfo {author} {\bibfnamefont {K.~J.}\ \bibnamefont
  {M{\aa}l{\o}y}}, \bibinfo {author} {\bibfnamefont {G.}~\bibnamefont
  {Sch{\"a}fer}},\ and\ \bibinfo {author} {\bibfnamefont {R.}~\bibnamefont
  {Toussaint}},\ }\bibfield  {title} {\bibinfo {title} {Connectivity
  enhancement due to film flow in porous media},\ }\href
  {https://doi.org/10.1103/PhysRevFluids.4.094102} {\bibfield  {journal}
  {\bibinfo  {journal} {Physical Review Fluids}\ }\textbf {\bibinfo {volume}
  {4}},\ \bibinfo {pages} {094102} (\bibinfo {year} {2019})}\BibitemShut
  {NoStop}%
\bibitem [{\citenamefont {Gouyet}\ \emph {et~al.}(1988)\citenamefont {Gouyet},
  \citenamefont {Rosso},\ and\ \citenamefont {Sapoval}}]{gouyet1988fractal}%
  \BibitemOpen
  \bibfield  {author} {\bibinfo {author} {\bibfnamefont {J.-F.}\ \bibnamefont
  {Gouyet}}, \bibinfo {author} {\bibfnamefont {M.}~\bibnamefont {Rosso}},\ and\
  \bibinfo {author} {\bibfnamefont {B.}~\bibnamefont {Sapoval}},\ }\bibfield
  {title} {\bibinfo {title} {Fractal structure of diffusion and invasion fronts
  in three-dimensional lattices through the gradient percolation approach},\
  }\href {https://doi.org/10.1103/PhysRevB.37.1832} {\bibfield  {journal}
  {\bibinfo  {journal} {Physical Review B}\ }\textbf {\bibinfo {volume} {37}},\
  \bibinfo {pages} {1832} (\bibinfo {year} {1988})}\BibitemShut {NoStop}%
\bibitem [{\citenamefont {Clément}\ \emph {et~al.}(1985)\citenamefont
  {Clément}, \citenamefont {Baudet},\ and\ \citenamefont
  {Hulin}}]{clement1985}%
  \BibitemOpen
  \bibfield  {author} {\bibinfo {author} {\bibfnamefont {E.}~\bibnamefont
  {Clément}}, \bibinfo {author} {\bibfnamefont {C.}~\bibnamefont {Baudet}},\
  and\ \bibinfo {author} {\bibfnamefont {J.~P.}\ \bibnamefont {Hulin}},\
  }\bibfield  {title} {\bibinfo {title} {Multiple scale structure of
  non-wetting fluid invasion fronts in {3D} model porous media},\ }\href
  {https://doi.org/10.1051/jphyslet:0198500460240116300} {\bibfield  {journal}
  {\bibinfo  {journal} {Journal de Physique Lettres}\ }\textbf {\bibinfo
  {volume} {46}},\ \bibinfo {pages} {1163} (\bibinfo {year}
  {1985})}\BibitemShut {NoStop}%
\bibitem [{\citenamefont {Hulin}\ \emph {et~al.}(1988)\citenamefont {Hulin},
  \citenamefont {Cl\'ement}, \citenamefont {Baudet}, \citenamefont {Gouyet},\
  and\ \citenamefont {Rosso}}]{hulin1988}%
  \BibitemOpen
  \bibfield  {author} {\bibinfo {author} {\bibfnamefont {J.~P.}\ \bibnamefont
  {Hulin}}, \bibinfo {author} {\bibfnamefont {E.}~\bibnamefont {Cl\'ement}},
  \bibinfo {author} {\bibfnamefont {C.}~\bibnamefont {Baudet}}, \bibinfo
  {author} {\bibfnamefont {J.~F.}\ \bibnamefont {Gouyet}},\ and\ \bibinfo
  {author} {\bibfnamefont {M.}~\bibnamefont {Rosso}},\ }\bibfield  {title}
  {\bibinfo {title} {Quantitative analysis of an invading-fluid invasion front
  under gravity},\ }\href {https://doi.org/10.1103/PhysRevLett.61.333}
  {\bibfield  {journal} {\bibinfo  {journal} {Physical Review Letters}\
  }\textbf {\bibinfo {volume} {61}},\ \bibinfo {pages} {333} (\bibinfo {year}
  {1988})}\BibitemShut {NoStop}%
\bibitem [{\citenamefont {Chaouche}\ \emph {et~al.}(1994)\citenamefont
  {Chaouche}, \citenamefont {Rakotomalala}, \citenamefont {Salin},
  \citenamefont {Xu},\ and\ \citenamefont {Yortsos}}]{chaouche1994invasion}%
  \BibitemOpen
  \bibfield  {author} {\bibinfo {author} {\bibfnamefont {M.}~\bibnamefont
  {Chaouche}}, \bibinfo {author} {\bibfnamefont {N.}~\bibnamefont
  {Rakotomalala}}, \bibinfo {author} {\bibfnamefont {D.}~\bibnamefont {Salin}},
  \bibinfo {author} {\bibfnamefont {B.}~\bibnamefont {Xu}},\ and\ \bibinfo
  {author} {\bibfnamefont {Y.}~\bibnamefont {Yortsos}},\ }\bibfield  {title}
  {\bibinfo {title} {Invasion percolation in a hydrostatic or permeability
  gradient: Experiments and simulations},\ }\href
  {https://doi.org/10.1103/PhysRevE.49.4133} {\bibfield  {journal} {\bibinfo
  {journal} {Physical Review E}\ }\textbf {\bibinfo {volume} {49}},\ \bibinfo
  {pages} {4133} (\bibinfo {year} {1994})}\BibitemShut {NoStop}%
\end{thebibliography}

%

\end{document}